\documentclass[12pt]{amsart}

\usepackage{comment}
\usepackage{biblatex}
\usepackage{graphicx}
\usepackage[export]{adjustbox}
\usepackage{caption}
\usepackage{amssymb,amsmath}
\usepackage[normalem]{ulem}
\usepackage{enumitem} 
\usepackage{listings} 
\usepackage{listings}
\usepackage{xcolor}
\definecolor{maccolor}{rgb}{0.3,0.3,0.8}
\lstdefinelanguage{Macaulay2}
{
basicstyle={\ttfamily},
keywordstyle={\color{maccolor!80!black}},
commentstyle={\color{gray}},
stringstyle={\color{red!40!black}},
rulecolor=\color{maccolor},
basewidth={1.2ex}, 
sensitive=false,
morecomment=[l]{--},
morecomment=[s]{-*}{*-},
morestring=[b]",
escapechar={`},
escapebegin={\rmfamily},
morekeywords={about,abs,AbstractToricVarieties,accumulate,Acknowledgement,acos,acosh,acot,addCancelTask,addDependencyTask,addEndFunction,addHook,AdditionalPaths,addStartFunction,addStartTask,Adjacent,adjoint,AdjointIdeal,AffineVariety,AfterEval,AfterNoPrint,AfterPrint,agm,AInfinity,alarm,AlgebraicSplines,Algorithm,Alignment,all,AllCodimensions,allowableThreads,ambient,analyticSpread,Analyzer,AnalyzeSheafOnP1,ancestor,ancestors,ANCHOR,and,andP,AngleBarList,ann,annihilator,antipode,any,append,applicationDirectory,applicationDirectorySuffix,apply,applyKeys,applyPairs,applyTable,applyValues,apropos,argument,Array,arXiv,Ascending,ascii,asin,asinh,ass,assert,associatedGradedRing,associatedPrimes,AssociativeAlgebras,AssociativeExpression,atan,atan2,atEndOfFile,Authors,autoload,AuxiliaryFiles,backtrace,Bag,Bareiss,baseFilename,BaseFunction,baseName,baseRing,baseRings,BaseRow,BasicList,basis,BasisElementLimit,Bayer,BeforePrint,beginDocumentation,BeginningMacaulay2,Benchmark,benchmark,Bertini,BesselJ,BesselY,betti,BettiCharacters,BettiTally,between,BGG,BIBasis,Binary,BinaryOperation,Binomial,binomial,BinomialEdgeIdeals,Binomials,BKZ,BlockMatrix,BLOCKQUOTE,BODY,Body,BoijSoederberg,BOLD,Book3264Examples,Boolean,BooleanGB,borel,Boxes,BR,break,Browse,Bruns,cache,CacheExampleOutput,CacheFunction,CacheTable,cacheValue,CallLimit,cancelTask,capture,catch,Caveat,CC,CDATA,ceiling,Center,centerString,Certification,ChainComplex,chainComplex,ChainComplexExtras,ChainComplexMap,ChainComplexOperations,ChangeMatrix,char,CharacteristicClasses,characters,charAnalyzer,check,CheckDocumentation,chi,Chordal,class,Classic,clean,clearAll,clearEcho,clearOutput,close,closeIn,closeOut,ClosestFit,CODE,code,codim,CodimensionLimit,coefficient,CoefficientRing,coefficientRing,coefficients,Cofactor,CohenEngine,CohenTopLevel,CoherentSheaf,CohomCalg,cohomology,coimage,CoincidentRootLoci,coker,cokernel,collectGarbage,columnAdd,columnate,columnMult,columnPermute,columnRankProfile,columnSwap,combine,Command,commandInterpreter,commandLine,COMMENT,commonest,commonRing,comodule,CompactMatrix,compactMatrixForm,CompiledFunction,CompiledFunctionBody,CompiledFunctionClosure,Complement,complement,complete,CompleteIntersection,CompleteIntersectionResolutions,Complexes,ComplexField,components,compose,compositions,compress,concatenate,conductor,ConductorElement,cone,Configuration,ConformalBlocks,conjugate,connectionCount,Consequences,Constant,Constants,constParser,content,continue,contract,Contributors,ConvexInterface,conwayPolynomial,ConwayPolynomials,copy,copyDirectory,copyFile,copyright,Core,CorrespondenceScrolls,cos,cosh,cot,CotangentSchubert,cotangentSheaf,coth,cover,coverMap,cpuTime,createTask,Cremona,csc,csch,current,currentColumnNumber,currentDirectory,currentFileDirectory,currentFileName,currentLayout,currentLineNumber,currentPackage,currentString,currentTime,Cyclotomic,Database,Date,DD,dd,deadParser,debug,debugError,DebuggingMode,debuggingMode,debugLevel,DecomposableSparseSystems,Decompose,decompose,deepSplice,Default,default,defaultPrecision,Degree,degree,degreeLength,DegreeLift,DegreeLimit,DegreeMap,DegreeOrder,DegreeRank,Degrees,degrees,degreesMonoid,degreesRing,delete,demark,denominator,Dense,Density,Depth,depth,Descending,Descent,Describe,describe,Description,det,determinant,DeterminantalRepresentations,DGAlgebras,diagonalMatrix,diameter,Dictionary,dictionary,dictionaryPath,diff,DiffAlg,difference,dim,directSum,disassemble,discriminant,dismiss,Dispatch,distinguished,DIV,Divide,divideByVariable,DivideConquer,DividedPowers,Divisor,DL,Dmodules,do,doc,docExample,docTemplate,document,DocumentTag,Down,drop,DT,dual,eagonNorthcott,EagonResolution,echoOff,echoOn,EdgeIdeals,edit,EigenSolver,eigenvalues,eigenvectors,eint,EisenbudHunekeVasconcelos,elapsedTime,elapsedTiming,elements,Eliminate,eliminate,Elimination,EliminationMatrices,EllipticCurves,EllipticIntegrals,else,EM,Email,End,end,endl,endPackage,Engine,engineDebugLevel,EngineRing,EngineTests,entries,EnumerationCurves,environment,Equation,EquivariantGB,erase,erf,erfc,error,errorDepth,euler,EulerConstant,eulers,even,EXAMPLE,ExampleFiles,ExampleItem,examples,ExampleSystems,Exclude,exec,exit,exp,expectedReesIdeal,expm1,exponents,export,exportFrom,exportMutable,Expression,expression,Ext,extend,ExteriorIdeals,ExteriorModules,exteriorPower,Factor,factor,false,Fano,FastMinors,FastNonminimal,FGLM,File,fileDictionaries,fileExecutable,fileExists,fileExitHooks,fileLength,fileMode,FileName,FilePosition,fileReadable,fileTime,fileWritable,fillMatrix,findFiles,findHeft,FindOne,findProgram,findSynonyms,FiniteFittingIdeals,First,first,firstkey,FirstPackage,fittingIdeal,flagLookup,FlatMonoid,flatten,flattenRing,Flexible,flip,floor,flush,fold,FollowLinks,for,forceGB,fork,FormalGroupLaws,Format,format,formation,FourierMotzkin,FourTiTwo,fpLLL,frac,fraction,FractionField,frames,FrobeniusThresholds,from,fromDividedPowers,fromDual,Function,FunctionApplication,FunctionBody,functionBody,FunctionClosure,FunctionFieldDesingularization,fusePairs,futureParser,GaloisField,Gamma,gb,GBDegrees,gbRemove,gbSnapshot,gbTrace,gcd,gcdCoefficients,gcdLLL,GCstats,genera,GeneralOrderedMonoid,GenerateAssertions,generateAssertions,generator,generators,Generic,GenericInitialIdeal,genericMatrix,genericSkewMatrix,genericSymmetricMatrix,gens,genus,get,getc,getChangeMatrix,getenv,getGlobalSymbol,getNetFile,getNonUnit,getPrimeWithRootOfUnity,getSymbol,getWWW,GF,gfanInterface,Givens,GKMVarieties,GLex,Global,global,globalAssign,globalAssignFunction,GlobalAssignHook,globalAssignment,globalAssignmentHooks,GlobalDictionary,GlobalHookStore,globalReleaseFunction,GlobalReleaseHook,Gorenstein,GradedLieAlgebras,GradedModule,gradedModule,GradedModuleMap,gradedModuleMap,gramm,GraphicalModels,GraphicalModelsMLE,Graphics,graphIdeal,graphRing,Graphs,Grassmannian,GRevLex,GroebnerBasis,groebnerBasis,GroebnerBasisOptions,GroebnerStrata,GroebnerWalk,groupID,GroupLex,GroupRevLex,GTZ,Hadamard,handleInterrupts,HardDegreeLimit,hash,HashTable,hashTable,HEAD,HEADER1,HEADER2,HEADER3,HEADER4,HEADER5,HEADER6,HeaderType,Heading,Headline,Heft,heft,Height,height,help,Hermite,hermite,Hermitian,HH,hh,HigherCIOperators,HighestWeights,Hilbert,hilbertFunction,hilbertPolynomial,hilbertSeries,HodgeIntegrals,hold,Holder,Hom,homeDirectory,HomePage,Homogeneous,Homogeneous2,homogenize,homology,homomorphism,HomotopyLieAlgebra,hooks,horizontalJoin,HorizontalSpace,HR,HREF,HTML,html,httpHeaders,Hybrid,HyperplaneArrangements,Hypertext,hypertext,HypertextContainer,HypertextParagraph,icFracP,icFractions,icMap,icPIdeal,id,Ideal,ideal,idealizer,identity,if,IgnoreExampleErrors,ii,image,imaginaryPart,IMG,ImmutableType,importFrom,in,incomparable,Increment,independentSets,indeterminate,IndeterminateNumber,Index,index,indexComponents,IndexedVariable,IndexedVariableTable,indices,inducedMap,inducesWellDefinedMap,InexactField,InexactFieldFamily,InexactNumber,InfiniteNumber,infinity,info,InfoDirSection,infoHelp,Inhomogeneous,input,Inputs,insert,installAssignmentMethod,installedPackages,installHilbertFunction,installMethod,installMinprimes,installPackage,InstallPrefix,instance,instances,IntegralClosure,integralClosure,integrate,IntermediateMarkUpType,interpreterDepth,intersect,intersectInP,Intersection,intersection,interval,InvariantRing,inverse,InverseMethod,inversePermutation,Inverses,inverseSystem,InverseSystems,Invertible,InvolutiveBases,irreducibleCharacteristicSeries,irreducibleDecomposition,isAffineRing,isANumber,isBorel,isCanceled,isCommutative,isConstant,isDirectory,isDirectSum,isEmpty,isField,isFinite,isFinitePrimeField,isFreeModule,isGlobalSymbol,isHomogeneous,isIdeal,isInfinite,isInjective,isInputFile,isIsomorphism,isLinearType,isListener,isLLL,isMember,isModule,isMonomialIdeal,isNormal,isOpen,isOutputFile,isPolynomialRing,isPrimary,isPrime,isPrimitive,isPseudoprime,isQuotientModule,isQuotientOf,isQuotientRing,isReady,isReal,isReduction,isRegularFile,isRing,isSkewCommutative,isSorted,isSquareFree,isStandardGradedPolynomialRing,isSubmodule,isSubquotient,isSubset,isSupportedInZeroLocus,isSurjective,isTable,isUnit,isWellDefined,isWeylAlgebra,ITALIC,Iterate,Jacobian,jacobian,jacobianDual,Jets,Join,join,Jupyter,K3Carpets,K3Surfaces,Keep,KeepFiles,KeepZeroes,ker,kernel,kernelLLL,kernelOfLocalization,Key,keys,Keyword,Keywords,kill,koszul,Kronecker,KustinMiller,LABEL,last,lastMatch,LATER,LatticePolytopes,Layout,lcm,leadCoefficient,leadComponent,leadMonomial,leadTerm,Left,left,length,LengthLimit,letterParser,Lex,LexIdeals,LI,Licenses,LieTypes,lift,liftable,Limit,limitFiles,limitProcesses,Linear,LinearAlgebra,LinearTruncations,lineNumber,lines,LINK,linkFile,List,list,listForm,listLocalSymbols,listSymbols,listUserSymbols,LITERAL,LLL,LLLBases,lngamma,load,loadDepth,LoadDocumentation,loadedFiles,loadedPackages,loadPackage,Local,local,localDictionaries,LocalDictionary,localize,LocalRings,locate,log,log1p,LongPolynomial,lookup,lookupCount,LowerBound,LUdecomposition,M0nbar,M2CODE,Macaulay2Doc,makeDirectory,MakeDocumentation,makeDocumentTag,MakeHTML,MakeInfo,MakeLinks,makePackageIndex,MakePDF,makeS2,Manipulator,map,MapExpression,MapleInterface,markedGB,Markov,MarkUpType,match,mathML,Matrix,matrix,MatrixExpression,Matroids,max,maxAllowableThreads,maxExponent,MaximalRank,maxPosition,MaxReductionCount,MCMApproximations,member,memoize,memoizeClear,memoizeValues,MENU,merge,mergePairs,META,method,MethodFunction,MethodFunctionBinary,MethodFunctionSingle,MethodFunctionWithOptions,methodOptions,methods,midpoint,min,minExponent,mingens,mingle,minimalBetti,MinimalGenerators,MinimalMatrix,minimalPresentation,minimalPresentationMap,minimalPresentationMapInv,MinimalPrimes,minimalPrimes,minimalReduction,Minimize,minimizeFilename,MinimumVersion,minors,minPosition,minPres,minprimes,Minus,minus,Miura,MixedMultiplicity,mkdir,mod,Module,module,ModuleDeformations,modulo,MonodromySolver,Monoid,monoid,MonoidElement,Monomial,MonomialAlgebras,monomialCurveIdeal,MonomialIdeal,monomialIdeal,MonomialIntegerPrograms,MonomialOrbits,MonomialOrder,Monomials,monomials,MonomialSize,monomialSubideal,moveFile,multidegree,multidoc,multigraded,MultigradedBettiTally,MultiGradedRationalMap,multiplicity,MultiplicitySequence,MultiplierIdeals,MultiplierIdealsDim2,MultiprojectiveVarieties,mutable,MutableHashTable,mutableIdentity,MutableList,MutableMatrix,mutableMatrix,NAGtypes,Name,nanosleep,Nauty,NautyGraphs,NCAlgebra,NCLex,needs,needsPackage,Net,net,NetFile,netList,new,newClass,newCoordinateSystem,NewFromMethod,newline,NewMethod,newNetFile,NewOfFromMethod,NewOfMethod,newPackage,newRing,nextkey,nextPrime,nil,NNParser,NoetherianOperators,NoetherNormalization,NonminimalComplexes,nonspaceAnalyzer,NoPrint,norm,normalCone,Normaliz,NormalToricVarieties,not,Nothing,notify,notImplemented,NTL,null,nullaryMethods,nullhomotopy,nullParser,nullSpace,Number,number,NumberedVerticalList,numcols,numColumns,numerator,numeric,NumericalAlgebraicGeometry,NumericalCertification,NumericalImplicitization,NumericalLinearAlgebra,NumericalSchubertCalculus,numericInterval,NumericSolutions,numgens,numRows,numrows,odd,oeis,of,ofClass,OL,OldPolyhedra,OldToricVectorBundles,on,OneExpression,OnlineLookup,OO,oo,ooo,oooo,openDatabase,openDatabaseOut,openFiles,openIn,openInOut,openListener,OpenMath,openOut,openOutAppend,operatorAttributes,Option,OptionalComponentsPresent,optionalSignParser,Options,options,OptionTable,optP,or,Order,order,OrderedMonoid,orP,OutputDictionary,Outputs,override,pack,Package,package,PackageCitations,PackageDictionary,PackageExports,PackageImports,PackageTemplate,packageTemplate,pad,pager,PairLimit,pairs,PairsRemaining,PARA,Parametrization,parent,Parenthesize,Parser,Parsing,part,Partition,partition,partitions,parts,path,pdim,peek,PencilsOfQuadrics,Permanents,permanents,permutations,pfaffians,PHCpack,PhylogeneticTrees,pi,PieriMaps,pivots,PlaneCurveSingularities,plus,poincare,poincareN,Points,polarize,poly,Polyhedra,Polymake,PolynomialRing,Posets,Position,position,positions,PositivityToricBundles,POSIX,Postfix,Power,power,powermod,PRE,Precision,precision,Prefix,prefixDirectory,prefixPath,preimage,prepend,presentation,pretty,primaryComponent,PrimaryDecomposition,primaryDecomposition,PrimaryTag,PrimitiveElement,Print,print,printerr,printingAccuracy,printingLeadLimit,printingPrecision,printingSeparator,printingTimeLimit,printingTrailLimit,printString,printWidth,processID,Product,product,ProductOrder,profile,profileSummary,Program,programPaths,ProgramRun,Proj,Projective,ProjectiveHilbertPolynomial,projectiveHilbertPolynomial,ProjectiveVariety,promote,protect,Prune,prune,PruneComplex,pruningMap,Pseudocode,pseudocode,pseudoRemainder,Pullback,PushForward,pushForward,Python,QQ,QQParser,QRDecomposition,QthPower,Quasidegrees,QuaternaryQuartics,QuillenSuslin,quit,Quotient,quotient,quotientRemainder,QuotientRing,Radical,radical,RadicalCodim1,radicalContainment,RaiseError,random,RandomCanonicalCurves,RandomComplexes,RandomCurves,RandomCurvesOverVerySmallFiniteFields,RandomGenus14Curves,RandomIdeals,randomKRationalPoint,RandomMonomialIdeals,randomMutableMatrix,RandomObjects,RandomPlaneCurves,RandomPoints,RandomSpaceCurves,Range,rank,RationalMaps,RationalPoints,RationalPoints2,ReactionNetworks,read,readDirectory,readlink,readPackage,RealField,RealFP,realPart,realpath,RealQP,RealQP1,RealRoots,RealRR,RealXD,recursionDepth,recursionLimit,Reduce,reducedRowEchelonForm,reduceHilbert,reductionNumber,ReesAlgebra,reesAlgebra,reesAlgebraIdeal,reesIdeal,References,ReflexivePolytopesDB,regex,regexQuote,registerFinalizer,regSeqInIdeal,Regularity,regularity,relations,RelativeCanonicalResolution,relativizeFilename,Reload,remainder,RemakeAllDocumentation,remove,removeDirectory,removeFile,removeLowestDimension,reorganize,replace,RerunExamples,res,reshape,ResidualIntersections,ResLengthThree,Resolution,resolution,ResolutionsOfStanleyReisnerRings,restart,Result,resultant,Resultants,return,returnCode,Reverse,reverse,RevLex,Right,right,Ring,ring,RingElement,RingFamily,ringFromFractions,RingMap,rootPath,roots,rootURI,rotate,round,rowAdd,RowExpression,rowMult,rowPermute,rowRankProfile,rowSwap,RR,RRi,rsort,run,RunDirectory,RunExamples,RunExternalM2,runHooks,runLengthEncode,runProgram,same,saturate,Saturation,scan,scanKeys,scanLines,scanPairs,scanValues,schedule,schreyerOrder,Schubert,Schubert2,SchurComplexes,SchurFunctors,SchurRings,SCRIPT,scriptCommandLine,ScriptedFunctor,SCSCP,searchPath,sec,sech,SectionRing,SeeAlso,seeParsing,SegreClasses,select,selectInSubring,selectVariables,SelfInitializingType,SemidefiniteProgramming,Seminormalization,separate,SeparateExec,separateRegexp,Sequence,sequence,Serialization,serialNumber,Set,set,setEcho,setGroupID,setIOExclusive,setIOSynchronized,setIOUnSynchronized,setRandomSeed,setup,setupEmacs,sheaf,SheafExpression,sheafExt,sheafHom,SheafOfRings,shield,ShimoyamaYokoyama,short,show,showClassStructure,showHtml,showStructure,showTex,showUserStructure,SimpleDoc,simpleDocFrob,SimplicialComplexes,SimplicialDecomposability,SimplicialPosets,SimplifyFractions,sin,singularLocus,sinh,size,size2,SizeLimit,SkewCommutative,SlackIdeals,sleep,SLnEquivariantMatrices,SLPexpressions,SMALL,smithNormalForm,solve,someTerms,Sort,sort,sortColumns,SortStrategy,source,SourceCode,SourceRing,SPACE,SpaceCurves,SPAN,span,SparseMonomialVectorExpression,SparseResultants,SparseVectorExpression,Spec,SpechtModule,SpecialFanoFourfolds,specialFiber,specialFiberIdeal,SpectralSequences,splice,splitWWW,sqrt,SRdeformations,stack,stacksProject,Standard,standardForm,standardPairs,StartWithOneMinor,stashValue,StatePolytope,StatGraphs,status,stderr,stdio,step,StopBeforeComputation,stopIfError,StopWithMinimalGenerators,Strategy,String,STRONG,StronglyStableIdeals,STYLE,Style,style,SUB,sub,SubalgebraBases,sublists,submatrix,submatrixByDegrees,Subnodes,subquotient,SubringLimit,Subscript,subscript,SUBSECTION,subsets,substitute,substring,subtable,Sugarless,Sum,sum,SumOfTwists,SumsOfSquares,SUP,super,SuperLinearAlgebra,Superscript,superscript,support,SVD,SVDComplexes,switch,SwitchingFields,sylvesterMatrix,Symbol,symbol,SymbolBody,symbolBody,SymbolicPowers,symlinkDirectory,symlinkFile,symmetricAlgebra,symmetricAlgebraIdeal,symmetricKernel,SymmetricPolynomials,symmetricPower,synonym,SYNOPSIS,syz,Syzygies,SyzygyLimit,SyzygyMatrix,SyzygyRows,syzygyScheme,TABLE,Table,table,take,Tally,tally,tan,TangentCone,tangentCone,tangentSheaf,tanh,target,Task,taskResult,TateOnProducts,TD,temporaryFileName,tensor,tensorAssociativity,TensorComplexes,terminalParser,terms,TEST,Test,testExample,testHunekeQuestion,TestIdeals,TestInput,tests,TEX,tex,TeXmacs,texMath,Text,TH,then,Thing,ThinSincereQuivers,ThreadedGB,threadVariable,Threshold,throw,Time,time,times,timing,TITLE,TO,to,TO2,toAbsolutePath,toCC,toDividedPowers,toDual,toExternalString,toField,TOH,toList,toLower,top,top,topCoefficients,Topcom,topComponents,topLevelMode,Tor,TorAlgebra,Toric,ToricInvariants,ToricTopology,ToricVectorBundles,toRR,toRRi,toSequence,toString,TotalPairs,toUpper,TR,trace,transpose,TriangularSets,Tries,Trim,trim,Triplets,Tropical,true,Truncate,truncate,truncateOutput,Truncations,try,TSpreadIdeals,TT,tutorial,Type,TypicalValue,typicalValues,UL,ultimate,unbag,uncurry,Undo,undocumented,uniform,uninstallAllPackages,uninstallPackage,Unique,unique,Units,Unmixed,unsequence,unstack,Up,UpdateOnly,UpperTriangular,URL,urlEncode,Usage,use,UseCachedExampleOutput,UseHilbertFunction,UserMode,userSymbols,UseSyzygies,utf8,utf8check,validate,value,values,Variable,VariableBaseName,Variables,Variety,variety,vars,Vasconcelos,Vector,vector,VectorExpression,VectorFields,VectorGraphics,Verbose,Verbosity,Verify,VersalDeformations,versalEmbedding,Version,version,VerticalList,VerticalSpace,viewHelp,VirtualResolutions,VirtualTally,VisibleList,Visualize,wait,WebApp,wedgeProduct,weightRange,Weights,WeylAlgebra,WeylGroups,when,whichGm,while,width,wikipedia,Wrap,wrap,WrapperType,XML,xor,youngest,zero,ZeroExpression,zeta,ZZ,ZZParser}
}
\lstalias{Macaulay2output}{Macaulay2}

\definecolor{m2comment}{rgb}{0.4,0.6,0.4}
\definecolor{m2keyword}{rgb}{0.1,0.1,0.8}
\definecolor{m2function}{rgb}{0.6,0.2,0.2}

\lstdefinelanguage{Macaulay2}{
  morekeywords={ideal, map, kernel, mingens, netList, QQ, flatten, entries, matrix, minors, det, primaryDecomposition},
  sensitive=true,
  morecomment=[l]{--},
  morestring=[b]",
}

\definecolor{codebg}{rgb}{0.95,0.95,0.95}       
\definecolor{codeframe}{rgb}{0.8,0.8,0.8}       
\definecolor{codecomment}{rgb}{0.25,0.5,0.35}   
\definecolor{codekeyword}{rgb}{0.2,0.3,0.7}     
\definecolor{codestring}{rgb}{0.7,0.2,0.2}      
\definecolor{codeidentifier}{rgb}{0.2,0.2,0.2}  

\lstdefinestyle{fancym2}{
  language=Macaulay2,
  backgroundcolor=\color{codebg},
  basicstyle=\ttfamily\tiny\color{codeidentifier},
  commentstyle=\color{codecomment},
  keywordstyle=\bfseries\color{codekeyword},
  stringstyle=\color{codestring},
  frame=single,
  rulecolor=\color{codeframe},
  framerule=0.5pt,
  xleftmargin=1em,
  framexleftmargin=1em,
  framesep=0.5em,
  numbers=none,
  numberstyle=\tiny\color{blue},
  stepnumber=1,
  breaklines=true,
  breakatwhitespace=true,
  showstringspaces=false,
  tabsize=2,
  captionpos=b,
  columns=fixed,
}

\usepackage{fullpage}              
\usepackage{tikz}               
\usetikzlibrary{plotmarks}
\usetikzlibrary{shapes}
\usetikzlibrary{arrows}

\usepackage{hyperref}
\usepackage[nameinlink,capitalise]{cleveref}

\numberwithin{equation}{section}
\theoremstyle{plain}

\theoremstyle{definition}

\usepackage{xspace}
\newcommand{\cdd}{\hyperref[item:CDD]{CDD}\xspace}
\newcommand{\pdr}{\hyperref[item:PDR]{PDR}\xspace}
\newcommand{\pdm}{\hyperref[item:PDM]{PDM}\xspace}
\newcommand{\cdr}{\hyperref[item:CDR]{CDR}\xspace}
\newcommand{\cdm}{\hyperref[item:CDM]{CDM}\xspace}

\definecolor{RoyalBlue}{cmyk}{1, 0.50, 0, 0}

\def\R{\mathbb{R}} 
\def\E{\mathbb{E}} 
\def\P{\mathbb{P}} 

\definecolor{Purple}{rgb}{.5,0,.5}

\makeatletter
\renewcommand\subsubsection{\@secnumfont}{\bfseries}%
\renewcommand\subsubsection{\@startsection{subsubsection}{3}
\z@{.5\linespacing\@plus.7\linespacing}{-.5em}%
{\normalfont\bfseries\itshape}}
\makeatother

\begin{document}

\title{Methodological considerations for semialgebraic hypothesis testing with incomplete U-statistics} 
\author{}\thanks{All authors contributed to the initial conception and planning of this project and approved the final manuscript. MG-L and MH developed the final code for implementing the SDL method and conducting algebraic computations for the CFN model, building on preliminary code and exploratory work by several team members. DB and JZ conducted early simulations that informed the final simulations for Section 3, which were performed and summarized by JR. Section 4 simulations and explorations were performed by MG-L and MH. MG-L, EG, MH, BK, and JR collaboratively prepared the final manuscript.}

\author{David Barnhill}
\address{Department of Mathematics, United States Naval Academy}
\author{Marina Garrote-López}
\address{Department of Mathematics, KTH Royal Institute of Technology}
\author{Elizabeth Gross}
\address{Department of Mathematics, University of Hawai`i M\={a}noa}
\author{Max Hill}
\address{Department of Mathematics, University of Hawai`i M\={a}noa}
\author{Bryson Kagy}
\address{Department of Mathematics, North Carolina State University}
\author{John A. Rhodes}
\address{Department of Mathematics and Statistics, University of Alaska Fairbanks}
\author{Joy Z. Zhang}
\address{Center for Applied Mathematics, Cornell University}

\begin{abstract}
    Recently, Sturma, Drton, and Leung proposed a general-purpose stochastic method for hypothesis testing in models defined by polynomial equality and inequality constraints. Notably, the method remains theoretically valid even near irregular points, such as singularities and boundaries, where traditional testing approaches often break down. In this paper, we evaluate its practical performance on a collection of biologically motivated models from phylogenetics. While the method performs remarkably well across different settings, we catalogue a number of issues that should be considered for effective application.

\end{abstract}

\maketitle

\section{Introduction}
Statistical models are typically described by a map from a parameter space to a set of distributions. Often the parameter space $\Theta$ can be identified with a full-dimensional subset of $\mathbb R^d$ with submodels arising by restricting to a subset $\Theta_0\subset \Theta$. In many instances $\Theta_0$ is described by a set of polynomial equality and inequality constraints on $\mathbb R^d$, in which case we say the submodel is \emph{semialgebraic}. (An \emph{algebraic} model requires polynomial equality constraints only; the prefix \emph{semi-} allows for inequalities.) Semialgebraic models are common in statistics, encompassing many log-linear models \cite{Dobra2009}, latent class models \cite{AMR2009,GuXu2020}, discrete and Gaussian graphical models  \cite{LauritzenBook}, as well as phylogenetic models \cite{sullivant2018algebraic}. The underlying algebraic structure of semi-algebraic sets often yields valuable insights into model selection and inference \cite{boege2024colored, schkoda:2024, wang2019highdimensional, cai:2023}.


A semialgebraic set $\Theta_0$ may be geometrically quite complicated. Singularities can occur where the dimension of $\Theta_0$ collapses or it self-intersects. It may also have components of different dimensions, as well as boundaries. Such irregularities create difficulties for standard approaches to  hypothesis testing. For instance, a likelihood ratio test using a $\chi^2$ distribution
is only justified through approximating the model by a tangent space.
While some research has addressed such issues of model geometry \cite{drton2009likelihood,MAR2019,evans2020model}, it is common for empirical studies to simply ignore the the challenges irregularities pose due to the lack of available tools.

Recently,  Sturma, Drton, and Leung \cite{SDL2024}, building on previous work \cite{chen2018gaussian, ChenEtAl2019,SongEtAl2019}, proposed a general hypothesis testing procedure based on randomized incomplete U-statistics \cite{blom1976some, brown1978reduced, janson1984asymptotic} in order to overcome these problems. In addition to presenting the method and establishing its asymptotic behaviour, they provided a running example using the tetrad constraints of factor analysis and applied their method to a biological dataset, testing a semialgebraic Gaussian tree model. 

In this work, we investigate the practical performance of the Sturma, Drton, and Leung (SDL) method through several other models, drawn from evolutionary biology. In particular we study how implementation choices such as constraint specification, kernel order, and decomposition into reducible components affect test performance.  Our study offers practical insights for researchers applying the SDL method to semialgebraic models, particularly in biological settings where singularities are common.

 Our first example models come from phylogenomics---the inference of species relationships from genomic-scale sequence data. These models are used to test whether biological species relationships are sufficiently described by an evolutionary tree or whether more complex depictions involving hybridization or gene flow are needed. These 
 are semialgebraic submodels of the general trinomial model, allowing for 2-dimensional plotting of rejection regions, providing immediate visual insight into testing behaviour. 
 Although more traditional deterministic tests have been developed for such models (see \Cref{sec:dettests}), and we do not expect the  SDL methodology to supplant them, comparison with those methods allows for better judgment of SDL performance.
 
 We then consider the Cavender-Farris-Neyman (CFN) 2-state model of nucleotide substitution on a 4-taxon gene tree, a more complicated model in a higher dimensional space. After exploring the use of the SDL test for hypothesis testing when assuming a specific gene tree 
 topology, we then adapt the test to present 
 a novel inference procedure for topological gene trees. We emphasize that this procedure depends only on knowing semialgebraic descriptions of the models for different trees without performing any likelihood computation or optimization.
 
These examples allow us to examine not only the general applicability of the SDL test to biologically meaningful models, but also the practical implications of certain parameter choices that must be made in order to implement the method.
We explore the effects of user-specified options on statistical performance such as Type I and Type II errors.  We also investigate the stochasticity of the test under different parameter regimes. Since SDL $p$-values have some randomness due to the test procedure, 
it is desirable to limit their variation when possible. 
While \cite{SDL2024} suggests that the subsample size used in calculating the incomplete U-statistics should be small,
moderately increasing it can greatly reduce variation while still controlling error.

Another user choice examined here is the specific constraints defining the semialgebraic parameter space $\Theta_0$, as these are not uniquely determined. We show that constraint choice can have a significant effect on the test's rejection region, and that using a redundant set of constraints is often desirable. We offer one approach which automatically produces a redundant set of constraints through convex combinations, making the test less dependent on the initial constraint choice.
We also illustrate that redundant constraints not produced by our approach may be needed for better performance. A minimal set of constraints may lead to a highly conservative test, with performance improved by the introduction of valid but seemingly unrelated inequalities. 

In addition, the intrinsic geometry of the model also plays a role in unexpected ways. If a model can be decomposed into irreducible components, doing so and using an intersection-union framework with the SDL test on each component can increase the test's statistical power, as one of our examples shows.

Finally, the SDL test procedure depends on a kernel function that must be symmetrized, although this can be computationally prohibitive. However, we found that a partial symmetrization, applying surprisingly few random permutations, is a highly effective substitute and can give good performance.

We emphasize that we ultimately obtained excellent performance of the SDL method for all models we considered. However, we believe that naive use for a specific model of interest, without exploration of the issues we found, is unlikely to achieve the best performance possible. While we give no new theoretical results in this work, we advance awareness of potential pitfalls thereby guiding users to better application of the methodology.

 \medskip
 
 This article proceeds as follows. In \Cref{sec:Background}, we introduce relevant background and outline the methodology from \cite{SDL2024}.  In \Cref{sec:trinomial}, we introduce four basic submodels of the trinomial model, with details of their biological motivation deferred to \Cref{sec:Coalescent}.
 \Cref{sec:pitfalls} is the main section of the paper, presenting the issues and lessons learned through application of the hypothesis testing procedure to the four submodels. In \Cref{sec:CFN}, we apply the hypothesis test to the CFN model.

Our implementation of the test in \texttt{R} with the \texttt{Rcpp} package \cite{CRAN} is adapted from code used in the \texttt{TestGGM} package \cite{TESTGGM} shared by N. Sturma.  Our code is freely available on the GitHub repository \cite{ourGithub}.

\section{The SDL Test} 
\label{sec:Background}
We first outline the hypothesis testing methodology of \cite{SDL2024} for semialgebraic models, henceforth referred to as the \emph{SDL test}.

\subsection{Semialgebraic models and hypothesis testing}\label{ssec:hyptest}
A statistical model 
$$\mathcal{M}:= \left\{P_{\theta}: \theta\in \Theta\right\}$$ is \emph{semialgebraic} if its parameter space $\Theta$ is a semialgebraic subset of $\mathbb R^d$, i.e., a finite union of sets, referred to as \emph{basic} semialgebraic sets, defined by finitely many polynomial equalities and inequalities. 
 
Semialgebraic statistical models arise frequently in applications.
For example, the classical Hardy-Weinberg model for two alleles in equilibrium can be described by a single parameter $\theta\in (0,1)$, with a parametrization map
defined by $$\phi(\theta)=(\theta^2,\ 2\theta(1-\theta),\ (1-\theta)^2),$$
possibly composed with a multinomial map for multiple samples.
Alternatively, one may define the model by taking $\Theta$ to be the image of $\phi$ in the probability simplex $\Delta^2$. In this case, $\Theta$ is implicitly defined by the constraint $y^2-4xz=0$, together with the linear constraints  that define $\Delta^2$ (namely, $x,y,z\geq0$ and $x+y+z=1$), and is thus semialgebraic.

To set notation in a hypothesis testing framework, we consider a model with parameter space $\Theta\subseteq \mathbb R^d$ (which need not be semialgebraic) and a semialgebraic submodel with parameter space $\Theta_0\subset\Theta$. Following \cite{SDL2024}, we assume throughout $\Theta_0$ is a basic semialgebraic set.  Noting that an equality is equivalent to two inequalities, we assume
\begin{equation}\label{eq:definition-of-Theta0}
  \Theta_{0} 
  := \left\{\theta\in \R^d: f_{i}(\theta) \leq 0\text{ for all }i=1,\ldots,p\right\},
\end{equation}
where the $f_i$ are polynomials. Given data consisting of $n$ independent and identically distributed (i.i.d.)~samples, assuming
\begin{equation*}
  X_1,\ldots,X_n \sim P_{\theta}
\end{equation*} for some $\theta\in \Theta$, 
we define null and alternative hypotheses
\begin{equation}
  \label{eq:hypothesis-testing}
  H_0: \theta \in \Theta_0 \quad \text{and} \quad H_1: \theta \in \Theta\backslash\Theta_0.
\end{equation}

\subsection{Overview of the SDL test} 
The SDL test uses randomized incomplete U-statistics and a Gaussian multiplier bootstrap approximation of the test distribution to perform hypothesis testing in the setting described in \Cref{ssec:hyptest}. 
We outline the main objects and steps of the method, focusing on computations. For full justification, see \cite{SDL2024}. 

\subsubsection{The kernel function}
\label{sec:the-kernel-function}
The incomplete $U$-statistic is defined using a kernel function to coarsely approximate $f(\theta)$.
Let $f:\Theta\to \R^p$, 
  $f(\theta):=(f_1(\theta),\ldots,f_p(\theta)),$
where the $f_i$ are the constraint polynomials of \cref{eq:definition-of-Theta0}.
For some $m\ge 1$, let $h:\R^{m}\to \R^{p}$ be a \emph{kernel function}, i.e., a measurable symmetric function satisfying
$\E\left[h(X_{1},\ldots,X_{m}) \right] = f(\theta)$ for i.i.d.~$X_i\sim P_{\theta}$. \Cref{sec:kernel} gives details about the specific construction of such an $h$.

The quantity $m$---called the \emph{order} of the kernel---is a user-specified choice of a subsample size. Given a random subsample $X_{i_1},\ldots,X_{i_m}$ of the data, $h(X_{i_1},\ldots,X_{i_m})$ estimates  $f(\theta)$, though perhaps poorly if $m$ is small. The SDL method averages many such estimates to construct a better one: the randomized incomplete U-statistic.

\subsubsection{The incomplete U-statistic and the SDL test statistic}
\label{sec:the-incomplete-U-statistic}
Now that we have defined the kernel function, we can define the SDL test statistic. Let $I_{n,m}$ be the set of $m$-element subsets  of $[n]=\{1,2,\dots,n\}$, viewed as ordered $m$-tuples,
\begin{equation*}
  I_{n,m}:=\left\{ \left( i_{1},\ldots,i_{m} \right)\in \mathbb{Z}^m: 1 \leq i_{1}<\cdots<i_{m}\leq n\right\}.
\end{equation*} 
Choose a \emph{computational budget parameter} $N\le\binom nm$. 
For each $\iota\in I_{n,m}$ let $Z_{\iota}\sim \text{Bernoulli}(N/\binom{n}{m})$, and define $\widehat{N}:=\sum_{\iota \in I_{n,m}}Z_{\iota}$. The \emph{randomized incomplete
  U-statistic} is
\begin{equation}
  \label{eq:definition-of-randomized-incomplete-u-statistic}
  U'_{n,N} := \frac{1}{\widehat{N}}\sum_{\iota \in  I_{n,m}}Z_{\iota} h(X_{\iota}),
\end{equation}
where $X_{\iota}:=(X_{i_{1}},\ldots,X_{i_{m}})$ if $\iota=(i_{1},\ldots,i_{m})$.

The \emph{SDL test statistic}, $\mathcal{T}$, is the
maximum component of the studentization of  $U_{n,N}'$:
\begin{equation}
  \label{eq:def-of-test-statistic}
  \mathcal{T}:=\max_{1 \leq j \leq p} \frac{\sqrt{n}U'_{n,N,j}}{\widehat{\sigma}_{j}}, 
\end{equation}
where $\widehat{\sigma}_{j}^{2}$ is a stochastic approximation of $\sigma_{j}^{2}$, the variance of the $j$-th coordinate of $U'_{n,N}$ (see \Cref{sec:studentization} for details on the computation of $\widehat{\sigma}_{j}^{2}$).

\subsubsection{The critical threshold}
\label{sec:critical-threshold}
A large value of $\mathcal{T}$ is interpreted as evidence against $H_{0}$. More precisely, $\mathcal{T}$ is judged using an approximate distribution of a related statistic,
\begin{equation} \label{eq:def-of-T_c}
  \mathcal{T}_{\rm c} := \max_{1 \leq j \leq p}\frac{\sqrt{n}\left( U'_{n,N,j}-f_{j}(\theta) \right) }{\widehat{\sigma}_{j}}.
\end{equation}
Since $\E \left[U'_{n,N}\right]=f(\theta)$ for all $\theta\in\Theta_0$, $\mathcal{T}_{\rm c}$ differs from $\mathcal{T}$ only in centring. Moreover, since the functions $f_j$ are non-positive on the null model, 
$\mathcal{T}\leq \mathcal{T}_{\rm c}$ whenever $\theta\in \Theta_{0}$.
Thus, using the distribution of $\mathcal{T}_{\rm c}$ to assess $\mathcal T$ would yield a conservative test.
Although the exact distribution of $\mathcal{T}_{\rm c}$ is unknown, it can be approximated, as we describe next. 

Let $U^{\#}_{n,n_{1}}$ be the Gaussian multiplier bootstrap of ${\sqrt{n}\left( U'_{n,N}-f(\theta) \right) }$  presented in detail in the next section.  The bootstrap statistic $U^{\#}_{n,n_{1}}$
has two independent sources of randomness: 
(1) the collection $\mathcal{D}_{n} = \left\{X_{1},\ldots,X_{n}\right\}\cup \left\{ Z_{\iota}: \iota\in I_{n,m} \right\}$ and 
(2) a sample from  $\binom{n}{m}+n_1$ independent standard normal random variables
\begin{equation*}
   R= \left\{\xi_{\iota}': \iota\in I_{n,m}\right\} \cup \left\{\xi_{i_{1}}:i_{1}\in S_{1}\right\},
\end{equation*}
where $S_{1}$ is a pre-specified subset of
$[n]$ and $n_1=|S_1|$.  Now let
\begin{equation}
    \label{eq:def-of-W}
    W: = \max_{1 \leq j \leq p}\frac{U^{\#}_{n,n_{1},j}}{\widehat{\sigma}_{j}}.
\end{equation}
To estimate a $p$-value, we fix a large number $A$ (chosen by the user), and then generate a sequence of random variables $W^{(1)},\ldots,W^{(A)}$ by evaluating $W$ on each of $A$ independent copies of $R$. The resulting $p$-value estimate is 

\begin{equation*}
    \widehat{p} := \frac{\#\left\{i \in [A]: W^{(i)}\geq \mathcal{T}\right\}}{A}.
\end{equation*}

\subsubsection{The Gaussian bootstrap approximation.} \label{sec:GaussianBoots}

The above procedure for estimating $p$-values is justified by \cite[Corollary 2.10]{SDL2024}, which shows that, under technical assumptions, the conditional law of $W$ given $\mathcal{D}_n$ approximates $\mathcal{T}_{\rm c}$ for large $n$. As a consequence, the SDL test is asymptotically conservative \cite[Corollary 3.1]{SDL2024}. Nonetheless, it is important to understand how the approximation of $\mathcal{T}_{\rm c}$ depends on user-specified test parameters when $n$ is bounded, as this can affect the $p$-value distribution and hence the statistical properties of the SDL test in practice.

The approximation proceeds in two steps: first the quantity $\sqrt{n}\left( U'_{n,N}-f(\theta) \right)$  from \Cref{sec:critical-threshold} is approximated by a Gaussian random vector $Y$, and subsequently $Y$ is approximated by a Gaussian bootstrap $U^{\#}_{n,n_1}$ defined in this section. By \cite[Theorem 2.4]{SDL2024},
the expression $\sqrt{n}\left( U'_{n,N}-f(\theta) \right)$ is well approximated asymptotically
by the $p$-variate Gaussian
\begin{equation}\label{eq:the-gaussian-approximation-Y}
  Y\sim \mathcal{N}_{p}\left( 0,m^{2}\Gamma_{g}+\alpha_{n}\Gamma_{h} \right),
\end{equation}
with 
\begin{equation*}
\alpha_{n}:=\frac{n}{N},\quad
  \Gamma_{h}
  := \text{Cov}\left[ h(X_{1},\ldots,X_{m}) \right],
  \quad \text{and} \quad
  \Gamma_{g}
  := \text{Cov}\left[ g(X_{1})  \right],
\end{equation*}
where $g:=\E\left[h(x,X_{2},\ldots,X_{m}) \right]$ is the Hájek projection of $h$.

While the covariance matrix $m^{2}\Gamma_{g}+\alpha_{n}\Gamma_{h}$ of $Y$ is typically unknown, since $Y = mY_{g}+\sqrt{\alpha_{n}} Y_{h}$ for independent $Y_{g}\sim \mathcal{N}_{p}(0,\Gamma_{g})$ and $Y_{h}\sim \mathcal{N}_{p}(0,\Gamma_{h})$, an approximation of $Y$ can be obtained from approximating the distribution of these two normal random variables:

\begin{enumerate}
  \item[$Y_{h}$:]  To approximate $Y_h$, let
  $\left\{\xi'_{\iota}: \iota \in I_{n,m}\right\}$ be a collection of
  independent standard normal variables, and define the multiplier bootstrap
  \begin{equation*}
    U^{\#}_{n,h} 
    := \frac{1}{\sqrt{\widehat{N}}}\sum_{\iota\in I_{n,m}}
    \xi'_{\iota} \sqrt{Z_{\iota}} \left( h(X_{\iota})- U'_{n,N} \right).
  \end{equation*}
  The distribution of $U^{\#}_{n,h}$ is used to approximate $Y_{h}$. \\
  
  \item[$Y_{g}$:] Since $g$ is not explicitly known, approximating $Y_{g}$ is more complicated.
  Fix some $S_{1}\subseteq [n]$ and let $n_{1} = \left| S_{1} \right|$. For each
  $i_{1}\in S_{1}$, partition
  $[n]\backslash \left\{i_{1}\right\}$ into
  $K := \lfloor \frac{n-1}{m-1}\rfloor$ disjoint subsets of size $m-1$:
  $
    S_{2,1}^{(i_{1})},S_{2,2}^{(i_{1})},\ldots, S_{2,K}^{(i_{1})}$.
  For each $i_{1}\in S_{1}$, we estimate $g(X_{i_{1}})$ using the
  \emph{divide-and-conquer estimator}
  \begin{equation*}
    G_{i_{1}}:= \frac{1}{K}\sum_{k=1}^{K}h(X_{i_{1}},X_{S_{2,k}^{(i_{1})}}).
  \end{equation*}
  With $\overline{G}:= \frac{1}{n_{1}}\sum_{i_{1}\in S_{1}}G_{i_{1}}$, define
  \begin{equation}\label{eq:bootstrap-approximation-for-Y_g}
    U^{\#}_{n_{1},g} 
    := \frac{1}{\sqrt{n_{1}}}\sum_{i_{1}\in S_{1}}\xi_{i_{1}}
    \left( G_{i_{1}}-\overline{G} \right),
  \end{equation}
  where $\left\{\xi_{i_{1}}: i_{1}\in S_{1}\right\}$ is a collection of $n_1$
  independent standard normal variables. The distribution of $U^{\#}_{n_{1},g}$ is used to approximate $Y_{g}$.
\end{enumerate}

Finally, the combined Gaussian bootstrap used to approximate the distribution of $Y$ is
\begin{equation*}
  U^{\#}_{n,n_{1}} := mU^{\#}_{n_{1},g}+ \sqrt{\alpha_{n}}U_{n,h}^{\#}.
\end{equation*}

\subsubsection{Studentization.} 
\label{sec:studentization}
For studentization of the statistics $\mathcal T$ and $W$ (\cref{eq:def-of-test-statistic,eq:def-of-W}) we  estimate ${\sigma}_{1}^{2},\ldots,{\sigma}_{p}^{2}$.
From the previous subsection, these can be obtained as
$ \widehat{\sigma}_{j}^{2} 
  := m^{2} \widehat{\sigma}_{g,j}^{2}+\alpha_{n} \widehat{\sigma}_{h,j}^{2},
$
where
\begin{equation*}
  \widehat{\sigma}_{g,j}^{2} 
  := \frac{1}{n_{1}} \sum_{i_{1}\in S_{1}} \left( G_{i_{1},j}-\overline{G}_{j} \right)^2  
  \quad \text{and} \quad
  \widehat{\sigma}_{h,j}^{2} 
  := \frac{1}{\widehat{N}} \sum_{\iota \in I_{n,m}} Z_{\iota} \left( h_{j}(X_{\iota})- U'_{n,N,j} \right)^{2}.  
\end{equation*}

\subsection{Kernel construction}
\label{sec:kernel}
 Now that we have laid out all the components of the SDL test, we discuss particulars about constructing a kernel function that satisfies the requirements of \cref{sec:the-kernel-function}. For a semialgebraic model,  the following procedure for constructing a kernel $h$ is suggested in \cite[Section 4]{SDL2024}. 

For each polynomial inequality $f_{i}(\theta) \leq 0$, $i\in \left\{1,\ldots, p\right\}$ used in defining the model,
write
\begin{equation} \label{eq:weird-multi-index-representation-of-f_i}
  f_{i}(\theta) 
  = a_{0} + \sum_{r=1}^{s} \sum_{ \substack {j=(j_1,\dots, j_r) \\j_i \in\{1,\dots, d\}} } a_{j} \theta_{j_1}\cdots \theta_{j_r},
\end{equation}
with $a_{0},a_{j} \in \mathbb R$.
Then the following steps construct a symmetric, unbiased
estimator $h_i(X_1,\ldots,X_m)$ of $f_i(\theta)$ from independent $X_i\sim P_\theta$, $\theta\in \Theta_0$:

\begin{enumerate}
  \item For some $\eta \geq 1$, find functions
$\widehat{\theta}_{1},\ldots,\widehat{\theta}_{d}:\mathbb{R}^\eta \to \mathbb{R}$ with
$\E\left[\widehat{\theta}_j(X_1,\ldots,X_\eta) \right] = \theta_{j}.$

  \item With $m = \eta\cdot \max_{1\le i\le p}\{\deg(f_{i})\}$, 
an unbiased estimator of $f_i(\theta)$ is $\breve{h}_{i}(X_1,\ldots,X_m)$, where
  \begin{align*}
    \breve{h}_{i}(x_1,\ldots,x_m) 
      &:=a_{0}+\sum_{r=1}^{s} \sum_{j\in J_{r}} a_{j}\prod_{z=1}^{r}\widehat{\theta}_{j_z}\left(x_{(z-1)\eta+1},x_{(z-1)\eta+2},\ldots,x_{z\eta}\right).
  \end{align*}

  \item\label{step:symm} With $S_m$ the symmetric group, the components of a symmetric kernel $h:\mathbb{R}^m \to \mathbb{R}^p$ are given by:
  \begin{equation*}
    h_i(x_1,\ldots,x_m) := \frac{1}{m!}\sum_{\pi\in S_m}\breve{h}_i\left(x_{\pi(1)}, \ldots, x_{\pi(m)}\right).
  \end{equation*}
\end{enumerate}

Note the symmetrization of  step \ref{step:symm} is computationally expensive if $\deg(f_i)$ is large. In \Cref{ssec:symh} we discuss this issue further.

\subsection{SDL test parameters}\label{ssec:SDLparams} Finally, we catalogue the different parameters that are needed for the SDL test, as these parameter choices will be explored in the context of our applications below. In addition to a semialgebraic description of a model, the SDL testing procedure requires four parameter values. They are listed here along with suggested values from \cite{SDL2024}.
\begin{enumerate}
\item[$m$:] The order $m$ of the kernel $h$ is determined by the constraint degrees and the number of data points $\eta$ used to estimate the $\theta_i$.
For the theoretical analysis of error bounds in \cite[Theorem 2.4]{SDL2024}, it is assumed that
$2\le m\le \sqrt n$, while the bound itself depends quadratically on $m$.
The authors suggest that $m$ be small, as ``larger $m$ imply worse performance of the
Gaussian approximation in terms of the required sample size'' \cite[Remark 2.6]{SDL2024}.

\item[$N$:] The computational budget parameter $N $ specifies the average number of terms in the randomized incomplete U-statistic. The asymptotic error bounds of \cite{SDL2024} require $N/|I_{n,m}| < 1/2$, but choosing
$N = \mathcal O(n)$ is suggested as the error  bounds vanish asymptotically under certain circumstances. Simulations in \cite{SDL2024} suggest larger $N$ provides more statistical power, but the authors warn too large an $N$ may cause the test to perform poorly near model irregularities. Ultimately, they observe that $N=2n$
 was reasonable for their model simulations.
 
\item[$n_1$:] The parameter $n_1$ specifies the number of terms used in the sum in \cref{eq:bootstrap-approximation-for-Y_g} to estimate $Y_g$. In   \cite{SDL2024}, a suggested value of $n_1=n$, the maximum possible, is given so that bootstrap accuracy is maximized. 

\item[$A$:] The final parameter, $A$, governs the number of samples $W$ used in approximating their distribution via bootstrap, with a suggested value of $A=1000$.

\end{enumerate}

\section{Trinomial submodels}\label{sec:trinomial}

Here we explore the behaviour of the SDL test on
some simple null semialgebraic models that arise when considering the coalescent model in phylogenomics. Their small size, in terms of dimension, allow for rejection regions to be visualized and compared to those from other methodologies.

\subsection{Basic examples}

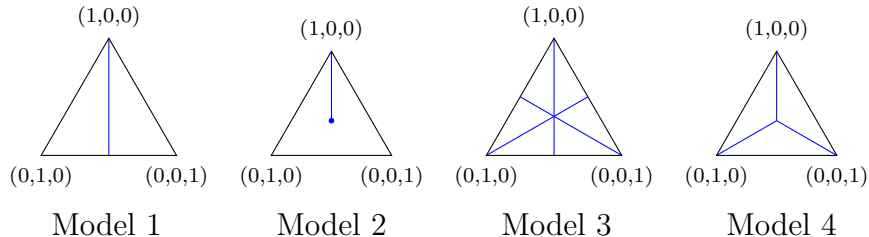
\begin{figure}[htb]
\begin{tikzpicture}[scale=.9]
      \coordinate[label=below :\text{\tiny(0,1,0)}] (A) at (0,0);
      \coordinate[label=below :\text{\tiny(0,0,1)}] (B) at (2,0); 
      \coordinate[label=above:\text{\tiny (1,0,0)}] (C) at (1,{1.732});
      \coordinate (D) at (1,0);
      \draw (A) -- (B) -- (C) -- cycle;
      \draw[blue] (C) -- (D);
\end{tikzpicture}
\begin{tikzpicture}[scale=.8]
      \coordinate[label=below :\text{\tiny(0,1,0)}] (A) at (0,0);
      \coordinate[label=below :\text{\tiny(0,0,1)}] (B) at (2,0);  
      \coordinate[label=above:\text{\tiny(1,0,0)}] (C) at (1,{1.732});
      \coordinate (center) at (barycentric cs:A=1,B=1,C=1);
      \draw (A) -- (B) -- (C) -- cycle;
      \draw[blue] (C) -- (center);
      \node[circle, fill=blue, inner sep=.75pt] at (center) {};
    \end{tikzpicture}
\begin{tikzpicture}[scale=.9]
      \coordinate[label=below :\text{\tiny(0,1,0)}] (A) at (0,0);
      \coordinate[label=below :\text{\tiny(0,0,1)}] (B) at (2,0); 
      \coordinate[label=above:\text{\tiny(1,0,0)}] (C) at (1,{1.732});
      \coordinate (D) at (1,0);
      \coordinate (E) at (.5,.866);
      \coordinate (F) at (1.5,.866);
      \draw (A) -- (B) -- (C) -- cycle;
      \draw[blue] (C) -- (D);
      \draw[blue] (A) -- (F);
      \draw[blue] (B) -- (E); 
    \end{tikzpicture}
\begin{tikzpicture}[scale=.8]
      \coordinate[label=below :\text{\tiny(0,1,0)}] (A) at (0,0);
      \coordinate[label=below :\text{\tiny(0,0,1)}] (B) at (2,0); 
      \coordinate[label=above:\text{\tiny(1,0,0)}] (C) at (1,{1.732});
      \coordinate (center) at (barycentric cs:A=1,B=1,C=1);
      \draw (A) -- (B) -- (C) -- cycle;
     \draw[blue] (A) -- (center);
     \draw[blue] (B) -- (center);
     \draw[blue] (C) -- (center);
    \end{tikzpicture}

\ Model 1\hskip .6in Model 2\hskip.6in Model 3\hskip.6in Model 4

\caption{Parameter spaces (blue line segments) of four submodels of the trinomial model, with parameter space $\Delta^2$. The submodels capture the form of the quartet Concordance Factor if the species relationships have specific features, as described in the text.}\label{fig:4models}
\end{figure}

Our first four example models are depicted inside the 2-simplex in \cref{fig:4models}. Each characterizes the frequencies of the three possible quartet gene tree topologies if four species are related by a tree or network with certain features. 
\Cref{ssec:msc}  provides a more complete explanation, but knowledge of the application is not necessary for a reader primarily interested in the SDL test for other uses.

 While each model is composed of line segments, they exhibit a variety of geometric features that may affect testing behaviours. Model 1 is regular. Model 2 has a boundary point in the interior of the simplex, causing a discontinuity in the asymptotic distribution of standard statistics.  Model 3 has no boundary points in the simplex but exhibits a singularity (in the sense of algebraic geometry) at the centroid, where 3 lines cross. Again this causes a discontinuity in the asymptotic distribution, with 
 slow convergence to it for parameters near the centroid.
In Model 4 the centroid is both a singularity and a boundary of each of the component lines.
 
 Because of their importance for testing whether biological data shows evidence for specific species relationships involving  hybridization or other lateral gene flow,
specialized test distributions for null hypotheses of  Model 2 and 4 are derived in \cite{MAR2019} and for Model 3 in \cite{ABMR2024}. Tests using these are implemented in the {\tt R}  package {\tt MSCquartets} \cite{MSCquartets_bioinformatics2020}.
These improve on a naive use of a standard distribution such as a $\chi^2$ that ignores the singularities and boundaries of the models. Model 1, of course, can be tested with a standard approach, as it lacks any irregularities. Thus for all these models we can compare SDL test behaviour to the behaviour of deterministic tests. 

We also consider several other semialgebraic trinomial submodels that we do not depict here. These are the Hardy-Weinberg equilibrium model for 2 alleles (a regular model for which good deterministic test methods are established) and two
artificial models chosen because of their specific algebraic nodal and cuspidal singularities. For these last two models we know of no other methods addressing their singularities, but they nonetheless illustrate important issues that may arise with general semialgebraic models.

\Cref{sec:dettests} presents rejection regions using current deterministic testing procedures for the null Models 1-4, as well as for the Hardy-Weinberg model, for a dataset of size 300.

\subsection{Semialgebraic descriptions of trinomial models}\label{ssec:semialg}

Each of the  models depicted in \cref{fig:4models} is easily given a semialgebraic description. 
With the parameter space $\Theta$ for each of the models viewed as a subset of $\Delta^2\subseteq \mathbb R^3$, we use coordinates $(x,y,z)$, with $x+y+z=1$, $x,y,z\ge 0$, for simplex points. 

\begin{description}
\item[Model 1] 
$y-z=0.$
\smallskip
\item[Model 2] 
$ y-z=0,\quad 1/3-x \le 0.$
\smallskip
\item[Model 3] 
 $(y-z)(x-y)(x-z)=0.$
 \smallskip
\item[Model 4] 
$(x-y)(x-z)(y-z)=0, \quad (x-z)^2(y-z)^2(1/3-x)\le 0,$\\
$(x-y)^2(y-z)^2(1/3-y)\le 0,\quad (x-y)^2(x-z)^2(1/3-z)\le 0.$

\end{description}

Note that other semialgebraic descriptions of these models exist, and although these are `simple' ones, we have no well-defined notion of a `simplest description' in general. For instance, the linear inequality given above in the description of Model 2 could be replaced by others and the effect of changing this description is one issue with the SDL test that we investigate in \Cref{ssec:convex}.

\subsection{SDL rejection regions for trinomial submodels}
\label{sec:CF-data}
One way to understand a hypothesis test is through its rejection region
at various test levels. For the models above, we considered all possible datasets (up to ordering) of size $n=300$, that is all collections of 300 vectors each of which is a standard basis vector in $\mathbb R^3$. 
The counts of the 3 basis vectors in such a dataset are then normalized (i.e., the mean of the vectors is computed) to give a point in the simplex. Applying the SDL test for a model to the dataset, this point can be coloured according to the dataset's $p$-value, indicating rejection at various levels. 

Note that rejection is based on the incomplete U-statistic of the data, which includes randomness, and the test distribution, which also includes randomness. Thus rejection region plots produced in this way may vary even though they are testing identical ``data'' and  there is no well-defined ``rejection region'' in the simplex.
Nonetheless, such plots, and the stochastic variation they show, give helpful insight into test behaviour.

In \Cref{sec:pitfalls} we follow this procedure to colour the simplex for various models using nominal test levels of $0.10$, $0.05$, and $0.01$ to delineate between purple, blue, green, and red colourings. Throughout, we use datasets of size $n=300$.  This size was chosen so that the rejection region plots were not overly pixilated, yet easily interpretable visually, since  for very large $n$ the size of the fail-to-reject region shrinks tightly around the model line segments.

\subsection{The SDL test of trinomial submodels}\label{sec:pitfalls}

For datasets of size $n=300$, we fix parameters of the SDL test to $N=1000$, $A=1000$, and $n_1=n=300$ throughout, but vary $m$. Our values of $A$ and $n_1$ follow suggestions of \cite{SDL2024}, since we only observed noticeable changes in performance with extreme variations from suggested values.
Varying $N$ or $m$ has more impact. However, we found  increasing $N$ to 1000 reduced the randomness in our $p$-values and, with appropriately chosen $m$, still allowed us to ensure our tests were conservative. We therefore only vary $m$ as \cite{SDL2024} already illustrated the effects of varying $N$.

\subsubsection{The order \emph{m} of the kernel}\label{ssec:m} 

\begin{figure}
    \centering
\includegraphics[width=0.24\linewidth]{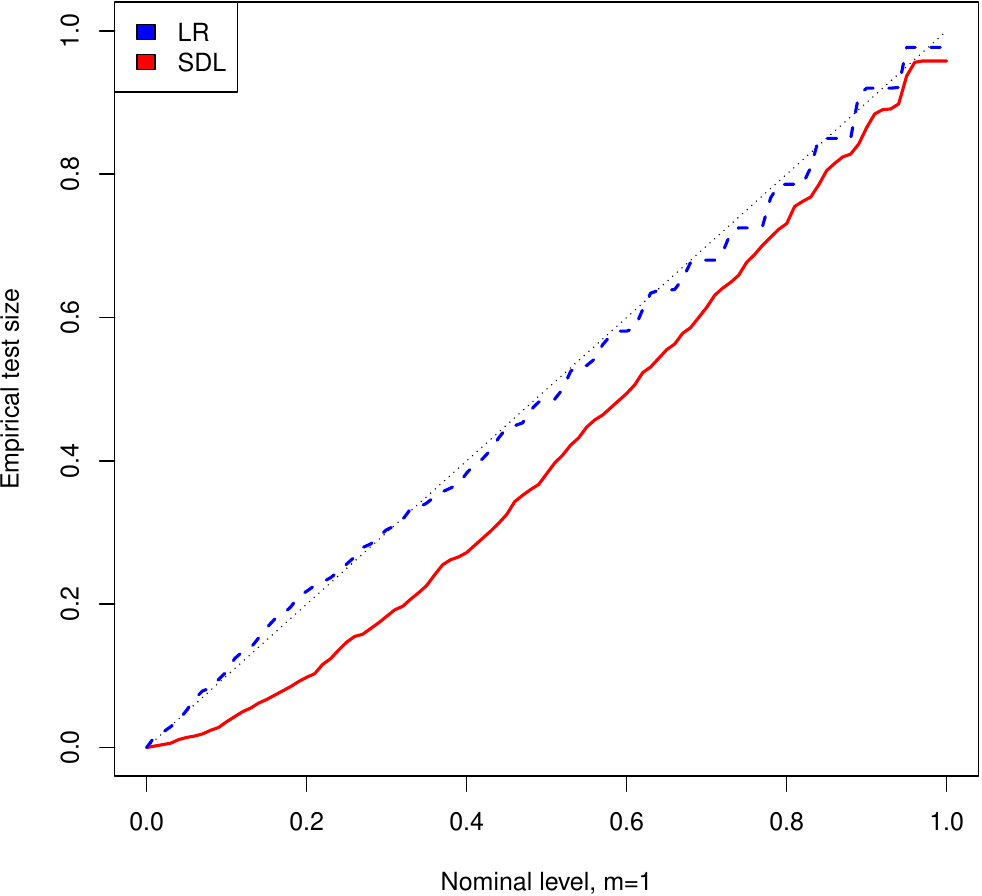}
\includegraphics[width=0.24\linewidth]{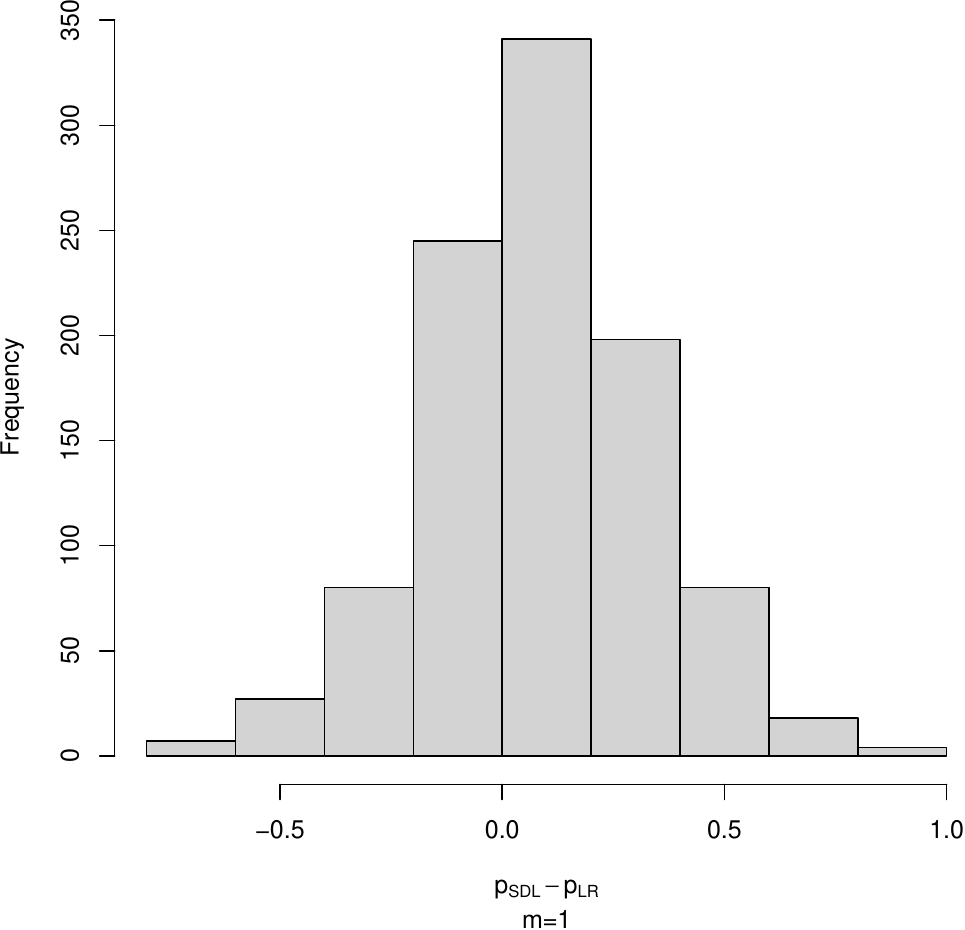}
\includegraphics[width=0.24\linewidth]{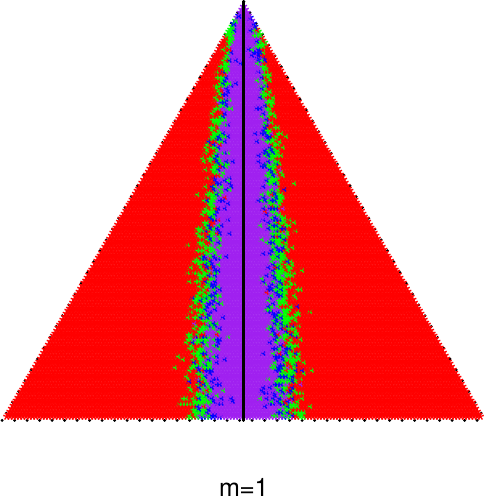}

\includegraphics[width=0.24\linewidth]{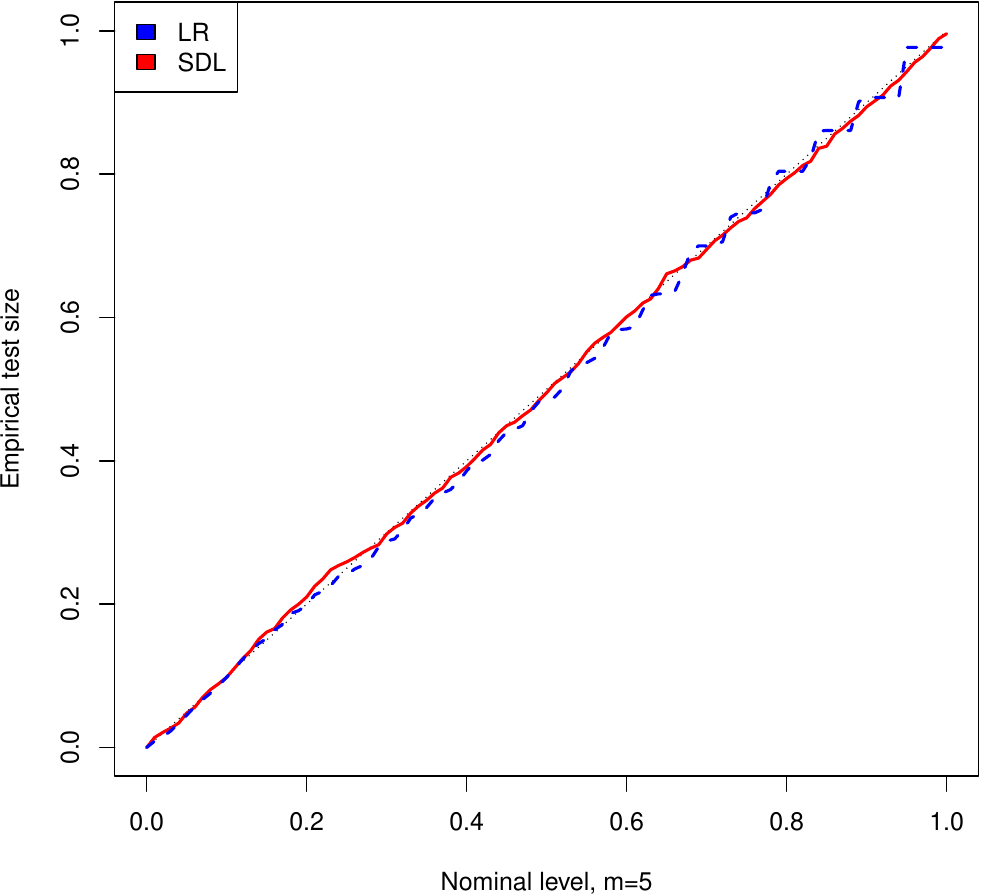}
\includegraphics[width=0.24\linewidth]{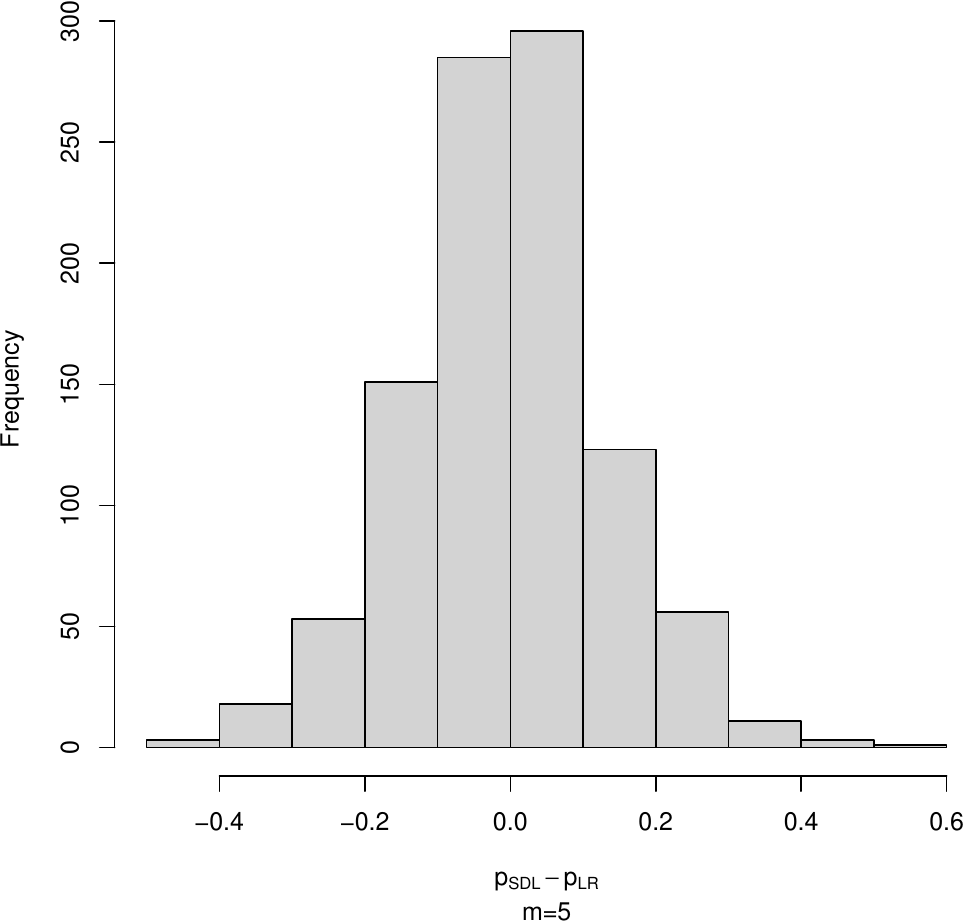}
\includegraphics[width=0.24\linewidth]{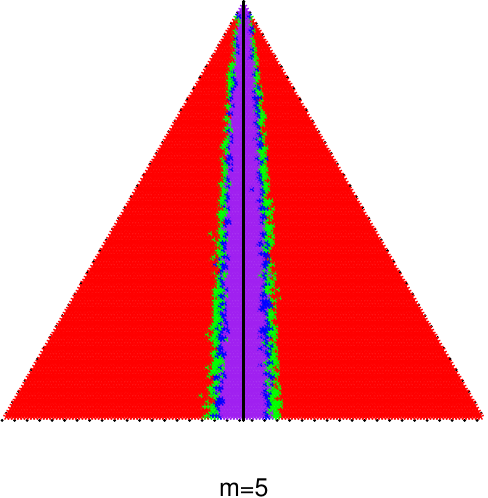}

\includegraphics[width=0.24\linewidth]{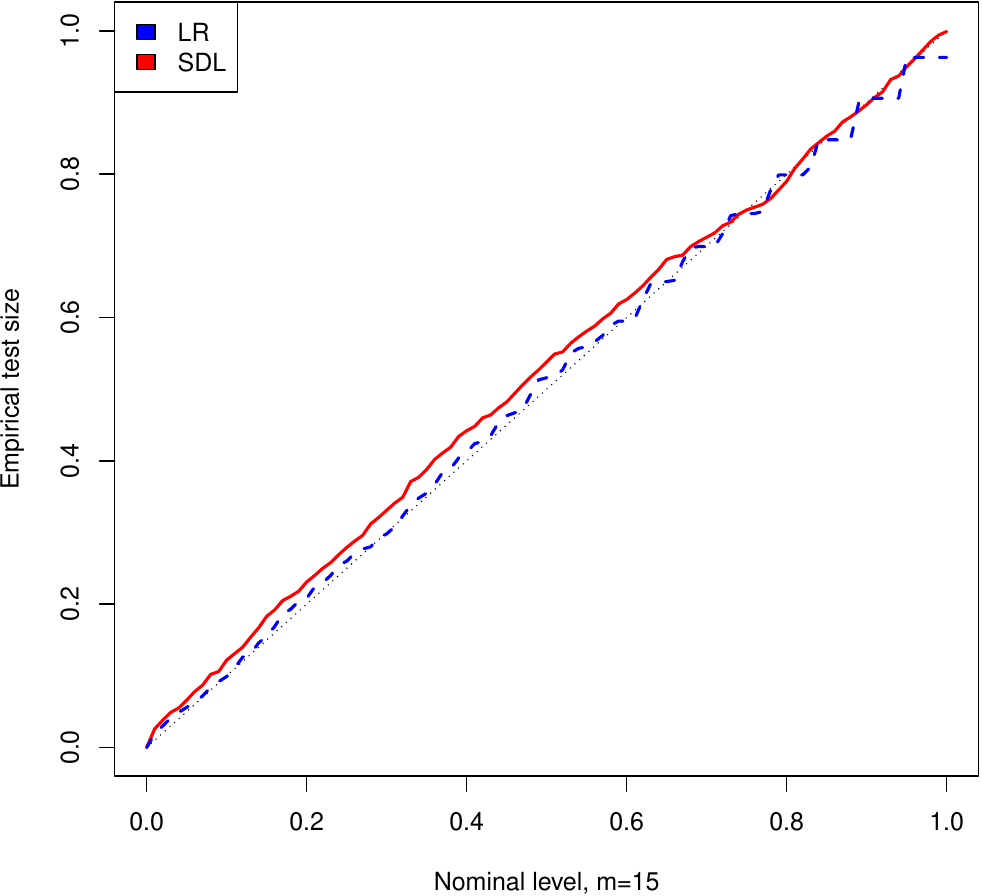}
\includegraphics[width=0.24\linewidth]{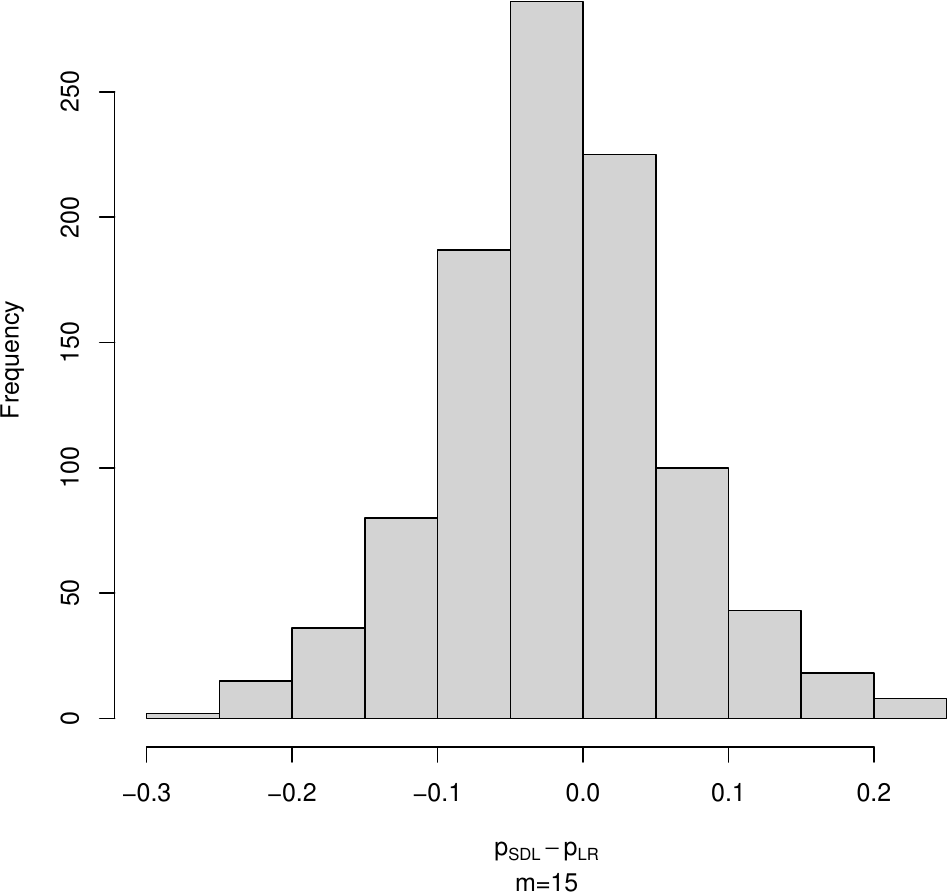}
\includegraphics[width=0.24\linewidth]{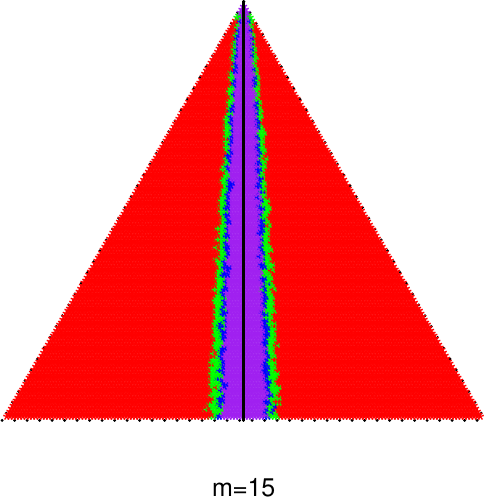}

\caption{SDL test behaviour for Model 1, with $m=1$, $5$, and $15$ (top row to bottom).
The left column shows nominal vs.~empirical sizes for the SDL and LR tests; the
middle, histograms of $p$-value differences; and the right, SDL rejection
regions.} \label{fig:m1}      
\end{figure}

As constructed in \Cref{sec:kernel}, the kernel function $h$ depends on
$m=\eta\cdot\max_{i} \left\{\deg (f_i) \right\}$ data points, with $\eta$ the
number per scalar parameter. While \cite{SDL2024} suggests that $m$ should be
chosen to be small, we experimented with different choices of $\eta$ and
found that choosing a minimal value was generally \emph{not} optimal as it
could lead to both lower statistical power and increased stochasticity of the
SDL $p$-values.

This conclusion is illustrated in \Cref{fig:m1}, which compares SDL $p$-values
from Model 1 (which has only regular points) using $m=1,5,$ and $15$ (top to
bottom), along with $p$-values from the standard Likelihood Ratio (LR) test.
Since Model 1 is a linear model we estimate the variables $x$, $y$, $z$ by taking means of $\eta=m$ data values.
The left column
of \Cref{fig:m1} compares the nominal level versus empirical test sizes of the
SDL test (red) and LR test (blue) from 1000 simulated datasets of size 300,
with model parameters $(1/3,1/3,1/3)$. The middle column histograms show the
differences between the approximate $p$-values of the SDL test and the
$p$-values computed with LR, for the same 1000 datasets. The right column
depicts the SDL rejection region for all datasets of size $n=300$.

Importantly, \Cref{fig:m1} illustrates the danger of choosing $m$ too large,
since it impacts the conservativity of the SDL test. For $\eta=m=1$, the test is
highly conservative (top left), with SDL $p$-values tending to be larger than
LR $p$-values (positive histogram mean). At $m=5$, the test retained an
acceptable size (middle left), and additional simulations with other
parameters (not shown) indicate that $m=5$ was a uniformly good choice. On the
other hand, $m>5$ resulted in invalid tests with an excess of small
$p$-values. \Cref{fig:m1} illustrates this for $m=15$, with the leftmost plot
exhibiting for most levels an excess in the test size, and the histogram a
negative mean.

Moreover, choosing $\eta=m$ very small (e.g., $m=1$) is also
suboptimal. For $m=1$, the rejection region plot (top right) has a smaller
rejection region than for the LR test (shown in \cref{fig:dettests} of
\Cref{sec:dettests}), and its $p$-values exhibit substantial random
variability. By contrast, increasing $m$ had the
benefit of increasing both the size of the rejection region and the precision
of the SDL $p$-values (right column), with the latter observation also evident
in the histograms, which concentrate with larger $m$. To quantify this, we
also computed the variance of the SDL $p$-values from 100 test applications
for each of 100 simulated datasets, and observed a decrease from $0.068$ for
$(m=1)$ to $0.030$ for $(m=5)$.

We note that while choosing $m$ minimally gave a conservative test here, in our examples below, and in \cite{SDL2024},
there are no theoretical assurances that this will be the case for all models. Regardless, varying $m$ in the models we explored suggests a clear tradeoff between increasing $m$ to reduce the stochasticity of $p$-values and type II errors, and keeping $m$ small to reduce type I errors. However, the value of $m$ at which the 
test size exceeded the nominal level is dependent on the specific model, constraints used to describe it, and the model parameter $\theta\in\Theta_0$, and we were unable to develop any general rules to apply. Simulation at a number of model points seems to be the most informative approach.

In the following subsections, we use the largest $m$ which simulations suggest gives a valid test size at a number of model points, including singularities and boundary points. For instance, we find that for Model 2 (discussed in the next subsection) $m=5$ gave good performance for the boundary parameter point (1/3,1/3,1/3), with empirical test size closely tracking the nominal level (plot not shown). However, for parameters (2/3,1/6,1/6), this choice of $m$ gives a conservative test for Model 2, and $m=20$ gives a more powerful yet valid test at that point. We nonetheless consider $m=5$ for Model 2 as the better choice overall.

\subsubsection{Choice of model constraints}\label{ssec:convex}  

Semialgebraic models may have many different semialgebraic descriptions in which the  polynomial equalities and inequalities differ. The choice of specific model constraints can impact the shape of the rejection region for the SDL test.

\begin{figure}[!htb]
    \centering
\includegraphics[width=.24\textwidth,valign=t]{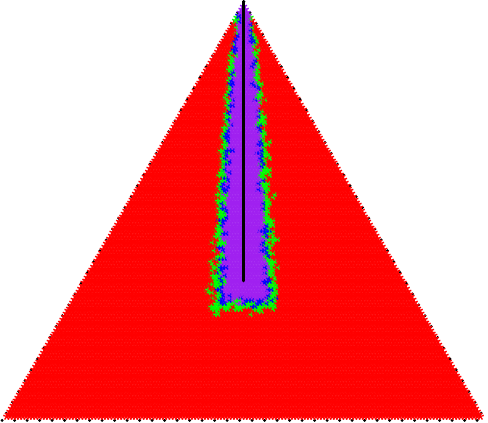} 
\includegraphics[width=.24\textwidth,valign=t]{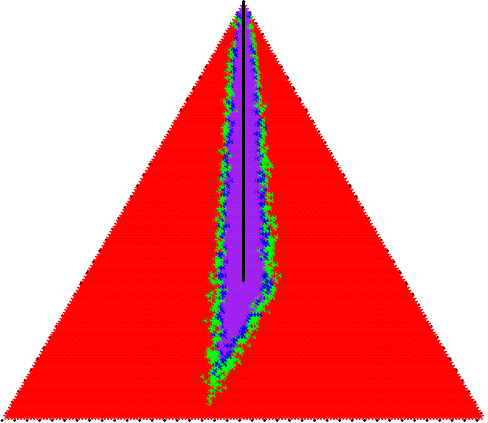}
\includegraphics[width=.24\textwidth,valign=t]{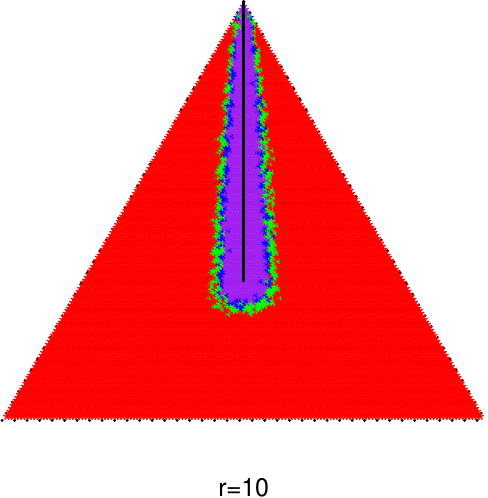}
\includegraphics[width=.24\textwidth,valign=t]{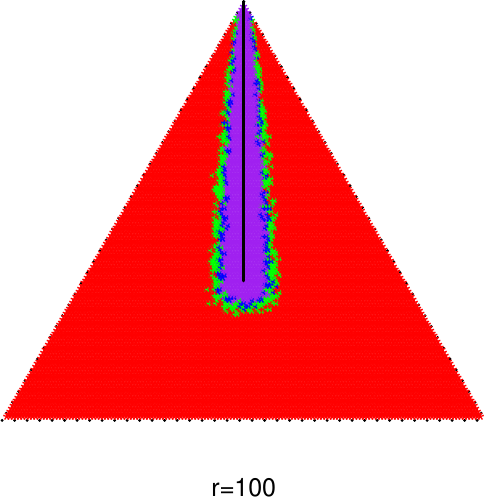}
\caption{Rejection regions for Model 2 under the SDL test using (L to R) a) the constraints $y-z\le 0$, $z-y\le 0$, and
$1/3-x \le 0$; b) replacing the last inequality by $2/3-x-y\le 0$; c) including $r=10$ random convex combinations of the inequalities of (a) ; and d) including $r=100$ random convex combinations.}
\label{fig:convexcomb}
\end{figure}

For Model 2 with $m=5$ data points in the kernel function, we illustrate this in
\cref{fig:convexcomb}. On the left we use the constraints given in the previous section. Note that the `flat bottom' of the purple region reflects the horizontal boundary from the constraint $1/3-x\le 0$. 

For the next plot in \cref{fig:convexcomb} the inequality $1/3-x\le 0$ is replaced by $2/3-x-y\le 0$, giving a different description of the same model. Again the shape of the rejection region reflects the choice of the constraint. While both of these regions are valid in the sense of ensuring an acceptable rejection rate for data generated by the model, the fact that an arbitrary choice of constraints determines the shape of the rejection region is undesirable.

To be agnostic in terms of semi-algebraic description, it would be preferable to simultaneously use all possible constraints for the model. But by including only a small number of additional model constraints in a redundant model description, we found we could approximate that situation for Model 2.
 
 In particular, after first converting the equality constraint to two inequalities, we created 10 and then 100
 random convex combinations of the original three inequalities and included them in the SDL procedure. This gave the two rightmost plots in \cref{fig:convexcomb}, with `rounded' bottoms, approximately reflecting all the linear constraints that might be used to truncate the model line at the centroid of the simplex. Using more random combinations more consistently smooths the boundary, but at additional computational cost. 
 
For this example, with a complete geometric view of the model in the ambient simplex, we could have chosen fewer specific
combinations for the same effect. In more general settings, however, choosing randomly has the advantage of not requiring any detailed geometric understanding of the model.
 
 However, it may be necessary to use many such combinations, especially when the model's co-dimension is large. For a simple example, a model that is a half-line in a $d$-dimensional simplex is minimally described by $d-1$ linear equalities and 1 linear inequality, or $2d-1$ inequalities. Rejection region  boundaries using such a set of constraints form a roughly polyhedral cylinder with opposite sides approximately parallel (due to the equality constraints), which is cut off by a hyperplane (from the inequality). If $d$ is large, an adequate number of combinations to approximate a full set of inequality constraints might be quite large, but would give a rounder boundary.
 
 \medskip

For work that follows, we introduce a new parameter, $r$, indicating the number of random convex combinations of the model's specified inequality constraints to include as new constraints in the SDL testing procedure. By `random' we mean that if the model is specified by $D$ inequality constraints then the convex sum weight $w$ for each new constraint is an independent random variable $w\sim \text{Dirichlet}(D;1,1,\dots,1)$, meaning that $w$ is drawn uniformly from $\Delta^{D-1}$.

In \Cref{ssec:highdegree} we consider a more complex situations in which supplying additional redundant  constraints may be desirable.

\subsubsection{Symmetrizing the kernel}\label{ssec:symh}

As described in \Cref{sec:kernel}, we construct our kernel function $h$ of $m$ data points from the semialgebraic model constraints by a process including symmetrization. Then the symmetrization occurs over the symmetric group $\mathcal S_m$.

For general semialgebraic models there is no upper bound on the degree of defining constraints, so even if $\eta$ may be chosen to be small, $m=\eta \max_i \deg f_i$ may be large. Moreover, as was discussed in \Cref{ssec:m}, performance of the method is sometimes improved by choosing $\eta$ larger than its theoretical minimum. Thus $m$ may be large in practice, and a full symmetrization may not be computationally feasible.

To investigate situations in which symmetrization of the kernel by summing over all data permutations is not feasible, we focus on Model 3 with $\eta=5$ so $m=15$. (Since this model is defined by a single equality constraint,
convex combinations of the resulting inequalities would have no effect.)
From the construction of the kernel we already have symmetry within the 5-element blocks of data points which are averaged to estimate each parameter. Thus full symmetrization would only require
\begin{equation*}
  \frac{15!}{(5!)^3}\approx 7.5\times 10^{5}
\end{equation*}
permutations, though this is already computationally excessive.
We therefore explore summing only over a relatively small
number, $s$, of permutations, chosen uniformly at random. We sample these permutations anew each time the kernel must be evaluated, both for computing the test statistic and for estimating the distribution by which it is judged.

\begin{figure}[!htb]
    \centering
\includegraphics[width=.24\textwidth]{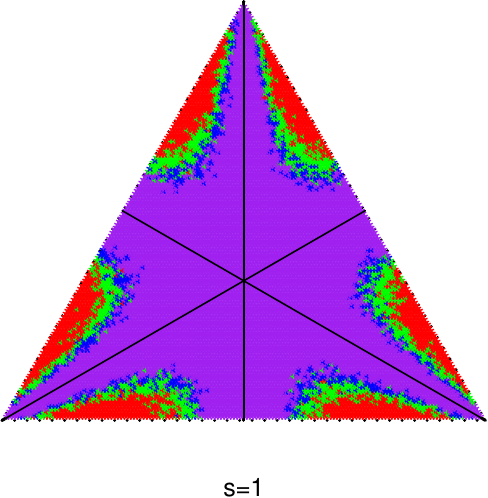}
\includegraphics[width=.24\textwidth]{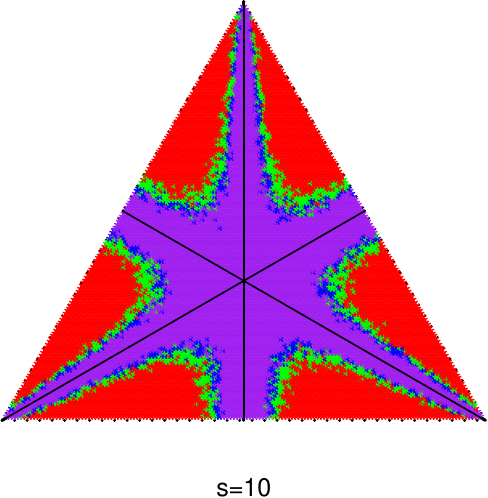}
\includegraphics[width=.24\textwidth]{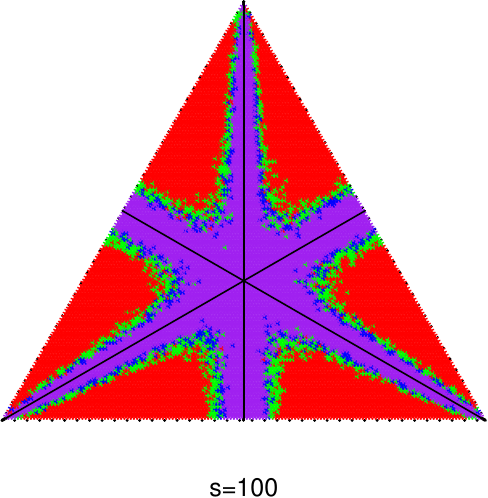}
\caption{Rejection regions for Model 3 under the SDL test using (L to R) $s=1,\ 10,$ and $100$ random permutations to partially symmetrize $\breve h$. For all, $m=15$.}
\label{fig:symm}
\end{figure}

In \cref{fig:symm} (left) we see that even a single ($s=1$) random permutation produces an appropriately symmetric rejection region, though that region is quite small. With even $s=10$ permutations used (middle) the rejection region grows considerably.
This trend continues through $s=100$ permutations,  although the gain between these last two is not large.

While our explorations indicate that this random partial symmetrization scheme can be effective, theory justifying its use is currently lacking. The incomplete U-statistics already incorporate two sources of randomness --- the data and the subsampling/bootstrapping of the test procedure --- and random partial symmetrization brings in a third which is not considered in \cite{SDL2024}. Moreover, our simulations are all low-dimensional and we did not explore thoroughly how increasing dimension may affect the number of random permutations needed. While in \Cref{sec:CFN} we explore one higher dimensional case, extension of the underlying theory of the SDL test is needed.

\subsubsection{Irreducible components and an intersection-union test}\label{ssec:irred}
 
Some natural semialgebraic statistical models are formed as the union of several components, such as the intersecting line segments that comprise Models 3 and 4. More specifically, in algebro-geometric terms, a model may be Zariski-dense in a variety with several irreducible components. Although for these examples the irreducible components are simply lines, more generally irreducible components may be higher degree but will have degree at most that of the full model. Computational algebra software can be used to calculate equality constraints of the components.

\begin{figure}[!ht]
        \centering
         \includegraphics[width=.24\textwidth]{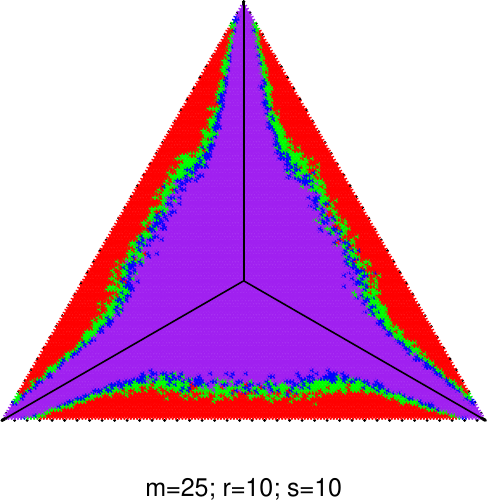}
        \includegraphics[width=.24\textwidth]{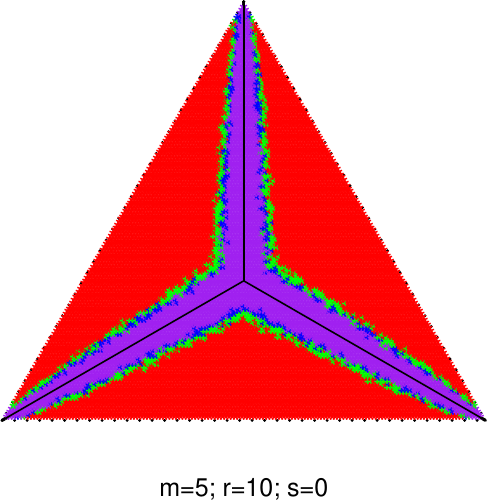}
    \caption{(L) Rejection region for Model 4 obtained from SDL test using semialgebraic description given above. (R) Rejection region for an Intersection-Union test using the SDL tests for the 3 irreducible components of Model 4 (each essentially Model 2).}
    \label{fig:irred}
\end{figure}

In addition to performing the SDL test directly for Model 4 using the constraints given above, we performed an intersection-union test by applying the SDL test to each irreducible component, rejecting the full null hypothesis if we reject it for each of the component null hypotheses. Thus we take the maximum of the $p$-values from the irreducible component tests as an overall $p$-value.

\cref{fig:irred} shows comparison plots for Model 4, using the standard SDL test and the intersection-union variant. In both we used $\eta=5$ data points to estimate individual model parameters, giving $m=25$ and $5$, respectively, due to the different degrees of the constraints. Note the intersection-union test led to both a larger rejection region and less randomness in its boundary. Indeed, the direct SDL test for Model 4 remained conservative for all values of $m$ we tried (up to 45) and in particular the null hypothesis  was never rejected in a very large central region of the simplex. In addition to having much greater power, the intersection-union test was faster to compute, and showed less random behaviour.

Model 3 can similarly be decomposed, with an SDL intersection-union test showing better performance than was obtained in \Cref{ssec:symh}. We suspect that similar gains can be achieved for other reducible models 

\subsubsection{Higher degree irreducible models}\label{ssec:highdegree} 

As seen for Models 3 and 4, the degree of the model's constraints seems to affect the power of the test, particularly around singularities, but somewhat for points far from these. If the model  can be decomposed into irreducible components of lower degree, an intersection-union approach may ameliorate the behaviour. To investigate the effect of degree further, we considered several irreducible models of degree 2 and 3.

\begin{figure}[!ht]
\centering
\includegraphics[width=0.24\linewidth]{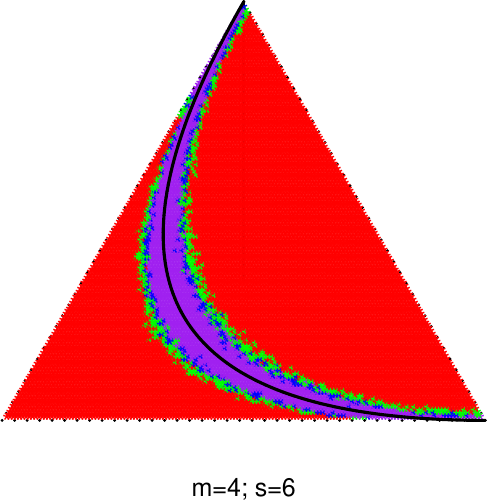}  
\includegraphics[width=0.24\linewidth]{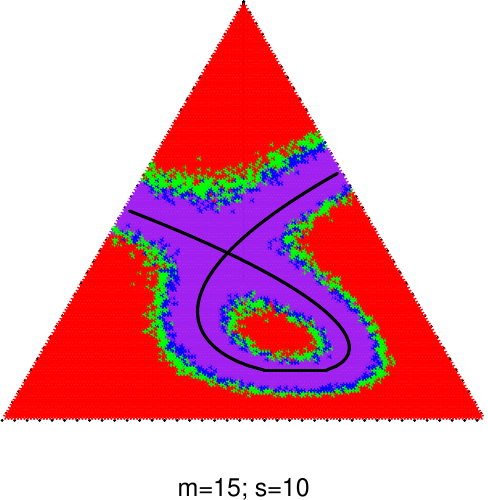}
\includegraphics[width=0.24\linewidth]{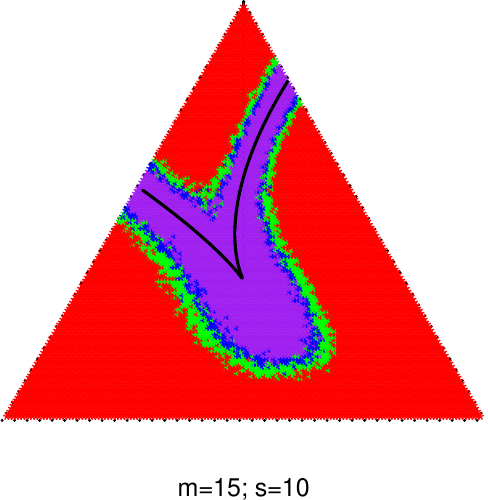}
\caption{Rejection regions for SDL tests of (L-R) (a) the Hardy-Weinberg 2-allele model defined by $y^2-4xz=0$, (b) a nodal cubic model defined by $(y-1/3)^2-6(x-2/5)^2(x-1/9)=0$, (c) a cuspidal cubic model, defined by
    $(y-1/3)^2- (x-1/3)^3=0$.}
    \label{fig:hd}
\end{figure}

The Hardy-Weinberg 2-allele model, whose SDL rejection region is shown in  \cref{fig:hd}(a), is a quadratic model with no irregularities. The rejection region for $n=300$ is close to that for the standard chi-squared test of the model (\cref{fig:dettests} of \Cref{sec:dettests}) with the added stochastic variation inherent in uses of SDL.  Note the low value of $m=4$ here; higher values produced excesses in small $p$-values

 \cref{fig:hd}(b) shows results for a nodal cubic model (chosen for its degree and geometry rather than any application) with a single crossing singularity. The higher degree seems to result in both less power than seen in previous models, and more stochastic variation at the boundary of the rejection region, at least for the same choices of test parameters used for previous models.

In \cref{fig:hd}(c) the SDL test is applied to a cuspidal cubic model. Note the large region (extending downward and right from the cusp) on which the test fails to reject the model. In that region the equality constraint is nearly met, with the polynomial taking on small values, resulting in an inability of the SDL approach to reject the model. This is an important feature to note, since it shows that a minimal set of model constraints may fail to adequately distinguish
between points on the model and some off the model for an SDL test.

\begin{figure}[!ht]
\centering
\includegraphics[width=0.24\linewidth]{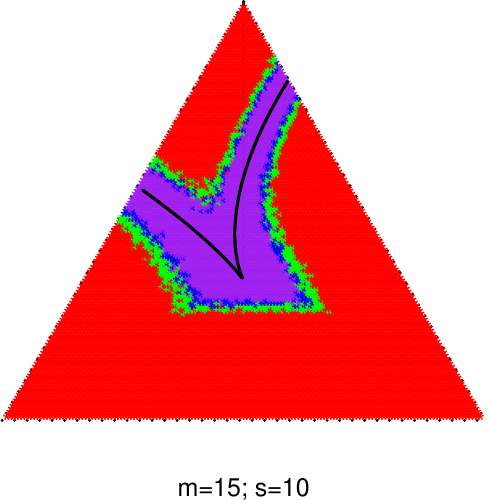}  
\includegraphics[width=0.24\linewidth]{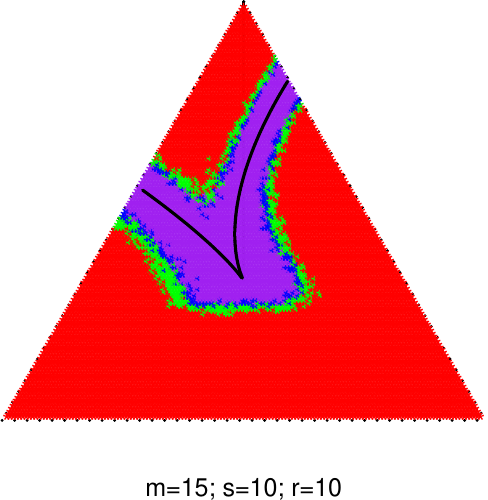} 
\includegraphics[width=0.24\linewidth]{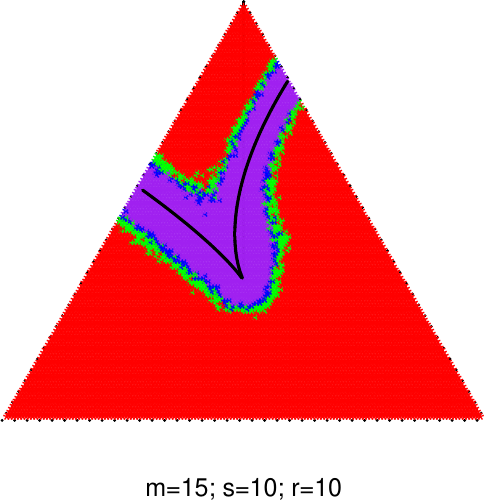}  
\caption{Rejection regions for SDL tests of the cuspidal cubic (L-R) with (a) constraints supplemented by $1/3-x\le 0$; (b) constraints supplemented by the inequality from (a) plus $r=10$ random convex combinations of inequalities, and (c) constraints supplemented by 3 linear inequalities as described in the text and $r=10$ random convex combinations of inequalities.
} \label{fig:hd2}
\end{figure}

In \cref{fig:hd2}(a) we see that adding a single linear inequality which is satisfied on the model expands the rejection region, and increases the test's power. This further reinforces the point of \Cref{ssec:convex} that ideally one would use all semialgebraic constraints satisfied on the model. However, the linear constraint introduced here is not suggested by the model's defining equation, and it is unclear how one might determine a good finite set of supplementary constraints in an automated way. Through human agency, doing so would be facilitated by a thorough understanding of the model geometry, but particularly in high dimensional settings that may be difficult to obtain.

\cref{fig:hd2}(b), which uses the same inequality constraint as in (a), illustrates an instance of the random convex combination approach of \Cref{ssec:convex} failing to have much impact. For \cref{fig:hd2}(c) we included two additional linear inequalities, with bounding lines stretching from the cusp to the points at which the model intersects the simplex boundary. These improve performance, though note the slight bulge in the
non-rejection region to the right of the cusp.
Adding additional non-linear constraints, with appropriate concavity, can remove this bulge, though such an approach is \emph{ad hoc.}

The conservative nature of the SDL test near model singularities 
may be partially explained by the vanishing of the gradients of the equality constraints at such points. This implies the constraints will be nearly satisfied at nearby points off the model, and (if there are only equality constraints) the incomplete U-statistics may be close to 0 as well. Notice this is quite different from the behaviour at non-singular boundary points of a model as in \Cref{ssec:convex}.

Finally, note in \cref{fig:hd2}(a) the reduced stochasticity of the rejection region boundary for the linear constraint vs.~the cubic. This suggests that using low degree  constraints (when possible) is preferable.


\section{Hypothesis Testing and Inference of Phylogenetic Trees}\label{sec:CFN} 

We next explore the performance of the SDL method for testing and inference of phylogenetic tree topologies through \emph{phylogenetic invariants}. Introduced in \cite{CavFel87,lake1987rate}, phylogenetic invariants are polynomials vanishing on pattern distributions in genetic alignments. They have have
been widely studied and used to establish parameter identifiability for various models \cite[e.g.,][]{ AllmanRhodes04, Sturmfels2005, AllmanRhodes06,  AllmanRhodes2008, Draisma2008, casanellas2008geometry}, 
and underlie several inference methods
\cite{casanellas2006, Chifman2014, fernandez2016, allman2017, Casanellas2023}. (See \cite{sullivant2018algebraic} for a general introduction.) Viewing invariants as equality constraints on the data distribution,
the SDL method offers a new statistical approach for their use.

\subsection{The CFN model and its semialgebraic descriptions}\label{ssec:CFNsemi}

We focus on the Cavender-Farris-Neyman (CFN) model for 2-state sequence evolution on 4-leaf binary trees \cite[Chapter 8]{semple2003phylogenetics}), a higher-dimensional model
than those considered in previous sections. The two states 0, 1 usually represent purines (A,G) and pyrimidines (C,T) in DNA sequences. 

\begin{figure}[htb!]
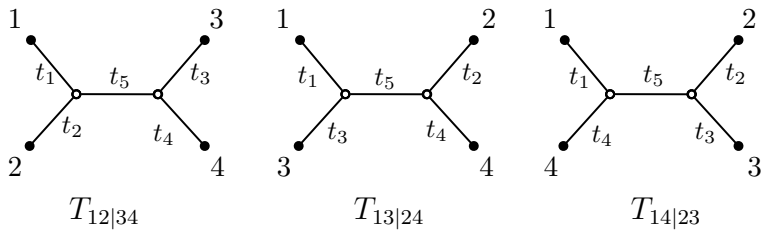

  \include{topologies}
  \vspace{-1cm}
  \caption{The 4-leaf binary tree topologies, with edge lengths $t_i
  $.  The names $T_{xy|zw}$ indicate the partition of  leaves induced by the central edge.}
  \label{fig:3topologies}
\end{figure}

Let $T$ be one of the leaf-labeled trees of \cref{fig:3topologies}. Arbitrarily introducing a tree root representing the common ancestor of $1,2,3,4$,
the CFN base substitution process on an edge of length $t$
 is given by a two-state, continuous-time, time-reversible Markov chain, with equal state transition rates and expected number of transitions $t$, proceeding from the parent to child node. Time reversibility ensures this model is independent of root location.

The
CFN model on $T$ is the marginal distribution of states on leaves, as internal tree states are hidden, represented by the $2\times 2\times 2\times 2$ tensor
\begin{equation*}
  p = (p_{ijkl})_{i,j,k,l\in \left\{0,1\right\} }, \ \ 
  p_{ijkl} = \P\left[X_{1}=i,X_{2}=j,X_{3}=k,X_{4}=l\right],
\end{equation*}
where $X_i$ is the state at leaf $i$. This models a single site in a sequence alignment arising from tree $T$, with all aligned sites viewed as i.i.d. samples.
Fixing the topology of $T$, but varying edge lengths gives a parametrized family of models on $T$.
Reparametrizing with $\theta_i:=e^{-2t_i}$ gives this family a polynomial parameterization:
\begin{equation*}\label{eq:param_probs} \phi_T:(0,1]^5 \to \Delta^{15}\subset \mathbb{R}^{16}.\end{equation*}
By the term \emph{CFN model} on a topological tree $T$, we mean the parametrized family of statistical models
given by the image $\mathcal M$
of this map. As the polynomial image of a semialgebraic set, $\mathcal{M}$ is a semialgebraic subset of
$\Delta^{15}$. The polynomials vanishing on this set, and thus all polynomial equality constraints for the model, form an ideal $I_T$, which can be computed using Gr\"obner basis techniques with computational algebra software such as \texttt{Macaulay2} \cite{M2}.

The set of points on which the polynomials in $I_T$ vanish form an algebraic variety $V_T\supset \mathcal M$. Both $V_T$ and $\mathcal M$ are of dimension 5, matching the number of numerical edge length parameters on $T$.
$I_T$ is finitely generated, and any choice of generators gives sufficient equality constraints to define $V_T$.

For $T=T_{12|34}$, one set of 
defining equations for $V_T$ is the 2 quadratic constraints: 
\begin{equation}
  \label{eq:BD}
  f_1 = \det\begin{pmatrix}
      q_{0000} & q_{0011} \\ q_{1100} & q_{1111}
  \end{pmatrix} = 0,  \qquad
  f_2 = \det\begin{pmatrix}
      q_{0101} & q_{1001} \\ q_{0110} & q_{1010}
  \end{pmatrix} = 0,
\end{equation}
{\scriptsize\begin{align*}
\text{where }  
    q_{0000} := \ & p_{0000} + p_{0001} + p_{0010} + p_{0011} + p_{0100} + p_{0101} + p_{0110} + p_{0111}, 
    \\
    q_{1111} :=\ &p_{0000} - p_{0001} - p_{0010} + p_{0011} - p_{0100} + p_{0101} + p_{0110} - p_{0111},\\
    q_{0011} :=\ &p_{0000} - p_{0001} - p_{0010} + p_{0011} + p_{0100} - p_{0101} - p_{0110} + p_{0111}, \\ 
    q_{1100} :=\ &p_{0000} + p_{0001} + p_{0010} + p_{0011} - p_{0100} - p_{0101} - p_{0110} - p_{0111}, \\
    q_{1010} :=\ &p_{0000} + p_{0001} - p_{0010} - p_{0011} + p_{0100} + p_{0101} - p_{0110} - p_{0111}, \\ 
    q_{0101} :=\ &p_{0000} - p_{0001} + p_{0010} - p_{0011}  - p_{0100} + p_{0101} - p_{0110} + p_{0111}, \\
    q_{0110} :=\ &p_{0000} + p_{0001} - p_{0010} - p_{0011} - p_{0100} - p_{0101} + p_{0110} + p_{0111}, \mbox{ and} \\ 
    q_{1001} :=\ & p_{0000} - p_{0001} + p_{0010} - p_{0011} + p_{0100} - p_{0101} + p_{0110} - p_{0111},
\end{align*}}
along with the 9 linear equations:
{\scriptsize\begin{equation}\label{eq:degree-1-generators}
  \begin{aligned}
    (p_{0000}+p_{0001}+p_{0010}+\ldots+p_{1111}) -1=0, \hskip .7in\ \\
    p_{0000}-p_{1111}=0, \quad p_{0001}-p_{1110}=0, \quad p_{0010}-p_{1101}=0, \quad p_{0011}-p_{1100}=0, \\
    p_{0100}-p_{1011}=0, \quad p_{0101}-p_{1010}=0, \quad p_{0110}-p_{1001}=0, \quad p_{0111}-p_{1000}=0.
  \end{aligned}
\end{equation}}
The linear polynomials are \emph{model invariants}, since they are zero for any of the 3 topological trees, and the quadratics are \emph{topology invariants}, as they are not zero for some tree \cite{Casanellas2017}. 

 Computation shows (see {Supplementary Materials}) that $V_T$'s singularities are  
\begin{equation*}
  (V_T)_{\rm sing} = \left\{\phi_{T}(\theta_1,\ldots, \theta_5): \theta_1,\ldots, \theta_5\in [0,1]\text{ and } \theta_1=\theta_2=0 \text{ or }\theta_3=\theta_4=0 \text{ or }\theta_5=0\right\}.
\end{equation*}
Since $\theta_i=0$ corresponds to $t_i=\infty$, which produce sequence data that is uncorrelated at the ends of an edge, such singularities are unlikely to be relevant to empirical analyses.

For the stochastic model $\theta_i\in (0,1]$, one finds $\mathcal{M}\subsetneq V_{T}\cap \Delta^{15}$, but imposing
additional polynomial inequalities restricts from $V_T$ to $\mathcal M$  \cite{matsen2008fourier, Kosta2018}.
 In particular, the quadratic inequality 
\begin{align}\label{eq:ineq_internal}
{q_{0101}q_{1010} + q_{1001}q_{0110} -2\left( q_{0011}q_{1100}\right) } \leq 0
\end{align}
expresses $t_5\ge 0$,
with similar inequalities for the pendant edges.
We consider only the inequality in \cref{eq:ineq_internal}, as it is the only one that changes for different tree topologies.

\medskip

While \cref{eq:BD,eq:degree-1-generators} gives one  set of equality constraints for $V_T$, others are equally natural. 
 We say that a topology invariant $F\in I_{T}$ is \emph{partially distinguishing} if there exists a tree $T'\neq T$ on the same taxa such that  $F\in I_{T'}$ as well. If $F$ is not partially distinguishing, we say that it is \emph{completely distinguishing}.
We consider the following five specific choices of quadratic topology invariants that, together with the linear invariants, generate $I_T$. Explicit formulas are given in \Cref{app:CFN-generators}.

\begin{enumerate}
  \item[(CDD)] \label{item:CDD} \emph{Completely Distinguishing Determinantal}: These are derived from the determinantal polynomials in \cref{eq:BD} together with \cref{eq:degree-1-generators} (see \Cref{app:CFN-generators} for the explicit construction).
  \item[(PDR)] \label{item:PDR} \emph{Partially Distinguishing Rank}: These constraints are indirectly obtained from $3\times 3$ minors of a certain flattening of the tensor $p$ described in \cite{AllmanRhodes2008}.
  \item[(PDM)] \label{item:PDM} \emph{Partially Distinguishing Minimal}: This is a minimal
  basis obtained by applying the \texttt{mingens} function of \texttt{Macaulay2} to the kernel of $\phi_T$.
  \item[(CDR)] \label{item:CDR} \emph{Completely Distinguishing Rank}: These two polynomials are the sum and difference of the polynomials of \pdr). 
  \item[(CDM)] \label{item:CDM} \emph{Completely Distinguishing Minimal}: These two polynomials are  the sum and difference of the invariants of \pdm.
\end{enumerate}

\subsection{Data simulation}\label{sec:CFN-data}
\begin{figure}[ht]
  \input{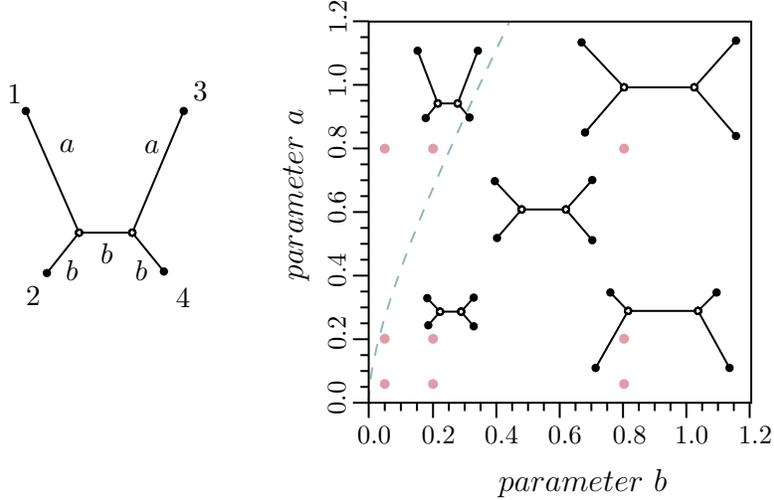}
  \caption{Left: The tree $T_{12|34}$ with edge lengths $t_1=t_3=a$ and $t_2=t_4=t_5=b$, in units of expected number of substitutions per site. 
Right: The tree space, with $a,b$ varying from $0$ to $1.2$. In red, nine parameter pairs with $a,b\in\{0.05,0.2,0.8\}$. The dashed blue curve is the lower boundary of the Felsenstein zone, defined by $\theta(b)^2-2\theta(a)+\theta(a)^2>0$ for $\theta(t)=e^{-2t}$ \cite{Felsenstein1978}.}
  \label{fig:treespace}
\end{figure}

To evaluate the SDL test on the CFN model, we focused on datasets from the trees studied in \cite{treespace95}, shown in \cref{fig:treespace} (left), where tree $T_{12|34}$ has edge lengths $t_1 = t_3 = a$ and $t_2 = t_4 = t_5 = b$, for varying $a,b>0$.

A dataset consists of $n$ independent samples drawn from the multinomial distribution with parameter 
\begin{equation}\label{eq:reduced_param}
\overline{p} = \left(\overline{p}_{xxxx}, \overline{p}_{xxxy}, \overline{p}_{xxyx}, \overline{p}_{xxyy}, \overline{p}_{xyxx}, \overline{p}_{xyxy}, \overline{p}_{xyyx}, \overline{p}_{xyyy}\right) \in\Delta^{7},
\end{equation}
where $x,y$ represent distinct states in $\left\{0,1\right\}$, and the coordinates of $\overline{p}$ are 
$\overline{p}_{xxxx} = p_{0000}+p_{1111}$, $\overline{p}_{xxxy} = p_{0001}+p_{1110}$, and so forth, where $p = \phi_T(\theta_1(a), \theta_2(b), \theta_3(a), \theta_4(b), \theta_5(b))$ and $\theta_i(t) = e^{-2t}$.
We thus assume \textit{a priori} that the linear constraints of \cref{eq:degree-1-generators} hold, allowing us to reduce the length of the data vector of length 16 to 8, and subsequently ignore those equalities.

We consider two  collections of datasets:
\begin{enumerate}

    \item \emph{Collection 1.} We generated $30$ datasets of size $n=1000$ site samples for each pair of parameters $(a,b)$ with $a$ and $b$ ranging from 0 to 1.2 in increments of 0.05.
    \item \emph{Collection 2.} We selected nine parameter pairs to be analysed in greater detail, with $a,b\in \left\{0.05,0.2,0.8\right\}$. We generated  $1000$ datasets for each choice of parameters, with each dataset consisting of $n=1000$ site samples.
\end{enumerate}

Collection 1 samples from throughout the tree space of \cref{fig:treespace} (right). The upper left region is the ``Felsenstein zone,'' leading to datasets susceptible to long branch attraction, which makes accurate tree inference by standard methods difficult \cite{Felsenstein1978,Hendy1989}. The nine parameter choices underlying Collection 2 are indicated in red dots in the figure.

\subsection{SDL test parameters and hypotheses}

To apply the SDL test we must choose its test parameters, $m, n_1,N,A$ as well as a partial symmetrization level $s$. For our data sets of size $n=1000$, preliminary investigations led us to use    
$$
m = 12, \quad
N = 1000, \quad
n_1 = 80, \quad
A = 5000, \quad
s = 100 
.$$
Large values of $n_{1}, N, s$ lead to substantial computation, but the values above gave a good balance between performance and runtime. For example, no major impact on the test results was observed compared with $n_1 = 500$ and $N=5000$.

We consider each of the five different sets of quadratic equality constraints presented in \Cref{ssec:CFNsemi}. We also increased the number of polynomial constraints by adding $r=20$ random convex combinations of the original ones. 

We denote by $H_{12|34}$ the hypothesis that the true tree topology is $T_{12|34}$, and similarly $H_{13|24}$ and $H_{14|23}$. 
Constraints for tests of $H_{13|24}$ and $H_{14|23}$ can be found by permuting taxon labels from those for $H_{12|34}$, and are given in the {Supplementary Materials.}
Since our simulated data is always sampled from a $T_{12|34}$ tree, in our experiments $H_{12|34}$ is always the true hypothesis and the other two are false.

\subsection{Hypothesis tests results} \label{sec:CFNpvalues}

We compute $p$-values from simulated data to test several different null hypotheses.

\subsubsection{Collection 1} \label{ssec:p-vals-Data-1}

As an initial exploration of the behaviour of the SDL test, we examined the  distribution of all $p$-values from Collection 1 for each of the three hypotheses $H_{12|34}, H_{13|24}$, and $H_{14|23}$. 
Aggregating $p$-values across a wide range of parameter values $(a,b)$ in a single histogram gives insight into the overall behaviour of the test.

Since varying the model constraints can affect test behaviour (\Cref{ssec:convex,ssec:highdegree}), we created histograms for the five sets of quadratic equality constraints in \Cref{ssec:CFNsemi}. No other constraints, including \cref{eq:ineq_internal}, were used.
For each constraint set, we also added $r=20$ random convex combinations of the resulting inequalities. 

\begin{figure}[ht]
  \centering
  \includegraphics[scale=0.12]{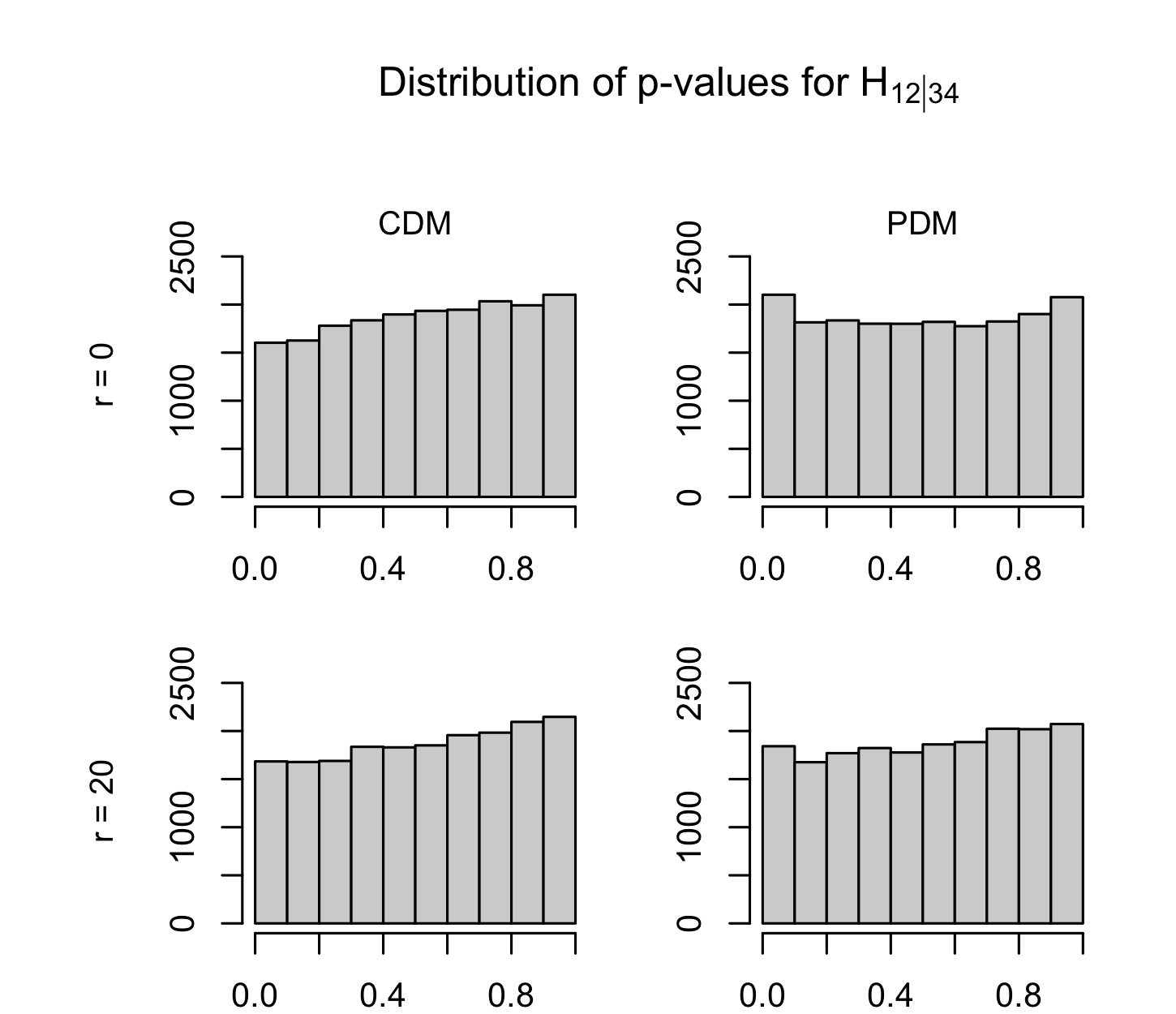}
    \includegraphics[scale=0.12]{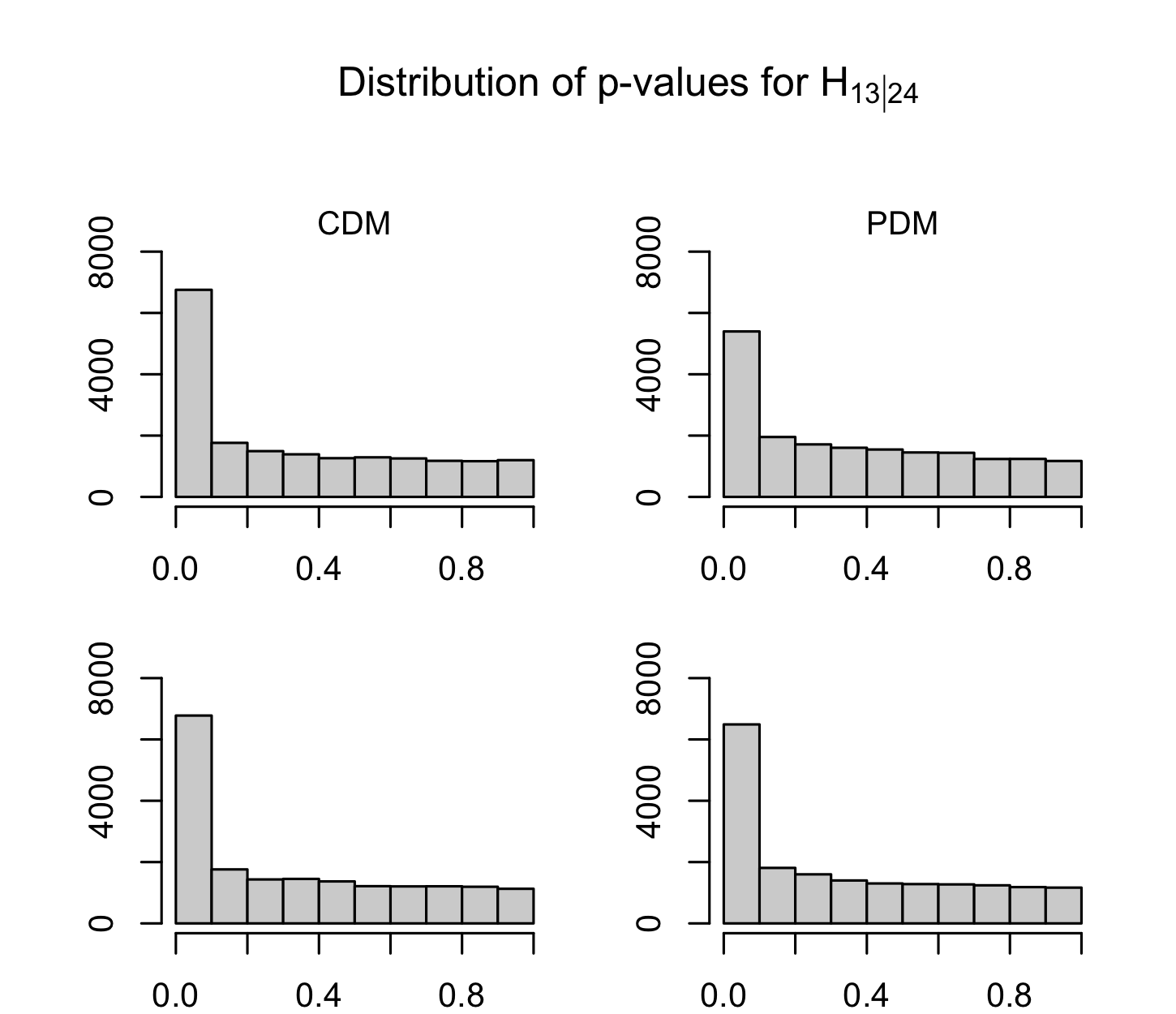}\\
  \caption{Aggregated $p$-values for a test of the true null hypothesis $H_{12|34}$ (left) and a false null hypothesis $H_{13|24}$ (right) for datasets in Collection 1. Constraints sets \cdm and \pdm, and number of convex combinations $r=0$ and 20 are varied.} 
  \label{fig:combined-pvalue-histogram-12-34}
\end{figure}

\cref{fig:combined-pvalue-histogram-12-34} presents aggregated $p$-values for each of 4 conditions (\cdm and \pdm, $r=0$ and 20), for
the true null hypothesis $H_{12|34}$ (left) and a false null hypothesis $H_{13|24}$ (right). 
(See \Cref{ssec_app:p-vals-Data-1} for all five constraint sets.)

For $r=0$  and the true $H_{12|34}$, the  \pdm set shows anti-conservative behaviour, with an excess of small $p$-values.  The \cdm constraints, on the other hand, shows conservative behaviour, with an excess of large $p$-values. 
For the false $H_{13|24}$, the \cdm constraints  gave a greater concentration of $p$-values near zero compared to the \pdm  constraints, suggesting greater power.

Increasing $r$ did not substantially change the behaviour of the test with the \cdm set.
However, for the \pdm set, increasing $r$ had two important and beneficial effects: first, it decreased the number of small $p$-values when testing $H_{12|34}$, and second, it increased the number of small $p$-values when testing $H_{13|23}$. This suggests for \pdm, the addition of convex combination constraints simultaneously made the test more conservative as well as increased its statistical power. 
However, the effect of adding convex combinations constraints highly depends on the choice of starting constraints, as we discuss in \cref{ssec_app:p-vals-Data-1}.

Although these effects of increasing $r$ might appear relatively small, they are based on aggregated $p$-values from a large parameter regime, and it is possible specific regions of the parameter space might exhibit more substantial effects. 
In \Cref{sec:disappearance-of-lba} we show this is the case, by analysing a particular choice of parameters $(a,b)$ within the Felsenstein zone (a region of particular interest for the phylogenetics community).

\subsubsection{Collection 2} 
\begin{figure}[ht]
  \includegraphics[scale=0.10]{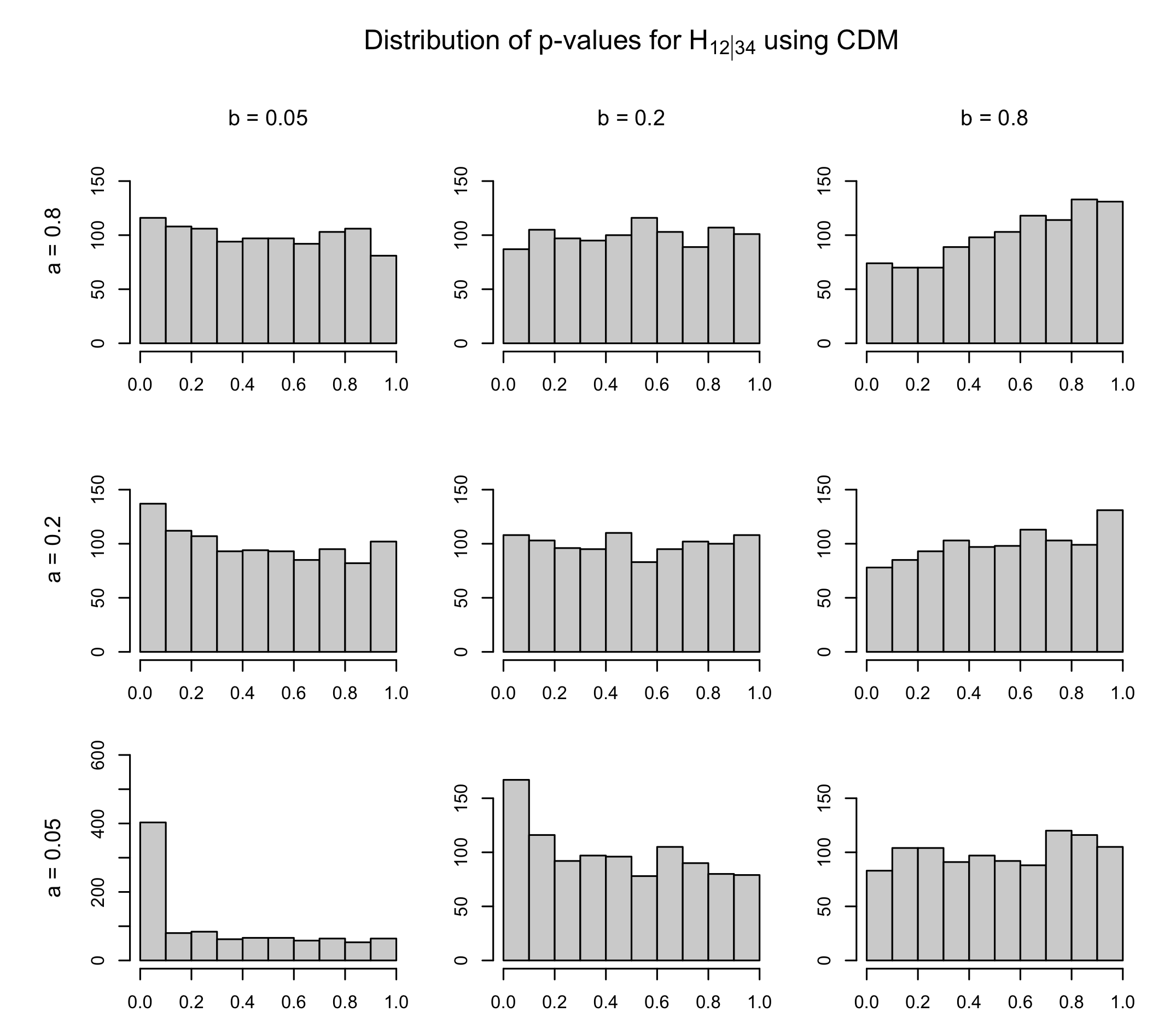}
  \includegraphics[scale=0.10]{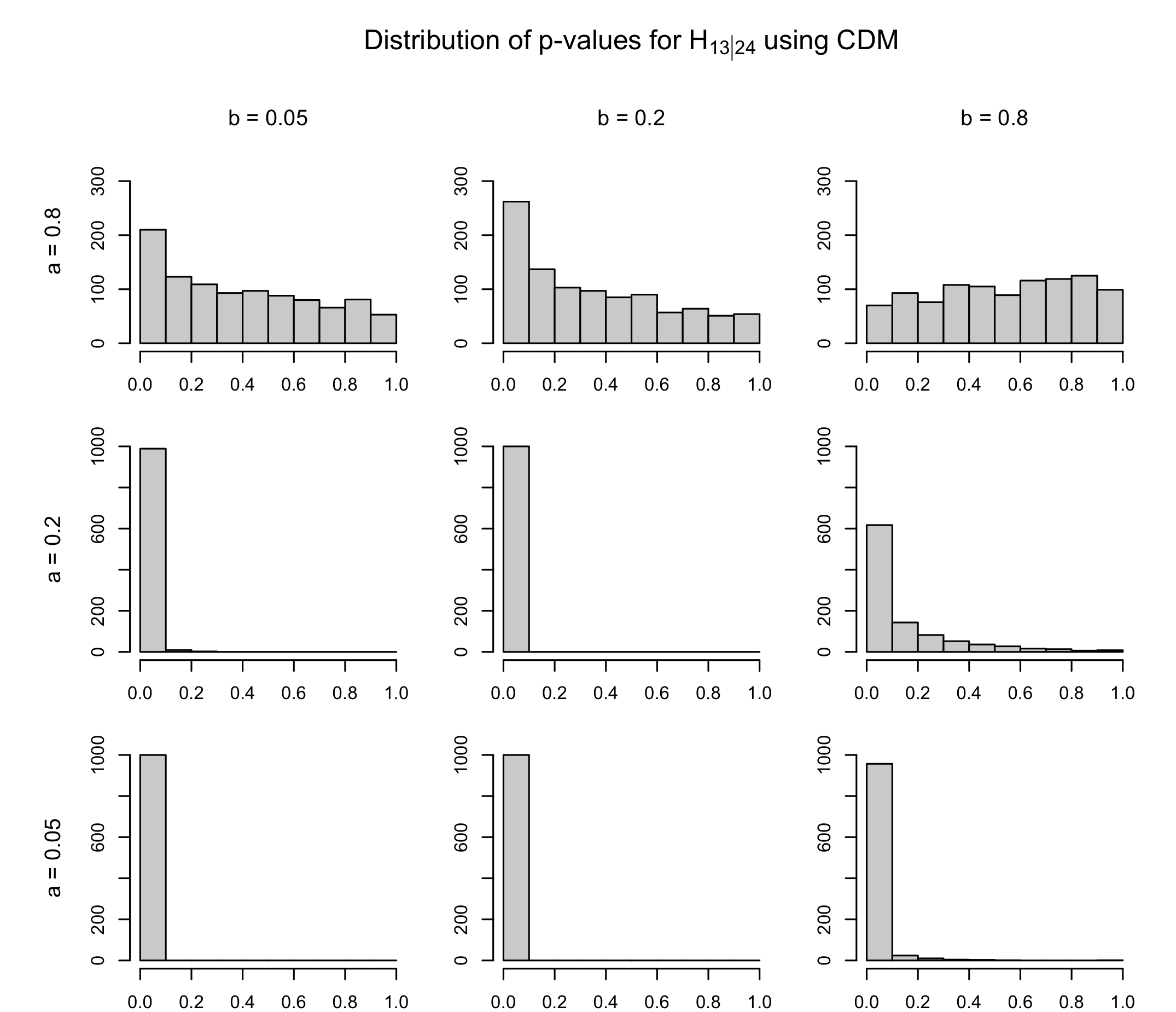}\\
  \vspace{0.2cm}
  \includegraphics[scale=0.10]{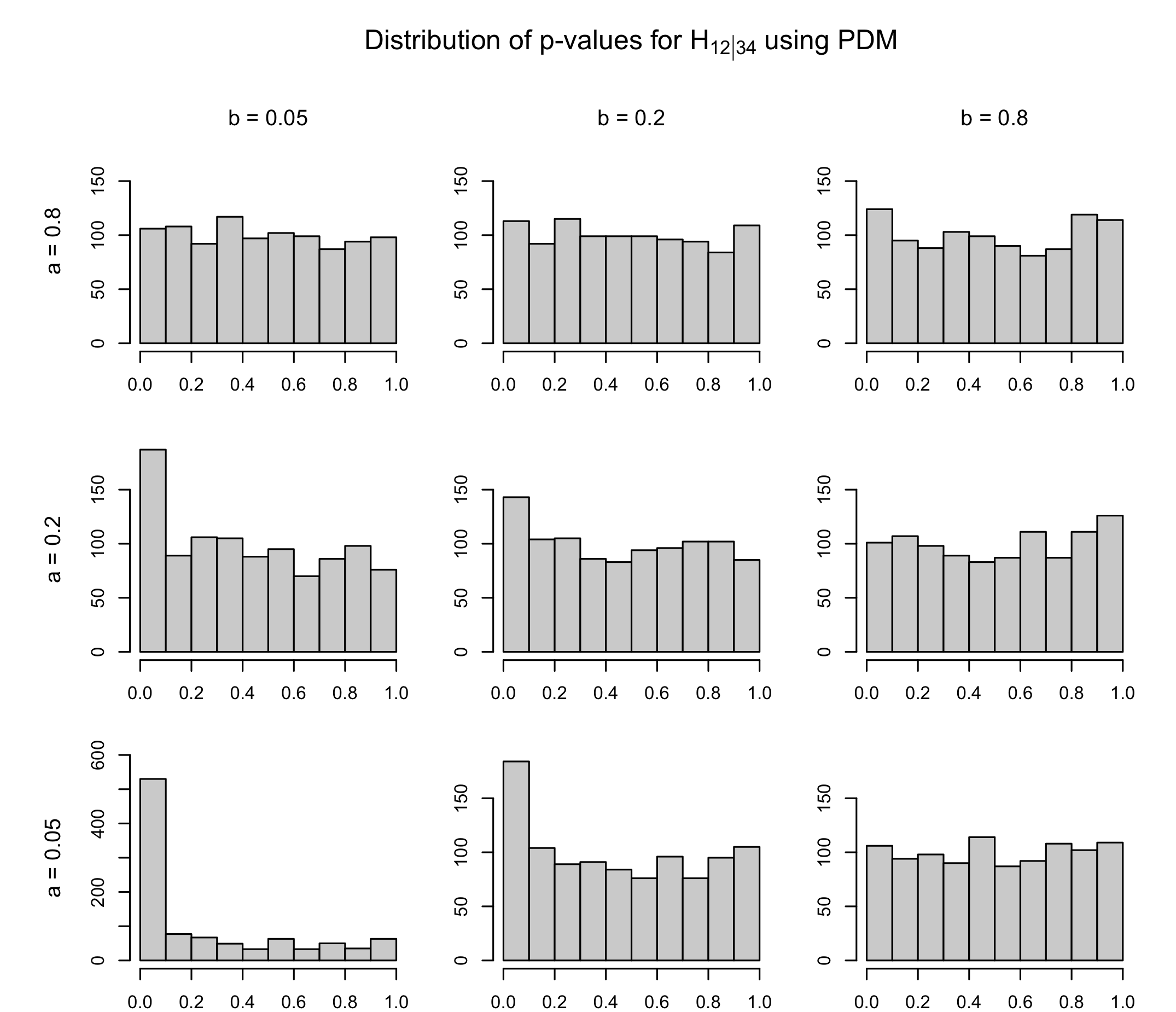}
  \includegraphics[scale=0.10]{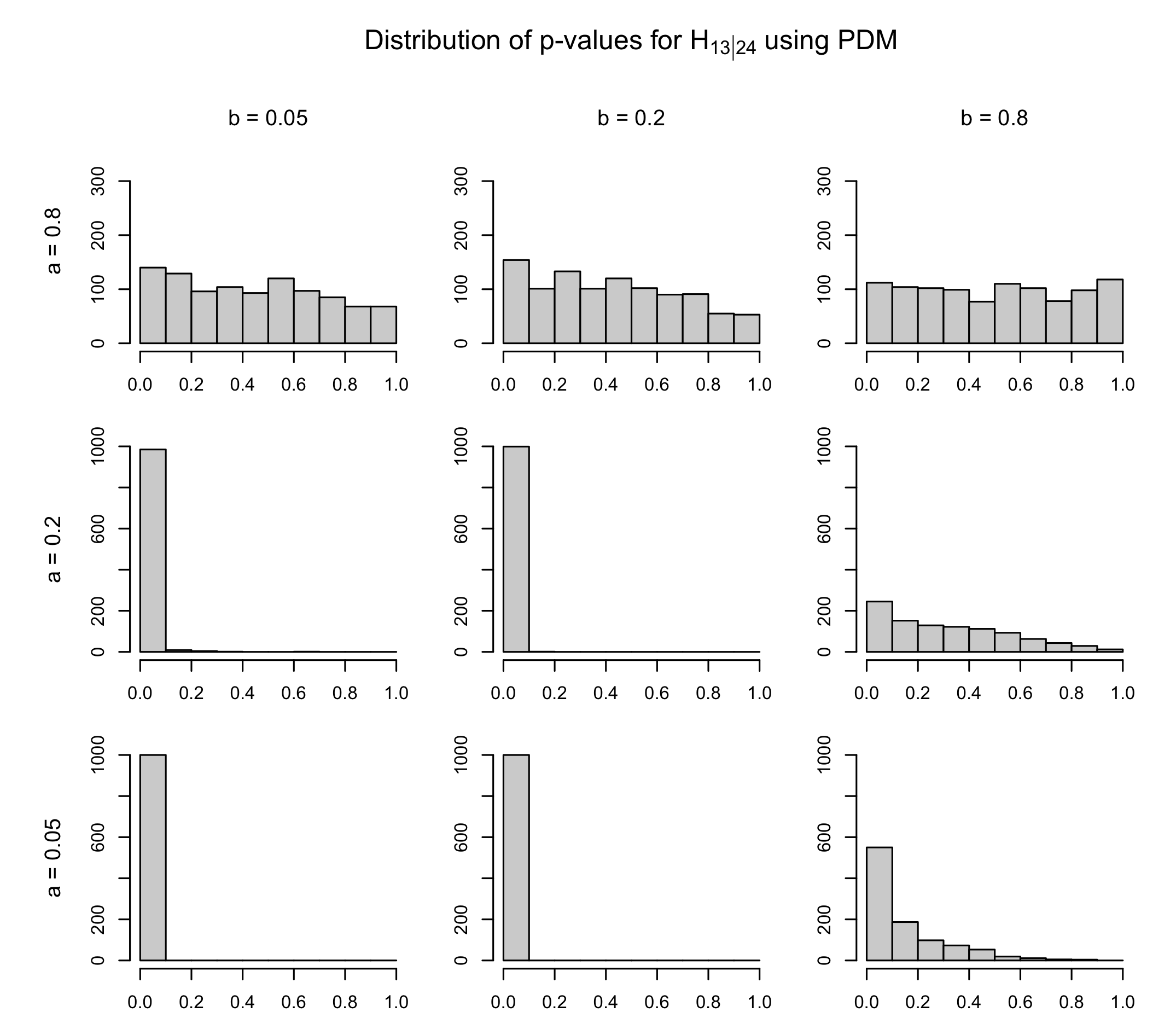}
  \caption{$p$-values obtained from the SDL test on Collection 2 for different constraint sets:  \cdm (top 3 rows) and \pdm (bottom 3 rows).  The hypotheses tested are $H_{12|34}$ (left 3 columns) and $H_{13|23}$ (right 3 columns), with $r=0$.
  }  \label{fig:$p$-values-for-particular-CFN-models--1234-and-1324}
\end{figure}

We next examine the performance of the SDL test more closely, for the 9 particular edge parameters shown in \cref{fig:treespace}.
\cref{fig:$p$-values-for-particular-CFN-models--1234-and-1324} shows histograms of 1000 $p$-values, with the test differing only in use of the topology constraints  \cdm and \pdm; in both cases the internal branch inequality \cref{eq:ineq_internal} is not used. 

Despite this seemingly small difference, the SDL test with the \cdm polynomials tends to be both more conservative and more powerful than when compared to the \pdm polynomials. \cref{fig:$p$-values-for-particular-CFN-models--1234-and-1324} illustrates that when testing the true hypothesis $H_{12|34}$ \pdm is more likely have have $p$-values close to zero for 8 out of the 9 choices of model parameters. On the other hand, when testing the false $H_{13|24}$, both \cdm and \pdm constraints produce small $p$-values for $a,b\in \left\{0.05,0.2\right\}$. However, for 4 of the remaining 5 choices for $(a,b)$, the test utilizing the \cdm constraints gave small $p$-values for the incorrect null hypothesis substantially more often than the test utilizing the \pdm constraints. Results for $H_{14|23}$ (not shown) were similar to those for  $H_{13|24}$.

The SDL test performed quite poorly when testing the correct model hypothesis $H_{12|34}$ for trees with short edge parameters. For example,  when $(a,b)=(0.05,0.05)$ the test produced far too many small $p$-values, regardless of whether the polynomials were the \cdm or \pdm sets, though worse for the second.

\subsubsection{Effect of internal edge constraint} \label{ssec:internal-branch-matters}

\begin{figure}[htb!]
 \centering
    \includegraphics[scale=.12]{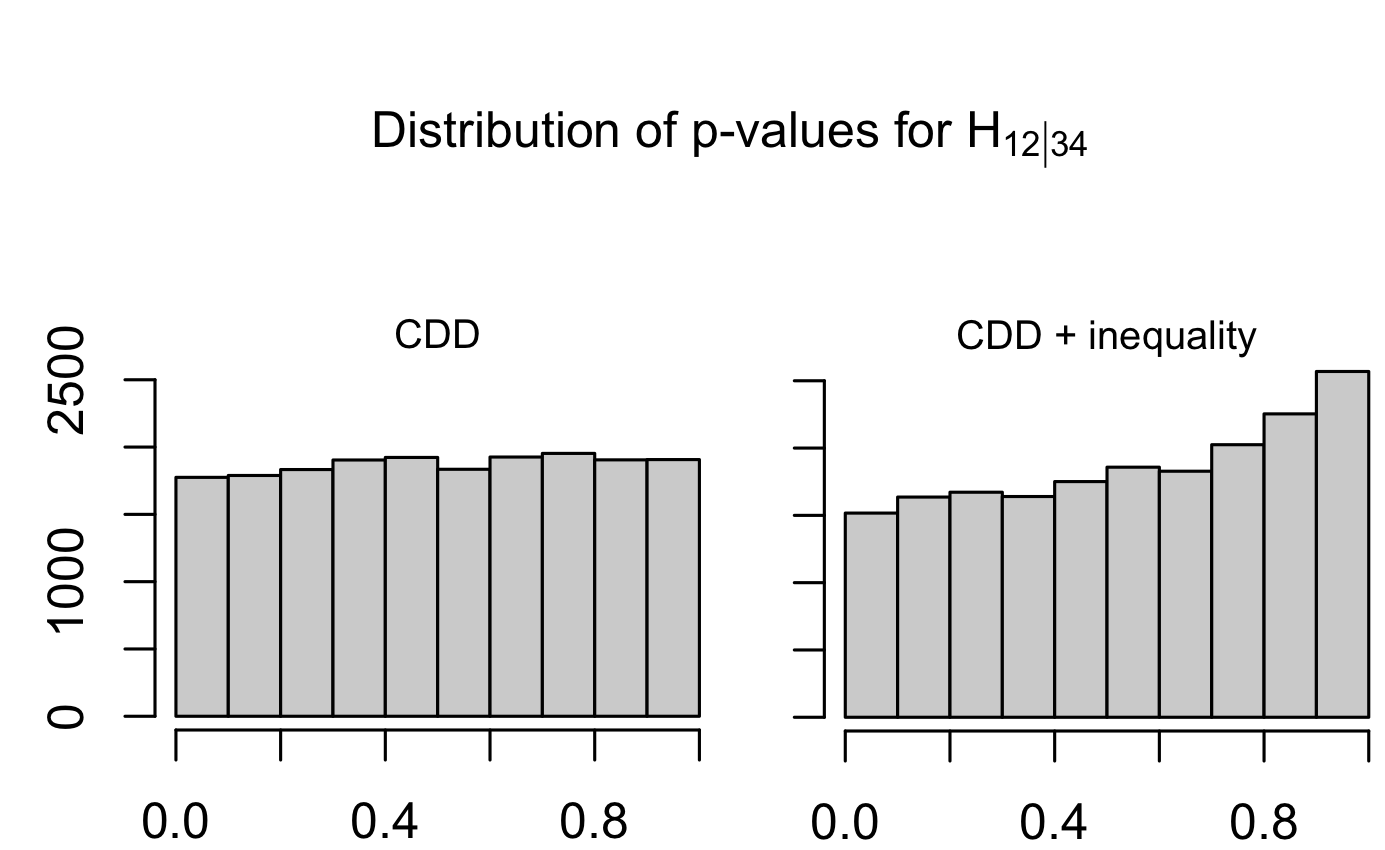}
    \includegraphics[scale=.12]{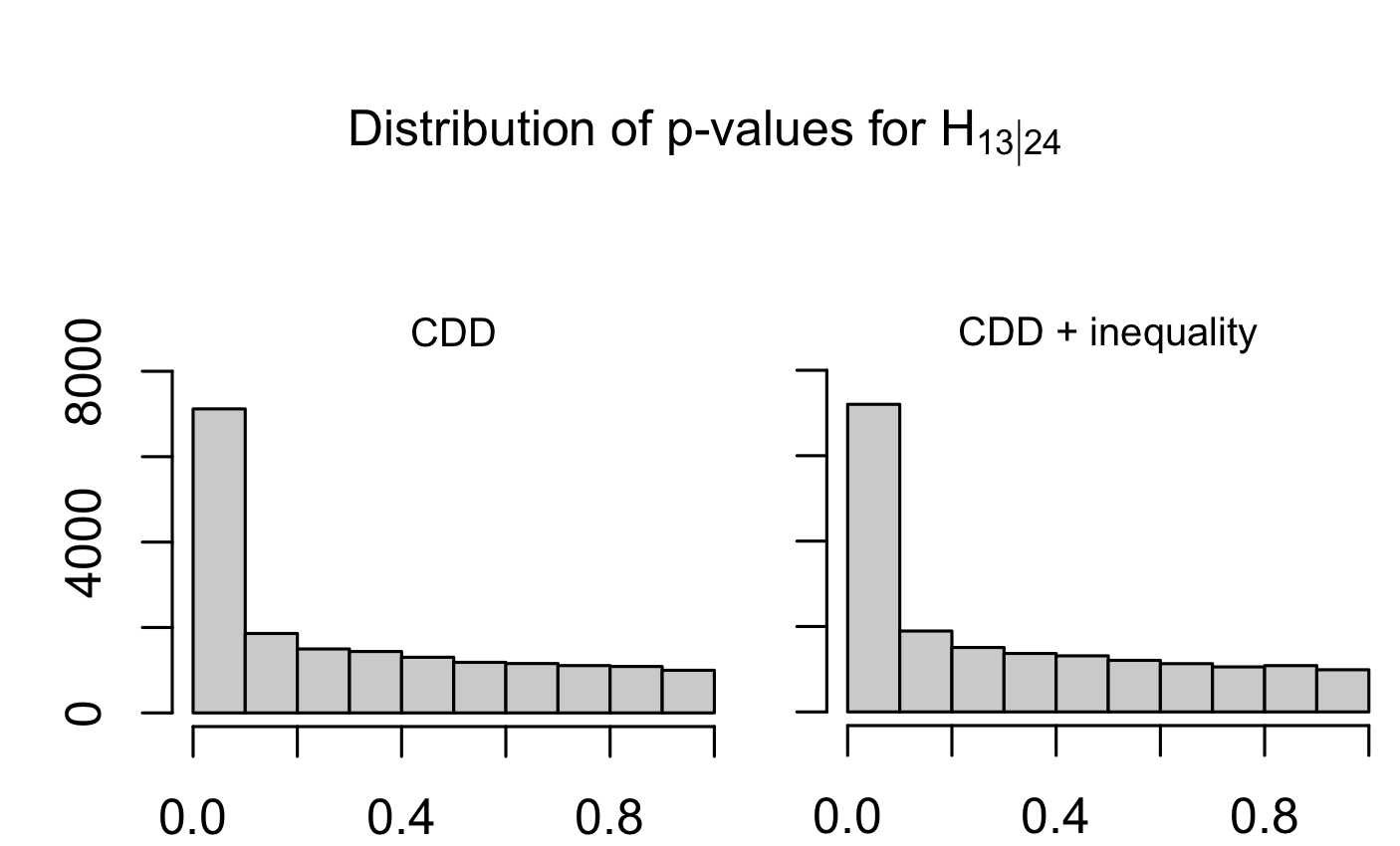}
    \includegraphics[scale=.12]{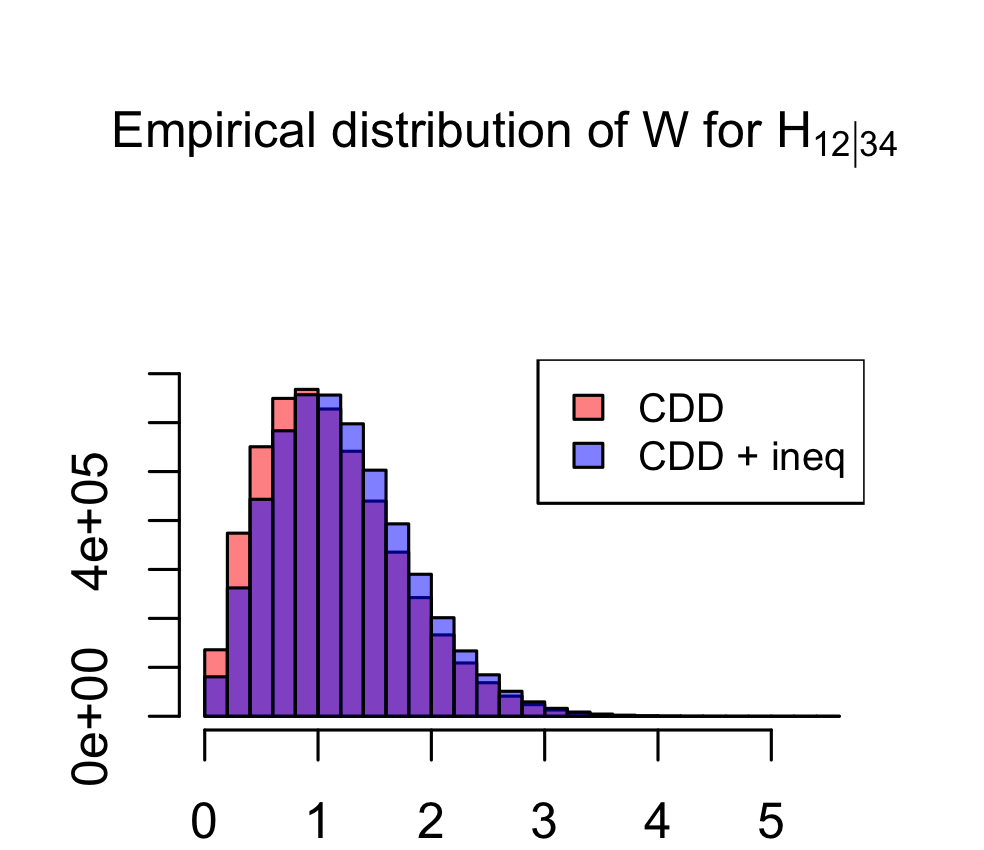}
    \caption{Histograms of $p$-values when testing $H_{12|34}$ (left two) and $H_{13|24}$ (middle two) showing the effect of including the internal edge constraint. The plot for $H_{14|23}$ is omitted because it is similar to that of $H_{13|24}$. 
 Right: Histogram of $W^{(1)},\ldots, W^{(A)}$, approximating the test distribution $\mathcal{T}_c$ for \cdd (red) and \cdd + inequality (blue) for aggregate data from $1000$ trees with random parameters $a,b\in(0,1.2]$.
    }
  \label{fig:combined-pvalue-histogram-with-internal-branch-constraint--for-1234-and-1324}
\end{figure}

We also investigated the effect on the SDL test of augmenting the \cdd set with the inequality of \cref{eq:ineq_internal}, expressing that the tree's internal edge length is non-negative. 
 \cref{fig:combined-pvalue-histogram-with-internal-branch-constraint--for-1234-and-1324} compares the distribution of $p$-values using the CDD generating set with the internal edge inequality verses without it, amalgamating all test results for Collection 1 on a true $H_{12|34}$ and false $H_{13|24}$null hypothesis. Tests of the false $H_{14|23}$ were similar, and are omitted. 

Based on the aggregated $p$-values, including the internal
edge inequality appears to make the test  more conservative, with no appreciable change in power. These results were essentially unchanged for $r=10$ and $20$.
An analogous analysis (not shown) considered test results for the datasets of Collection 2, not amalgamating over different parameters. 
We observed a similar behaviour as in \cref{fig:combined-pvalue-histogram-with-internal-branch-constraint--for-1234-and-1324}: Testing $H_{12|34}$ gave an increase in the number of large $p$-values and a decrease in the number of very small $p$-values. In fact, for 8 of the 9 parameters, there was a reduction in the number of $p$-values less than $0.1$, with the exceptional case, $(a,b)=(0.2,0.2)$, showing no difference. 

This effect of adding a constraint may seem counter-intuitive.
By further restricting the model, one might think
the test would be more inclined to reject a true hypothesis $H_{12|34}$. Indeed, the test statistic $\mathcal{T}$ is defined in \cref{eq:def-of-test-statistic} as a maximum over all constraints, so an additional constraint can only lead to larger $\mathcal{T}$ values. However, the critical threshold $\mathcal{T}_c$, as well as the quantities $W^{(1)},\ldots, W^{(A)}$ used to approximate it (see \cref{eq:def-of-T_c,eq:def-of-W}) also correspondingly increase. 
For our simulations, we did not observe a significant increase in the value of $\mathcal{T}$ when the new inequality was included, but we did observe a shift in the distribution of $W$ to larger values across many parameter choices. This is clearly shown in \cref{fig:combined-pvalue-histogram-with-internal-branch-constraint--for-1234-and-1324} (right) comparing the amalgamated distribution of $W$ with and without the internal branch inequality for aggregate data from 1000 trees drawn randomly from the treespace shown in \cref{fig:treespace}. Similar tests with data drawn from fixed trees support this conclusion.

For the false $H_{13|24}$, the aggregate histogram plots in \cref{fig:combined-pvalue-histogram-with-internal-branch-constraint--for-1234-and-1324} (middle two plots) shows no effect from including the internal branch inequality. However, in testing $H_{13|24}$ and $H_{14|23}$ on Collection 2 (not shown), we observed an effect dependent on the region of the parameter space. When $a,b\leq .2$, the inclusion of the internal edge inequality had no appreciable effect on the observed distribution of $p$-values, which were overwhelmingly concentrated near zero regardless. However, for $(a,b)\in \left\{(.8,.05), (.8,.2),  (.2,.8)\right\}$, including the internal branch inequality increased the number of small $p$-values. However $(a,b) = (.05,.8)$ with $H_{13|24}$ was exceptional, showing almost no difference.

\subsection{SDL-Based Phylogenetic Inference}\label{sec:CFNinference}

We next investigate the potential of the SDL test as an inference method for tree topology from sequence data. Standard statistical approaches for this depend on repeated calculation of a likelihood function depending not only on the tree topology but also its edge lengths, with a search or MCMC exploration over all these parameters. The SDL methodology suggests a procedure to avoid consideration of the edge length ``nuisance parameters'' and likelihood calculations by first calculating SDL $p$-values for each possible tree topology and then selecting the tree with the highest $p$-value. We implemented this procedure for the CFN model with 4 taxa.

\subsubsection{Performance for differing constraint sets}
\label{sec:performance-for-differeng-constraint-sets}

\begin{figure}[htb!]
  \includegraphics[scale=0.12]{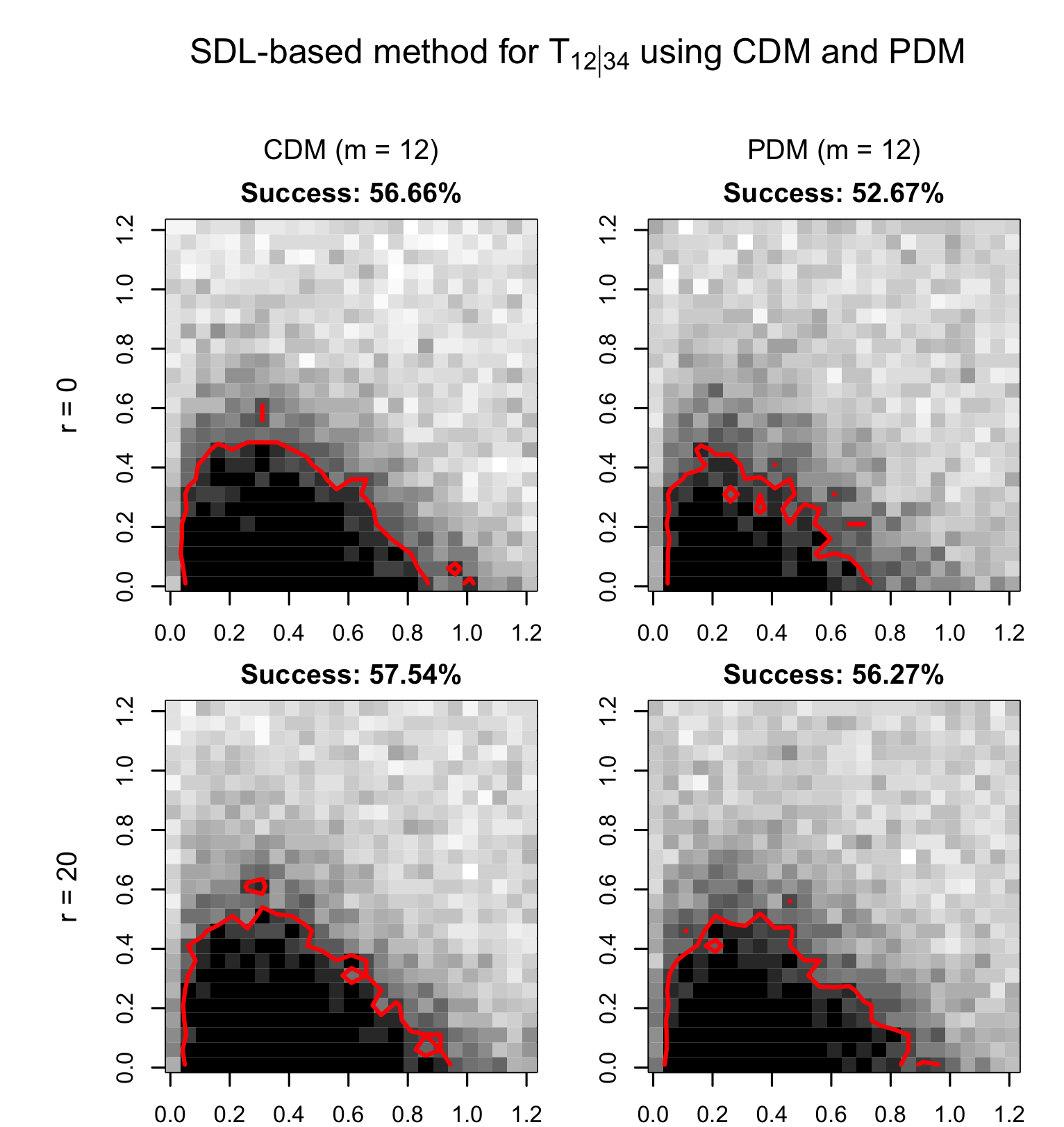}
  \includegraphics[scale=0.12]{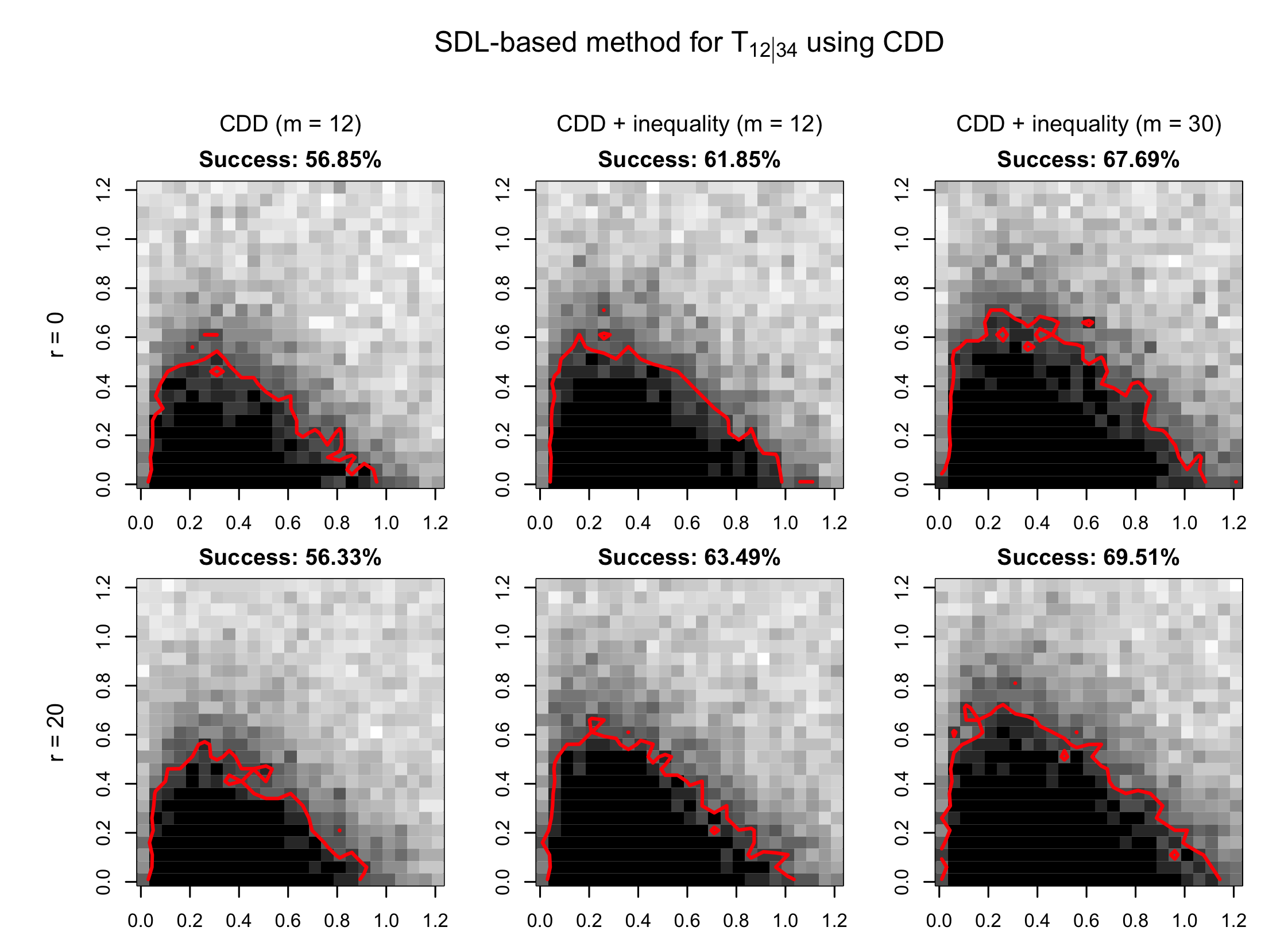}\\
  \caption{Performance of the SDL test for inferring the tree topology $T_{12|34}$ using different constraint sets and values of $m$.  Left: \cdm and \pdm constraints with $m=12$. Right: \cdd constraints with $m=12$, \cdd with the inequality of \cref{eq:ineq_internal} and $m=12$, \cdd with the inequality and $m=30$. Rows vary the number of convex
    combinations, $r=0$ and $r=20$. Grey levels represent the frequency of correctly inferring
    the topology for edge length pairs $(a,b)$ (black 100\%, white 0\%).}
  \label{fig:treespace_inequality}
\end{figure}

\cref{fig:treespace_inequality} shows results of this method applied to $p$-values from Collection 1, following a standard graphical depiction introduced in \cite{treespace95}. The columns of plots correspond to different choices of constraint sets, and the rows to $r=0$ and $20$. Within each plot, each pixel corresponds to a pair $(a,b)$ of edge length parameters, as in \cref{fig:treespace}. Grey levels indicate the frequency of inferring the true tree topology (black=100\%, white=0\%).
The red curves demarcate a region of good performance where correct inference occurs with frequency at least $90$\%. This region forms a 
right-skewed hump along the horizontal axis, similar to those produced by other well-performing methods \cite{treespace95, fernandez2016, Casanellas2023}. In this region, the SDL method performs well in part due to the $p$-values for $H_{13|24}$ and $H_{14|23}$ being highly concentrated near zero (see \cref{fig:combined-pvalue-histogram-12-34,fig:$p$-values-for-particular-CFN-models--1234-and-1324}).

The left two columns of \cref{fig:treespace_inequality} compare the use of the constraint sets \cdm and \pdm $(m=12)$, as in \cref{fig:combined-pvalue-histogram-12-34}. For $r=0$, \cdm outperformed \pdm both in terms of raw success percentage and overall shape and size of the dark region. However, this advantage was diminished with improved performance of \pdm when $r=20$ convex combination constraints were included. A similar pattern was observed for \cdr and \pdr as we show in \Cref{sec_app:model-selection-performance-for-differeng-constraint-sets}. 

This observation that for $r=0$, the use of \cdm gives better performance for model selection than \pdm is consistent with our conclusions from \Cref{ssec:p-vals-Data-1} on hypothesis testing. However, when we increase the number $r$ of convex combinations this performance gap almost entirely disappears, suggesting that the use of convex combinations may be a powerful general-purpose tool to improve performance of the SDL test, especially when the geometry of a model is not fully understood.

We next investigated whether the performance of the SDL-based inference method improved with the inclusion of the internal edge inequality, \cref{eq:ineq_internal}. \cref{fig:treespace_inequality} (right) presents results for the \cdd constraints, showing the more complete semi-algebraic model description expands the region of good performance. This reinforces previous observations about the importance of using the full semi-algebraic description for phylogenetic model selection \cite{Casanellas2021,Casanellas2021-SAQ}.  We also found that increasing $m$ from 12 to 30 resulted in a larger region of good performance. 
Despite theoretical reasons to prefer smaller values of $m$,  for model selection choosing $m=30$ resulted in a better performance, even though the $p$-value distributions showed little difference.

\medskip

\subsubsection{Comparison with other inference methods}\label{sec:comparison}
We compared the performance of SDL-based inference to that of two other phylogenetic reconstruction methods, Maximum Likelihood and the SVD method. The SVD approach is also motivated by polynomial model constraints, as it relies on the fact that a certain matrix flattening of the probability tensor $p$, determined by the tree topology, must have rank 2. Although based on essentially the same constraints as \pdr, it uses the Singular Value Decomposition of an estimate of $p$ to measure its closeness (in Frobenius norm) to one of rank 2, choosing the tree topology minimizing this. SVD-based inference has been exploited for empirical inference several in phylogenetic settings \cite{allman2017,Chifman2014,fernandez2016}. 

\begin{figure}
  \includegraphics[scale=0.2]{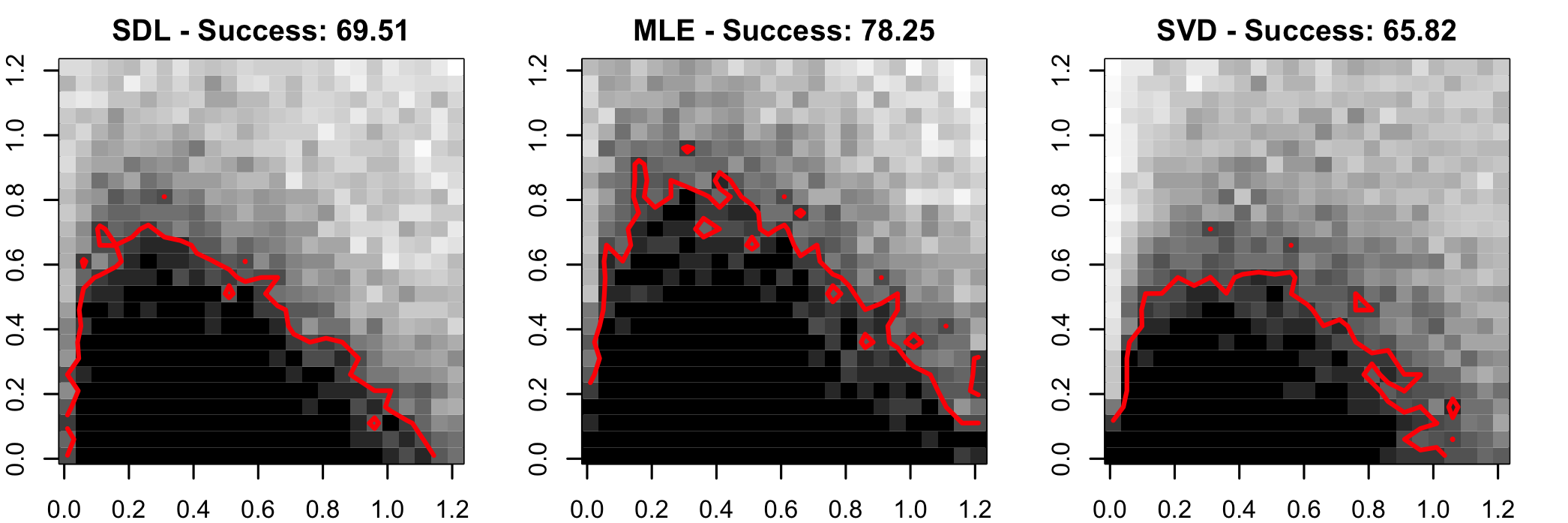}\\
  \caption{ Performance of 3 methods of topological tree inference on data from Collection 1: (left) the SDL-based inference method using the \cdd constraint set with the internal edge inequality, with $m=30$ and $r=20$, (middle) Maximum Likelihood \cite{Hill2024},(right) the SVD method.}
  \label{fig:classicalPR}
\end{figure}

\cref{fig:classicalPR} shows the performance of these three methods on identical simulated data. For the SDL approach we use the \cdd constraint set together with the internal edge inequality, $m=30$ and $r=20$. 
For the gold standard maximum likelihood estimation (MLE), calculations used the \texttt{Julia} package \texttt{FourLeafMLE.jl} \cite{Hill2024}. An important conclusion of \cref{fig:classicalPR} is that with well-chosen user-specified parameters, the SDL method can achieve overall performance approaching Maximum Likelihood, and better than the SVD approach most often used in algebraic approaches to inference.

Of special note is the performance of the SDL test for tree parameters in the Felsenstein zone (see \cref{fig:treespace}) in which correct inference is difficult for all methods. The SDL test achieved a success rate of {60.2\%}, compared to {71.65\%} for MLE and {37.13\%} for SVD. Thus while performance declined in this region, for SDL the decline was considerably less than for SVD.
We also observed that the SDL test substantially reduced (especially compared to SVD) the bias toward a specific false hypothesis (i.e., long branch attraction) in the Felsenstein zone, as is common for other methods. For more details see \Cref{sec:disappearance-of-lba}.





%

However the SDL approach is by far the most computationally intensive than the other two methods. The computational time producing this figure for the SDL-based approach was $12.57$ hours (using an \texttt{R} and \texttt{C++} implementation) versus $53.2$ minutes for MLE (in \texttt{Julia}) and $11.25$ seconds for SVD (in \texttt{R}) (see \Cref{sec:execution_time} for more details).


\section{Implementation Details and Computational Performance}\label{sec:execution_time}

The code used in our simulations is primarily written in \texttt{R} (version 4.2.2), with performance-critical parts implemented in \texttt{C++} and integrated using the \texttt{Rcpp} package (version 1.0.12). The code, which builds on the original implementation from \cite{SDL2024}, is available at:
\begin{center}
    \href{https://github.com/marinagarrote/Semialg-Hypothesis-Test-with-Incomplete-U-Stats}{github.com/marinagarrote/Semialg-Hypothesis-Test-with-Incomplete-U-Stats.}
\end{center}
All computations were performed with an Intel(R) Core(TM) i5-10400 CPU @ 2.90GHz Processor equipped with 64 GB RAM, running Debian 12.5.

Average computation times for the trinomial models presented in \Cref{sec:trinomial} are as follows. For Model 1 computing a single $p$-value took an average of {$0.21$} seconds when $m=1$, {$0.09$} seconds when $m=5$, and $0.06$ seconds when $m=15$. In the case of Model 2, the average time to compute a single $p$-value was {$0.09$} seconds for $r=0$, {$0.12$} seconds for $r=10$, and {$0.3$} seconds for $r=100$. Model 3 $p$-values required an average of {$0.11$} seconds for $s=1$, {$0.36$} seconds for $s=10$, and {$2.88$} seconds for $s=100$.

To efficiently run simulations for the CFN model in \Cref{sec:CFN}, we used the \texttt{parallel} package in \texttt{R} (version 4.2.2) and $6$ cores. 
For a fixed choice of parameters $(a,b)$ as defined in \Cref{sec:CFN-data}, the average runtime for a single $p$-value was approximately $0.78$ seconds when $r=0$ and $0.97$ seconds when $r=20$. The chosen constraint set of polynomials had negligible effect on these runtimes.

Finally, the MLE computations presented in \Cref{sec:comparison} were carried out using \texttt{Julia} (version {1.10.3}).

\section{Conclusion}\label{sec:conclusions}

The SDL method offers a general-purpose framework for hypothesis testing for models defined by polynomial constraints. It is a strong and much needed technique, especially in settings where traditional frameworks are not available, such as when models have singularities or boundaries. 
Indeed, as illustrated by the trinomial submodels in \cref{sec:trinomial}, the method's performance can closely match that of traditional deterministic tests, such as the likelihood ratio  or $\chi^2$, where they are justified, but is more widely applicable. By focusing on two well-studied types of algebraic models used in phylogenetic inference, our investigation confirms that the method performs well across different settings. 
While no alternative method matches its generality, our results emphasize that thoughtful implementation choices, particularly around the key elements of constraint specification, kernel construction, and symmetrization, are necessary to enhance test performance.

In the case of the multispecies coalescent trinomial submodels, the SDL method not only recovers rejection regions that closely match those of conventional tests when available, but also remains valid at boundary and singular points, such as line crossings and nodes.  However, our simulations show that near singularities, the complement of the rejection region widens, making rejection more difficult. This indicates that the behaviour of the test is influenced not just by the zero set of the defining polynomials, but also by the size of constraint polynomial values near that set. As we see in those models, adding redundant constraints, especially near singularities and boundaries, can increase the power of the test, but how to choose these in a general manner requires further investigation. 

For the CFN model, we illustrated how the SDL method can be used both for hypothesis testing and selection among non-nested models.  This is especially useful in situations such as phylogenetic tree inference where the three possible four-leaf topologies give rise to intersecting semi-algebraic sets.  In addition, the CFN model highlighted how the choice of generating polynomials for the defining ideal plays a key role. Generators that are completely distinguishing for the tree topology of interest lead to better-calibrated tests than partially distinguishing ones.  Furthermore, this case study illustrated how the addition of convex combinations or extra constraints can have mixed effects, sometimes improving and sometimes degrading performance. Both of these issues raises the question of whether it is possible to develop a principled method for constraint choice. 

Both types of models that we explore in this paper are relatively small, in terms of ambient dimension, in terms of the number of constraints, and in terms of the constraint degrees. The method presents computational challenges for moving to higher dimensional, and higher degree, settings. In particular, full symmetrization is infeasible for large degree constraints, which requires larger $m$, but our results indicate that partial symmetrization using a modest number of permutations performs well in practice.  This raises an important theoretical question: How many permutations are sufficient to approximate the fully symmetrized kernel, and how does this number scale with dimension and degree?

While our case studies were chosen from evolutionary biology, they highlight that the SDL method fills a critical methodological gap in statistics for any semi-algebraic model. However, its performance is intimately tied to both algebraic and geometric aspects of the model. Future work under the lens of algebraic geometry would be helpful to develop a more complete theoretical understanding of how types of singularities and constraint choices influence the behaviour of the method, especially in higher dimensional settings in which visualisation is difficult. Such developments would further enhance the utility of the SDL method for both hypothesis testing and model selection in phylogenetics and other fields.  

\subsection*{Acknowledgements} This research began while the authors were visiting the Institute for Mathematical and Statistical Innovation (IMSI), Fall 2023 Semester Program on Algebraic Statistics and Our Changing World, supported by the National Science Foundation under Grant No. DMS-1929348. It continued at the Institute for Computational and Experimental Mathematics (ICERM), under NSF Grant No. DMS-1929284, while some of the authors were in residence at the Fall 2024 Semester Program on Theory, Methods, and Applications of Quantitative Phylogenomics.  EG was supported by National Science Foundation grant DMS-1945584. JAR was supported by National Science Foundation grant DMS-2051760. The views expressed in this article are those of the author(s) and do not reflect the official policy or position of the U.S. Naval Academy, Department of the Navy, the Department of Defense, or the U.S. Government.

\newpage
\appendix

\section{Coalescent models}\label{sec:Coalescent}

\subsection{The multispecies coalescent model}\label{ssec:msc}

\begin{figure}[ht]
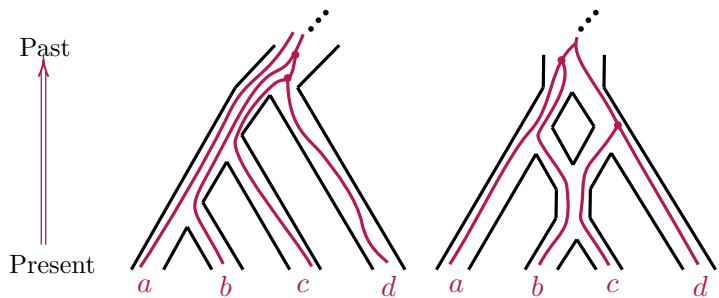

\centering
     \include{coalescent_tree_network}
     \vspace{-1cm}
\caption{ Gene trees (in red) form within a species tree and network (black `tubes')}\label{fig:ILS}   \vspace{-0.4cm} 
\end{figure}

The (network) multispecies coalescent (MSC) \cite{Pamilo1998,Meng2009} models the formation of gene trees within species trees or networks, for example as in \cref{fig:ILS}. A gene tree describes the history of a single genetic locus drawn from individuals in several extant species, as lineages trace back through individuals in the ancestral species populations, coalescing at common ancestors. While constrained by the species relationships, a gene tree may differ from them significantly, due to multiple gene lineages
remaining distinct in an ancestral population until coalescence between less closely related species becomes possible. This effect, called \emph{incomplete lineage sorting,} is most pronounced when edges in the species tree or network are short (in number of generations) or population sizes are large (since bottlenecks promote coalescence).

Considering only trees or networks relating four species, a \emph{quartet Concordance Factor (CF)} for a fixed network is the vector of probabilities of the 3 possible unrooted topological gene trees shown in \cref{fig:3topologies} that may arise under the coalescent model. To be precise, we fix the order
$$CF=(p_{12|34},p_{13|24},p_{14|23}),$$
for some fixed designation of species $1,2,3,4$.

Under the MSC model, the form of CFs arising from metric species networks with certain topological structures has been studied in several papers,  leading to the four submodels of $\Delta^2$ depicted in \cref{fig:4models}. Model 1 is all CFs that may arise from a species network with a cut edge separating species $1,2$ from $3,4$ \cite{Allman2011}. Model 2 is all CFs that arise from a species tree with the same species separation \cite{Allman2011, MAR2019}.
Model 3 is all CFs from a network with a cut edge separating the species into some pair of sets of two \cite{ABMR2022}, and Model 4 all CFs from a tree with such a cut edge \cite{MAR2019}. Models 3 and 4 are obtained from Models 1 and 2 by considering the union of models obtained by permuting CF entries. It is also known that all points in $\Delta^2$ arise as CFs of some networks \cite{Banos2019}, so rejecting these models in a hypothesis test is a natural way to find evidence for gene flow or hybridization \cite{BjornerEtAl2024}.

From genomic sequences, one may infer many gene trees and from them estimate frequencies of the three possible quartet gene tree topologies. A hypothesis test with one of the above null models can then, give insight into an unknown network structure. For instance, rejection of Model 3 suggests that the data did not arise on a tree, so hybridization or introgression occurred among the species.
Specialized test distributions for null hypotheses of  Model 2 and 4 are derived in \cite{MAR2019} and for Model 3 in \cite{ABMR2024} that improve upon a naive use of a standard distribution that ignores the singularities and boundaries of the models. (Model 1 can be tested with a standard distribution, as it lacks any irregularities.) However, these models are all semialgebraic, and the SDL approach offers an alternative testing framework without the need for such detailed work for each model.

\section{Deterministic tests}\label{sec:dettests}
For comparison to the rejection region plots produced by the SDL tests in \Cref{sec:pitfalls} we show those for deterministic tests for models 1-4 and Hardy-Weinberg with sample size $n=300$. For Model 1 this is a standard Likelihood Ratio test; for Models 2, 3, and 4 we use the tests implemented in \texttt{MSCquartets} \cite{MSCquartets_bioinformatics2020} as
``T1'', ``cut'', and ``T3''. These last all use non-standard test distributions for the Likelihood ratio statistic, to deal with the boundaries and singularities of these models. For the Hardy-Weinberg 2-allele model we use a standard chi-squared test.

\begin{figure}[!htb]
    \centering
\includegraphics[width=.2\textwidth]{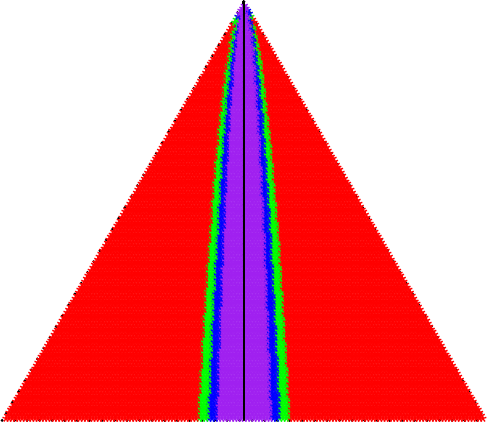}\ \ 
\includegraphics[width=.2\textwidth]{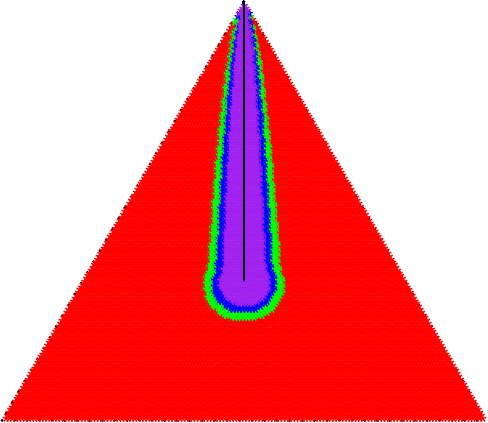}\ \ 
\includegraphics[width=.2\textwidth]{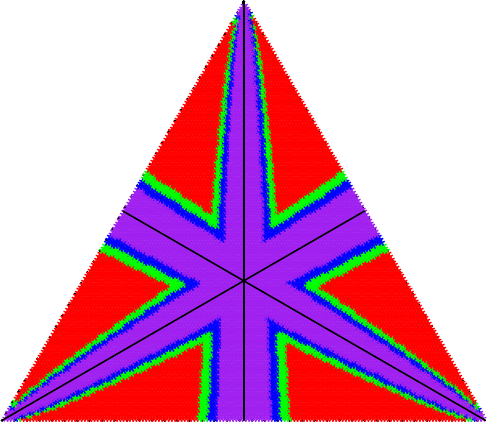} 

\ 

\includegraphics[width=.2\textwidth]{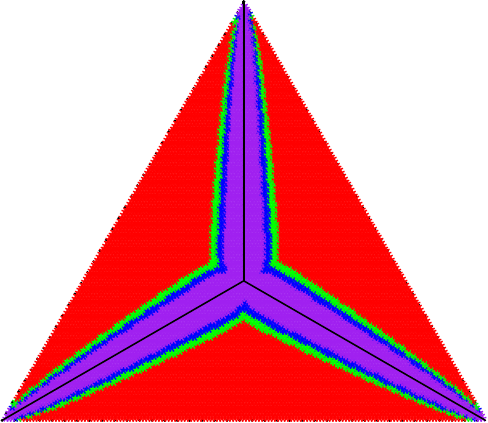}\ \ 
\includegraphics[width=.2\textwidth]{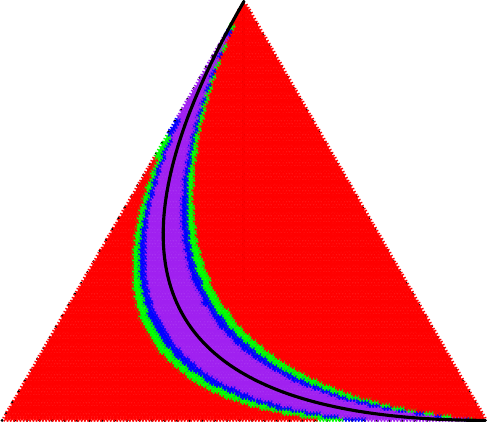}
\caption{Rejection regions for Models 1, 2, 3, 4, and Hardy-Weinberg 2-alleles, using deterministic tests, as described in text, with sample size $n=300$.}
\label{fig:dettests}
\end{figure}

\section{Additional Details on the CFN model}
\subsection{Generating sets for the CFN ideal}\label{app:CFN-generators}

This section details the derivation of generating sets for the ideal associated with the 4-taxon Cavender-Farris-Neyman (CFN) model presented in \Cref{ssec:CFNsemi,sec:CFN-data}, and provides explicit formulas for them. 

Although initially presented in $2^4=16$ dimensions, using pairwise equalities of certain pattern probabilities, the model can also be presented  in 8 dimensions. Specifically, let
\begin{itemize}
    \item $I_T \subset {\mathbb{C}}[p_{ijkl}]$ be the full phylogenetic ideal in the ring of $16$ site pattern probabilities $p_{ijkl}$, $i,j,k,l \in \{0,1\}$.
    \item $\overline{I}_{T} \subset {\mathbb{C}}[\overline{p}_{xyzt}]$, the ideal in the ring of $8$ symmetrized pattern probabilities $\overline{p}_{xyzt}$ (e.g., $\overline{p}_{xxxx} = p_{0000} + p_{1111}$). 
\end{itemize}
Our data consists of $n$ independent multinomial samples, with parameter
\begin{equation}\label{eq:reduced_param_revised} 
\overline{p} = \left(\overline{p}_{xxxx}, \overline{p}_{xxxy}, \overline{p}_{xxyx}, \overline{p}_{xxyy}, \overline{p}_{xyxx}, \overline{p}_{xyxy}, \overline{p}_{xyyx}, \overline{p}_{xyyy}\right) \in\Delta^{7},
\end{equation}
and we seek generators of $\overline{I}_{T}$ in the $\overline{p}$ coordinates.

With ${\mathbb{C}}[p]:={\mathbb{C}}[p_{ijkl}]$, and ${\mathbb C}[{\overline{p}}]: = {\mathbb{C}}[\overline{p}_{xyzt}]$,
the relationship between $I_T$ and $\overline{I}_{T}$ is given by the ring homomorphism $\psi: {\mathbb{C}}[{\overline{p}}] \to {\mathbb{C}}[p]$ that substitutes each $\overline{p}_{xyzt}$ with its definition as a sum of two $p_{ijkl}$:
\begin{equation*}
    \psi(\overline{p}_{xxxx}) = p_{0000} + p_{1111}, \quad
    \psi(\overline{p}_{xxxy}) = p_{0001} + p_{1110}, \quad \ldots \quad
    \psi(\overline{p}_{xyyy}) = p_{0111} + p_{1000}.
\end{equation*}
Let $L_{symm} \subset {\mathbb{C}}[p]$ be the ideal generated by the 8 linear symmetry relations,  $$p_{0000} - p_{1111} = 0,\  p_{0001} - p_{1110} = 0,\ \ldots, \ p_{0111} - p_{1000} = 0.$$ Then
\begin{equation}\label{eq:ideal_relation_IT_Ipbar}
I_T = \psi(\overline{I}_{T}) + L_{symm}.
\end{equation}

We will show that $\overline{I}_{T}$ is generated by the linear polynomial, $\ell := (\overline{p}_{xxxx} + \overline{p}_{xxxy} + \overline{p}_{xxyx} + \overline{p}_{xxyy} + \overline{p}_{xyxx} + \overline{p}_{xyxy} + \overline{p}_{xyyx} + \overline{p}_{xyyy}) - 1$, along with a set of quadratic polynomials. It then follows from Equation \ref{eq:ideal_relation_IT_Ipbar} that $I_T$ is generated by the symmetry ideal $L_{symm}$, the linear polynomial $\psi(\ell) = (\sum_{ijkl} p_{ijkl}) - 1$, and the $\psi$-images of the aforementioned quadratic polynomials (which remain quadratic in the $p_{ijkl}$ coordinates). The sets \cdd, \cdm, \cdr, \pdm and \pdr consist of variations of these quadratics in the $\overline{p}_{ijkl}$ coordinates.

For the tree $T=T_{12|34}$, we calculate generators for $\overline{I}_T$ using \texttt{Macaulay2} (version 1.21).
Below, we code the parametrization of the $\overline p_{xyzt}$ in terms of transformed edge lengths $\theta_i=e^{-2t_i}$, with
 \texttt{pxxxx}, \texttt{pxxxy}, etc., corresponding to the coordinates $\overline{p}_{xxxx}$, $\overline{p}_{xxxy}$, etc.. 

\begin{center} 
\begin{minipage}{\textwidth}
\begin{lstlisting}[language=Macaulay2, mathescape=true]
i1 : R  = QQ[$\theta$1, $\theta$2, $\theta$3, $\theta$4, $\theta$5]
i2 : Sp = QQ[pxxxx, pxxxy, pxxyx, pxxyy, pxyxx, pxyxy, pxyyx, pxyyy]

i3 : $\beta$ = $\theta$ -> ($\textcolor{codekeyword}{\mathtt{1}}$-$\theta$)/$\textcolor{codekeyword}{\mathtt{2}}$
i4 : $\alpha$ = $\theta$ -> ($\textcolor{codekeyword}{\mathtt{1}}$+$\theta$)/$\textcolor{codekeyword}{\mathtt{2}}$

i5 :  Pxxxx = $\alpha$($\theta$1)*$\alpha$($\theta$2)*$\alpha$($\theta$3)*$\alpha$($\theta$4)*$\alpha$($\theta$5) + $\alpha$($\theta$1)*$\alpha$($\theta$2)*$\beta$($\theta$3)*$\beta$($\theta$4)*$\beta$($\theta$5) + 
              $\alpha$($\theta$3)*$\alpha$($\theta$4)*$\beta$($\theta$1)*$\beta$($\theta$2)*$\beta$($\theta$5) + $\alpha$($\theta$5)*$\beta$($\theta$1)*$\beta$($\theta$2)*$\beta$($\theta$3)*$\beta$($\theta$4) -- p0000 + p1111
i6 :  Pxxxy = $\alpha$($\theta$1)*$\alpha$($\theta$2)*$\alpha$($\theta$3)*$\alpha$($\theta$5)*$\beta$($\theta$4) + $\alpha$($\theta$1)*$\alpha$($\theta$2)*$\alpha$($\theta$4)*$\beta$($\theta$3)*$\beta$($\theta$5) + 
              $\alpha$($\theta$3)*$\beta$($\theta$1)*$\beta$($\theta$2)*$\beta$($\theta$4)*$\beta$($\theta$5) + $\alpha$($\theta$4)*$\alpha$($\theta$5)*$\beta$($\theta$1)*$\beta$($\theta$2)*$\beta$($\theta$3) -- p0001 + p1110
i7 :  Pxxyx = $\alpha$($\theta$1)*$\alpha$($\theta$2)*$\alpha$($\theta$3)*$\beta$($\theta$4)*$\beta$($\theta$5) + $\alpha$($\theta$1)*$\alpha$($\theta$2)*$\alpha$($\theta$4)*$\alpha$($\theta$5)*$\beta$($\theta$3) + 
              $\alpha$($\theta$3)*$\alpha$($\theta$5)*$\beta$($\theta$1)*$\beta$($\theta$2)*$\beta$($\theta$4) + $\alpha$($\theta$4)*$\beta$($\theta$1)*$\beta$($\theta$2)*$\beta$($\theta$3)*$\beta$($\theta$5) -- p0010 + p1101
i8 :  Pxxyy = $\alpha$($\theta$1)*$\alpha$($\theta$2)*$\alpha$($\theta$3)*$\alpha$($\theta$4)*$\beta$($\theta$5) + $\alpha$($\theta$1)*$\alpha$($\theta$2)*$\alpha$($\theta$5)*$\beta$($\theta$3)*$\beta$($\theta$4) + 
              $\alpha$($\theta$3)*$\alpha$($\theta$4)*$\alpha$($\theta$5)*$\beta$($\theta$1)*$\beta$($\theta$2) + $\beta$($\theta$1)*$\beta$($\theta$2)*$\beta$($\theta$3)*$\beta$($\theta$4)*$\beta$($\theta$5) -- p0011 + p1100
i9 :  Pxyxx = $\alpha$($\theta$1)*$\alpha$($\theta$3)*$\alpha$($\theta$4)*$\alpha$($\theta$5)*$\beta$($\theta$2) + $\alpha$($\theta$1)*$\beta$($\theta$2)*$\beta$($\theta$3)*$\beta$($\theta$4)*$\beta$($\theta$5) + 
              $\alpha$($\theta$2)*$\alpha$($\theta$3)*$\alpha$($\theta$4)*$\beta$($\theta$1)*$\beta$($\theta$5) + $\alpha$($\theta$2)*$\alpha$($\theta$5)*$\beta$($\theta$1)*$\beta$($\theta$3)*$\beta$($\theta$4) -- p0100 + p1011
i10 : Pxyxy = $\alpha$($\theta$1)*$\alpha$($\theta$3)*$\alpha$($\theta$5)*$\beta$($\theta$2)*$\beta$($\theta$4) + $\alpha$($\theta$1)*$\alpha$($\theta$4)*$\beta$($\theta$2)*$\beta$($\theta$3)*$\beta$($\theta$5) + 
              $\alpha$($\theta$2)*$\alpha$($\theta$3)*$\beta$($\theta$1)*$\beta$($\theta$4)*$\beta$($\theta$5) + $\alpha$($\theta$2)*$\alpha$($\theta$4)*$\alpha$($\theta$5)*$\beta$($\theta$1)*$\beta$($\theta$3) -- p0101 + p1010
i11 : Pxyyx = $\alpha$($\theta$1)*$\alpha$($\theta$3)*$\beta$($\theta$2)*$\beta$($\theta$4)*$\beta$($\theta$5) + $\alpha$($\theta$1)*$\alpha$($\theta$4)*$\alpha$($\theta$5)*$\beta$($\theta$2)*$\beta$($\theta$3) + 
              $\alpha$($\theta$2)*$\alpha$($\theta$3)*$\alpha$($\theta$5)*$\beta$($\theta$1)*$\beta$($\theta$4) + $\alpha$($\theta$2)*$\alpha$($\theta$4)*$\beta$($\theta$1)*$\beta$($\theta$3)*$\beta$($\theta$5) -- p0110 + p1001
i12 : Pxyyy = $\alpha$($\theta$1)*$\alpha$($\theta$3)*$\alpha$($\theta$4)*$\beta$($\theta$2)*$\beta$($\theta$5) + $\alpha$($\theta$1)*$\alpha$($\theta$5)*$\beta$($\theta$2)*$\beta$($\theta$3)*$\beta$($\theta$4) + 
              $\alpha$($\theta$2)*$\alpha$($\theta$3)*$\alpha$($\theta$4)*$\alpha$($\theta$5)*$\beta$($\theta$1) + $\alpha$($\theta$2)*$\beta$($\theta$1)*$\beta$($\theta$3)*$\beta$($\theta$4)*$\beta$($\theta$5) -- p0111 + p1000

i13 : P = {Pxxxx, Pxxxy, Pxxyx, Pxxyy, Pxyxx, Pxyxy, Pxyyx, Pxyyy};
i14 : f = map(R, Sp, P);

\end{lstlisting}
\end{minipage}
\end{center}


\paragraph{\textbf{Minimal generating sets}}
To compute the \emph{Partially Distinguishing Minimal} (\pdm) generating set for the ideal $\overline{I}_T$ in the $\overline{p}_{xyzt}$ coordinates, we compute a minimal generating set for the kernel of the homomorphism $\texttt{f}$.

\begin{center} 
\begin{minipage}{\textwidth}
\begin{lstlisting}[language=Macaulay2, mathescape=true]
i15 : I = kernel f;
i16 : M = mingens I;
i17 : netList flatten entries M
      +----------------------------------------------------------------------------+
o17 = |pxxxx + pxxxy + pxxyx + pxxyy + pxyxx + pxyxy + pxyyx + pxyyy - $\textcolor{codekeyword}{\mathtt{1}}$            |
      +----------------------------------------------------------------------------+
      |pxxyx*pxyxx - pxxyy*pxyxy + pxxxy*pxyyx + pxxyx*pxyyx + pxxyy*pxyyx +       |
      |pxyxx*pxyyx + pxyxy*pxyyx + pxyyx*pxyyx + pxxxy*pxyyy + pxyyx*pxyyy - pxyyx |
      +----------------------------------------------------------------------------+
      |pxxxy*pxyxx + pxxxy*pxyxy + pxxyx*pxyxy + pxxyy*pxyxy + pxyxx*pxyxy +       | 
      |pxyxy*pxyxy - pxxyy*pxyyx + pxyxy*pxyyx + pxxyx*pxyyy + pxyxy*pxyyy - pxyxy |
      +----------------------------------------------------------------------------+
\end{lstlisting}
\end{minipage}
\end{center}

The kernel computation yields three generators for $\overline{I}_T$: one linear, $\ell = \sum \overline{p}_{xyzt} - 1$, and two quadratic. The \pdm set consists of the quadratics:
\begin{align*}
  {\overline{h}_1 }&\ {=}\quad   \overline{p}_{xxyx}\overline{p}_{xyxx} - \overline{p}_{xxyy}\overline{p}_{xyxy} + \overline{p}_{xxxy}\overline{p}_{xyyx} + \overline{p}_{xxyx}\overline{p}_{xyyx} + \overline{p}_{xxyy}\overline{p}_{xyyx} + \overline{p}_{xyxx}\overline{p}_{xyyx} +\\
        &\quad\ \overline{p}_{xyxy}\overline{p}_{xyyx} + \overline{p}_{xyyx}\overline{p}_{xyyx} + \overline{p}_{xxxy}\overline{p}_{xyyy} + \overline{p}_{xyyx}\overline{p}_{xyyy} - \overline{p}_{xyyx} , \text{\ and}\\
  {\overline{h}_2} &\ {=}\quad  \overline{p}_{xxxy}\overline{p}_{xyxx} + \overline{p}_{xxxy}\overline{p}_{xyxy} + \overline{p}_{xxyx}\overline{p}_{xyxy} + \overline{p}_{xxyy}\overline{p}_{xyxy} + \overline{p}_{xyxx}\overline{p}_{xyxy} + \overline{p}_{xyxy}\overline{p}_{xyxy} -\\
        &\quad\ \overline{p}_{xxyy}\overline{p}_{xyyx} + \overline{p}_{xyxy}\overline{p}_{xyyx} + \overline{p}_{xxyx}\overline{p}_{xyyy} + \overline{p}_{xyxy}\overline{p}_{xyyy} - \overline{p}_{xyxy}.
\end{align*}

The \emph{Completely Distinguishing Minimal} (\cdm) generating set is formed by the linear combinations $\overline{h}_1 + \overline{h}_2$ and $\overline{h}_1 - \overline{h}_2$.

\medskip

\paragraph{\textbf{Completely Distinguishing Determinantal generating set}}
For group-based models such as the CFN, applying a linear change of coordinates (a Fourier or Hadamard transformation \cite{Hendy1989, Sturmfels2005}) is often advantageous. The new coordinates $q_{xyzt}$ simplify the parametrization and the description of $\overline{I}_T$. For the CFN model on the tree $T = T_{12|34}$, this change of coordinates is as follows:
 
\begin{center} 
\begin{minipage}{\textwidth}
\begin{lstlisting}[language=Macaulay2, mathescape=true]
i18 : Sq = QQ[qxxxx, qxxyy, qxyxy, qxyyx, qyxxy, qyxyx, qyyxx, qyyyy];

i19 : Qxxxx = Pxxxx + Pxxxy + Pxxyx + Pxxyy + Pxyxx + Pxyxy + Pxyyx + Pxyyy
o19 = $\textcolor{codekeyword}{\mathtt{1}}$
i20 : Qxxyy = Pxxxx - Pxxxy - Pxxyx + Pxxyy + Pxyxx - Pxyxy - Pxyyx + Pxyyy
o20 = $\theta$3*$\theta$4
i21 : Qxyxy = Pxxxx - Pxxxy + Pxxyx - Pxxyy - Pxyxx + Pxyxy - Pxyyx + Pxyyy
o21 = $\theta$2*$\theta$4*$\theta$5
i22 : Qxyyx = Pxxxx + Pxxxy - Pxxyx - Pxxyy - Pxyxx - Pxyxy + Pxyyx + Pxyyy
o22 = $\theta$2*$\theta$3*$\theta$5
i23 : Qyxxy = Pxxxx - Pxxxy + Pxxyx - Pxxyy + Pxyxx - Pxyxy + Pxyyx - Pxyyy
o23 = $\theta$1*$\theta$4*$\theta$5
i24 : Qyxyx = Pxxxx + Pxxxy - Pxxyx - Pxxyy + Pxyxx + Pxyxy - Pxyyx - Pxyyy
o24 = $\theta$1*$\theta$3*$\theta$5
i25 : Qyyxx = Pxxxx + Pxxxy + Pxxyx + Pxxyy - Pxyxx - Pxyxy - Pxyyx - Pxyyy
o25 = $\theta$1*$\theta$2
i26 : Qyyyy = Pxxxx - Pxxxy - Pxxyx + Pxxyy - Pxyxx + Pxyxy + Pxyyx - Pxyyy
o26 = $\theta$1*$\theta$2*$\theta$3*$\theta$4
\end{lstlisting}
\end{minipage}
\end{center}

The generating set for the ideal $\overline{I}_T$ in the $q_{xyzt}$ coordinates is found by computing the kernel of  $\texttt{g}$.

\begin{center} 
\begin{minipage}{\textwidth}
\begin{lstlisting}[language=Macaulay2, mathescape=true]
i27 : Q = {Qxxxx, Qxxyy, Qxyxy, Qxyyx, Qyxxy, Qyxyx, Qyyxx, Qyyyy};
i28 : g = map(R, Sq, Q);
i29 : netList entries gens kernel g
      +---------+-------------------+-------------------------+
o29 = |qxxxx - $\textcolor{codekeyword}{\mathtt{1}}$|qxxyy*qyyxx - qyyyy|qxyyx*qyxxy - qxyxy*qyxyx|
      +---------+-------------------+-------------------------+
\end{lstlisting}
\end{minipage}
\end{center}

Transforming back to the $\overline{p}_{xyzt}$ probability coordinates, the linear polynomial $q_{xxxx} - 1$ becomes  $\ell = \sum \overline{p}_{xyzt} - 1$,
and the two quadratics yield the \emph{Completely Distinguishing Determinantal} (\cdd) set. 

\begin{center}
\begin{minipage}{\textwidth}
\begin{lstlisting}[language=Macaulay2, mathescape=true]
i30 : qxxxx = $\textcolor{codekeyword}{\mathtt{1}}$; -- pxxxx + pxxxy + pxxyx + pxxyy + pxyxx + pxyxy + pxyyx + pxyyy
i31 : qxxyy = pxxxx - pxxxy - pxxyx + pxxyy + pxyxx - pxyxy - pxyyx + pxyyy;
i32 : qxyxy = pxxxx - pxxxy + pxxyx - pxxyy - pxyxx + pxyxy - pxyyx + pxyyy;
i33 : qxyyx = pxxxx + pxxxy - pxxyx - pxxyy - pxyxx - pxyxy + pxyyx + pxyyy;
i34 : qyxxy = pxxxx - pxxxy + pxxyx - pxxyy + pxyxx - pxyxy + pxyyx - pxyyy;
i35 : qyxyx = pxxxx + pxxxy - pxxyx - pxxyy + pxyxx + pxyxy - pxyyx - pxyyy;
i36 : qyyxx = pxxxx + pxxxy + pxxyx + pxxyy - pxyxx - pxyxy - pxyyx - pxyyy;
i37 : qyyyy = pxxxx - pxxxy - pxxyx + pxxyy - pxyxx + pxyxy + pxyyx - pxyyy;


i38 : M1 = matrix{{qxxxx, qxxyy},
                  {qyyxx, qyyyy}}

i39 : M2 = matrix{{qxyxy, qyxxy},
                  {qxyyx, qyxyx}}

i40 : F1 = det(M1)
o40 = - pxxxx^$\textcolor{codekeyword}{\mathtt{2}}$ + pxxxy^$\textcolor{codekeyword}{\mathtt{2}}$ + $\textcolor{codekeyword}{\mathtt{2}}$pxxxy*pxxyx + pxxyx^$\textcolor{codekeyword}{\mathtt{2}}$ - $\textcolor{codekeyword}{\mathtt{2}}$pxxxx*pxxyy - pxxyy^$\textcolor{codekeyword}{\mathtt{2}}$ - $\textcolor{codekeyword}{\mathtt{2}}$pxxxy*pxyxx - 
        $\textcolor{codekeyword}{\mathtt{2}}$pxxyx*pxyxx + pxyxx^$\textcolor{codekeyword}{\mathtt{2}}$  + $\textcolor{codekeyword}{\mathtt{2}}$pxxxx*pxyxy + $\textcolor{codekeyword}{\mathtt{2}}$pxxyy*pxyxy - pxyxy^$\textcolor{codekeyword}{\mathtt{2}}$  + $\textcolor{codekeyword}{\mathtt{2}}$pxxxx*pxyyx + 
        $\textcolor{codekeyword}{\mathtt{2}}$pxxyy*pxyyx - $\textcolor{codekeyword}{\mathtt{2}}$pxyxy*pxyyx - pxyyx^$\textcolor{codekeyword}{\mathtt{2}}$  - $\textcolor{codekeyword}{\mathtt{2}}$pxxxy*pxyyy - $\textcolor{codekeyword}{\mathtt{2}}$pxxyx*pxyyy + $\textcolor{codekeyword}{\mathtt{2}}$pxyxx*pxyyy + 
        pxyyy^$\textcolor{codekeyword}{\mathtt{2}}$  + pxxxx - pxxxy - pxxyx + pxxyy - pxyxx + pxyxy + pxyyx - pxyyy

i41 : F2 = det(M2)
o41 = - $\textcolor{codekeyword}{\mathtt{4}}$(pxxxy*pxyxx - pxxyx*pxyxx - pxxxx*pxyxy + pxxyy*pxyxy + 
          pxxxx*pxyyx - pxxyy*pxyyx - pxxxy*pxyyy + pxxyx*pxyyy)
\end{lstlisting}
\end{minipage}
\end{center}
%

%


Note that for the $F_i$ defined in lines \texttt{i38} and \texttt{i39} of the code above
$\langle\psi(F_i)\rangle +L_{\text{symm}} = \langle f_i\rangle+L_{\text{symm}}$,  where $f_1,f_2$ are the polynomials of \cref{eq:BD}. 
In other words, up to the symmetries in $L_{\text{symm}}$ and a constant factor,  $\psi(F_i)$ is the same as $f_i$, $i=1,2$.

\medskip
\paragraph{\textbf{Rank generating sets}}
The  probabilities $p_{ijkl}$ for the tree $T$ can be arranged into a $4\times 4$ matrix according to the partition $12|34$ of its leaves, where rows are indexed by the states of leaves $1,2$ and columns by the states of $3,4$: 

\begin{equation*}
    \mathrm{Flat}_{12|34}(p) =  \begin{pmatrix}
       p_{0000} & p_{0001} & p_{0010} & p_{0011} \\
       p_{0100} & p_{0101} & p_{0110} & p_{0111} \\
       p_{1000} & p_{1001} & p_{1010} & p_{1011} \\
       p_{1100} & p_{1101} & p_{1110} & p_{1111}
    \end{pmatrix}= 
    \frac{1}{2} \begin{pmatrix}
       \overline{p}_{xxxx} & \overline{p}_{xxxy} & \overline{p}_{xxyx} & \overline{p}_{xxyy} \\
       \overline{p}_{xyxx} & \overline{p}_{xyxy} & \overline{p}_{xyyx} & \overline{p}_{xyyy} \\
       \overline{p}_{xyyy} & \overline{p}_{xyyx} & \overline{p}_{xyxy} & \overline{p}_{xyxx} \\ 
       \overline{p}_{xxyy} & \overline{p}_{xxyx} & \overline{p}_{xxxy} & \overline{p}_{xxxx}
    \end{pmatrix}.
\end{equation*}
The matrix $\mathrm{Flat}_{12|34}(p)$ has rank at most $2$,  reflecting a conditional independence statement holding for leaves separated by the central edge of $T$ \cite{AllmanRhodes2008}. Therefore, its $3 \times 3$ minors are polynomials in the CFN ideal $I_T$. We use this to construct the \emph{Partially Distinguishing Rank} (\pdr) set, working from  the matrix $2 \cdot \mathrm{Flat}_{12|34}(p)$ expressed in the $\overline{p}$ variables.

\begin{center} 
\begin{minipage}{\textwidth}
\begin{lstlisting}[language=Macaulay2, mathescape=true]
i44 : Flat1234 = matrix{{pxxxx, pxxxy, pxxyx, pxxyy},
                        {pxyxx, pxyxy, pxyyx, pxyyy},
                        {pxyyy, pxyyx, pxyxy, pxyxx},
                        {pxxyy, pxxyx, pxxxy, pxxxx}};

i45 : I = minors($\textcolor{codekeyword}{\mathtt{3}}$, Flat1234);
i46 : netList primaryDecomposition I
      +-------------------------------------------------------------------+
o46 = |ideal (pxxyx*pxyxx - pxxyy*pxyxy - pxxxx*pxyyx + pxxxy*pxyyy,      |
      |      (pxxxy*pxyxx - pxxxx*pxyxy - pxxyy*pxyyx + pxxyx*pxyyy)      |
      +-------------------------------------------------------------------+
      |ideal (pxyxy - pxyyx, pxyxx - pxyyy, pxxxy - pxxyx, pxxxx - pxxyy) |
      +-------------------------------------------------------------------+
      |ideal (pxyxy + pxyyx, pxyxx + pxyyy, pxxxy + pxxyx, pxxxx + pxxyy) |
      +-------------------------------------------------------------------+
\end{lstlisting}

\end{minipage}
\end{center}

The ideal generated by all $3 \times 3$ minors of $2 \cdot \mathrm{Flat}_{12|34}(p)$ (\texttt{Flat1234} in the code) is not prime. The first component in the primary decomposition \texttt{o46} corresponds to the CFN model. The quadratic polynomials from this component form the \pdr set:
\begin{align*}
   {\overline{g}_1 }&\ {=}\ \overline{p}_{xxyx}\overline{p}_{xyxx} - \overline{p}_{xxyy}\overline{p}_{xyxy} - \overline{p}_{xxxx}\overline{p}_{xyyx} + \overline{p}_{xxxy}\overline{p}_{xyyy},\\
   {\overline{g}_2 }&\ {=}\ \overline{p}_{xxxy}\overline{p}_{xyxx} - \overline{p}_{xxxx}\overline{p}_{xyxy} - \overline{p}_{xxyy}\overline{p}_{xyyx} + \overline{p}_{xxyx}\overline{p}_{xyyy}.
\end{align*}
{The \emph{Completely Distinguishing Rank} (\cdr) set consists of the polynomials $\overline{g}_1 + \overline{g}_2$ and $\overline{g}_1 - \overline{g}_2$.
}

\subsection{Additional results for Collection 1: Comparison of different constraint sets}
We provide additional results on the performance of the SDL test on data from Collection 1, supplementing \Cref{ssec:p-vals-Data-1,sec:performance-for-differeng-constraint-sets} of the main text.

\subsubsection{Aggregated $p$-value histograms}\label{ssec_app:p-vals-Data-1}

\begin{figure}[ht]
  \centering
  \includegraphics[scale=0.13]{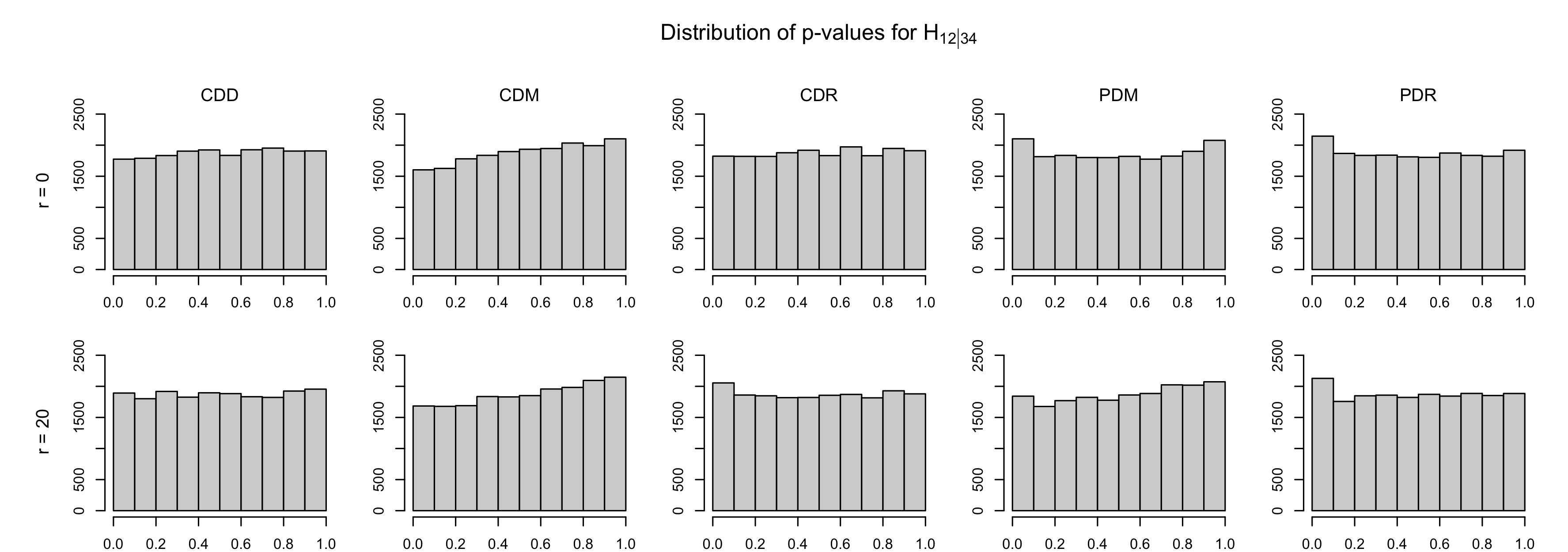}
  \caption{Aggregated $p$-values for a test of {the true null hypothesis} $H_{12|34}$ from datasets in Collection 1. Columns correspond to choices of defining polynomials. Rows correspond to the value of $r$.} 
  \label{fig_app:combined-pvalue-histogram-12-34}
\end{figure}

\begin{figure}[ht]
  \centering
  \includegraphics[scale=0.13]{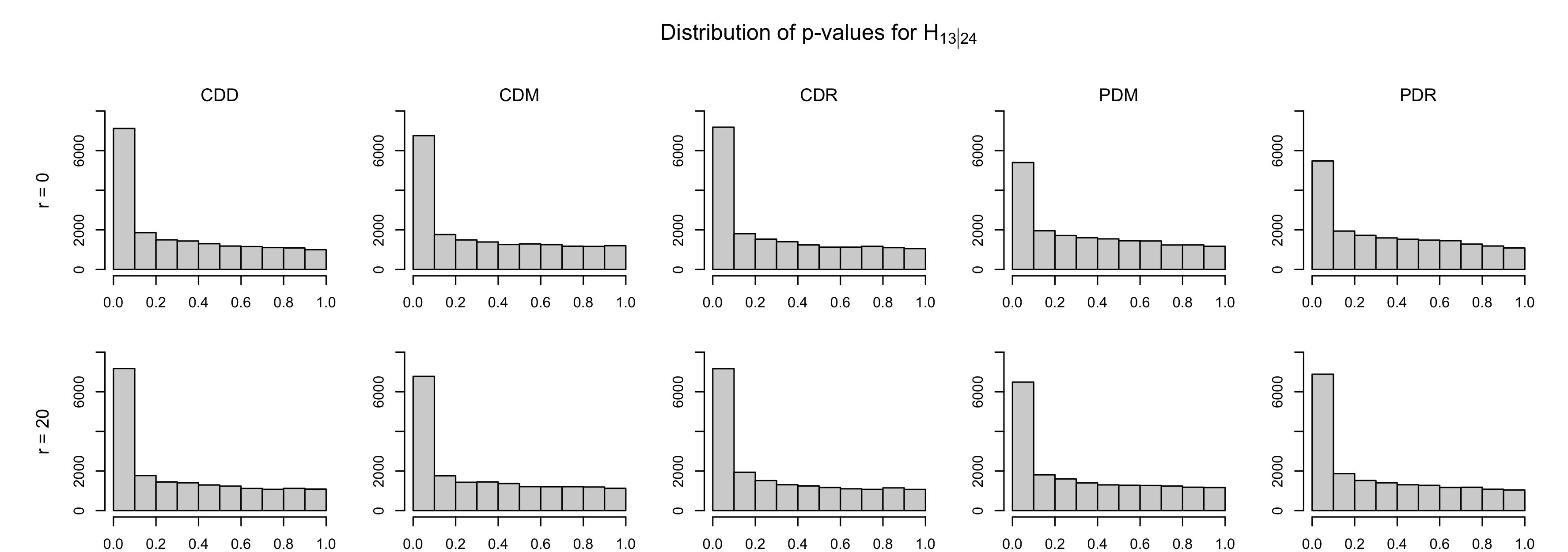}\\
  \caption{Aggregated $p$-values for a a test of $H_{13|24}$ (a false null hypothesis) from datatsets in Collection 1. Columns correspond to choices of defining polynomials. Rows correspond to the value of $r$.
   The test of $H_{14|23}$ produced similar results.
  }
  \label{fig_app:combined-pvalue-histogram-13-24}
\end{figure}

We analyse the performance of the five different choices of model constraints introduced in \cref{ssec:CFNsemi} by aggregating $p$-values across Collection 1.
\cref{fig_app:combined-pvalue-histogram-12-34} and \cref{fig_app:combined-pvalue-histogram-13-24} are analogous to the left and right parts of \cref{fig:combined-pvalue-histogram-12-34} in the main text, but also include the \cdd, \cdr and \pdr constraints.
These figures further support that the test behaviour is affected by the choice of model description.

In the $r=0$ case, \cref{fig_app:combined-pvalue-histogram-12-34} shows that the partially distinguishing sets \pdr and \pdm do not produce conservative tests due to an excess of small $p$-values when testing $H_{12|34}$. On the other hand, for both \cdd and \cdr, the $p$-value distribution appears to be close to uniform, and \cdm gives an especially conservative test, with an excess of large $p$-values (as previously seen in \Cref{ssec:p-vals-Data-1}).
Overall, completely distinguishing polynomials seem to produce a conservative test when $r=0$. 
%
%
In \cref{fig_app:combined-pvalue-histogram-13-24} we observe that for $r=0$, the completely distinguishing constraints had slightly greater concentration of $p$-values near zero compared to the partially distinguishing constraints, similarly to what we observed in \Cref{ssec:p-vals-Data-1}.

The effects of varying $r$ in both figures are relatively minor, and whether the effect was beneficial or not depended on whether the initial choice of constraints was partially distinguishing or completely distinguishing. When only partially distinguishing constraints were used, adding convex combinations improved performance by increasing the number of small $p$-values when testing the wrong model parameter (see \cref{fig_app:combined-pvalue-histogram-13-24}). The beneficial effect observed in \cref{fig:combined-pvalue-histogram-12-34} that increasing $r$ made the test more conservative for \pdm was not similarly observed for \pdr.
%
%
On the other hand, when completely distinguishing constraints were used, 
adding convex combinations constraints risks negatively affecting the quality of the $p$-values. Evidence for this can be seen in \cref{fig_app:combined-pvalue-histogram-12-34}, which shows that for \cdr, the test appears to be conservative when $r=0$, but not when $r=20$ due to an increased proportion of small $p$-values.
%


\subsubsection{SDL-based Phylogenetic Inference}
\label{sec_app:model-selection-performance-for-differeng-constraint-sets}
\begin{figure}[ht]
  \includegraphics[scale=0.12]{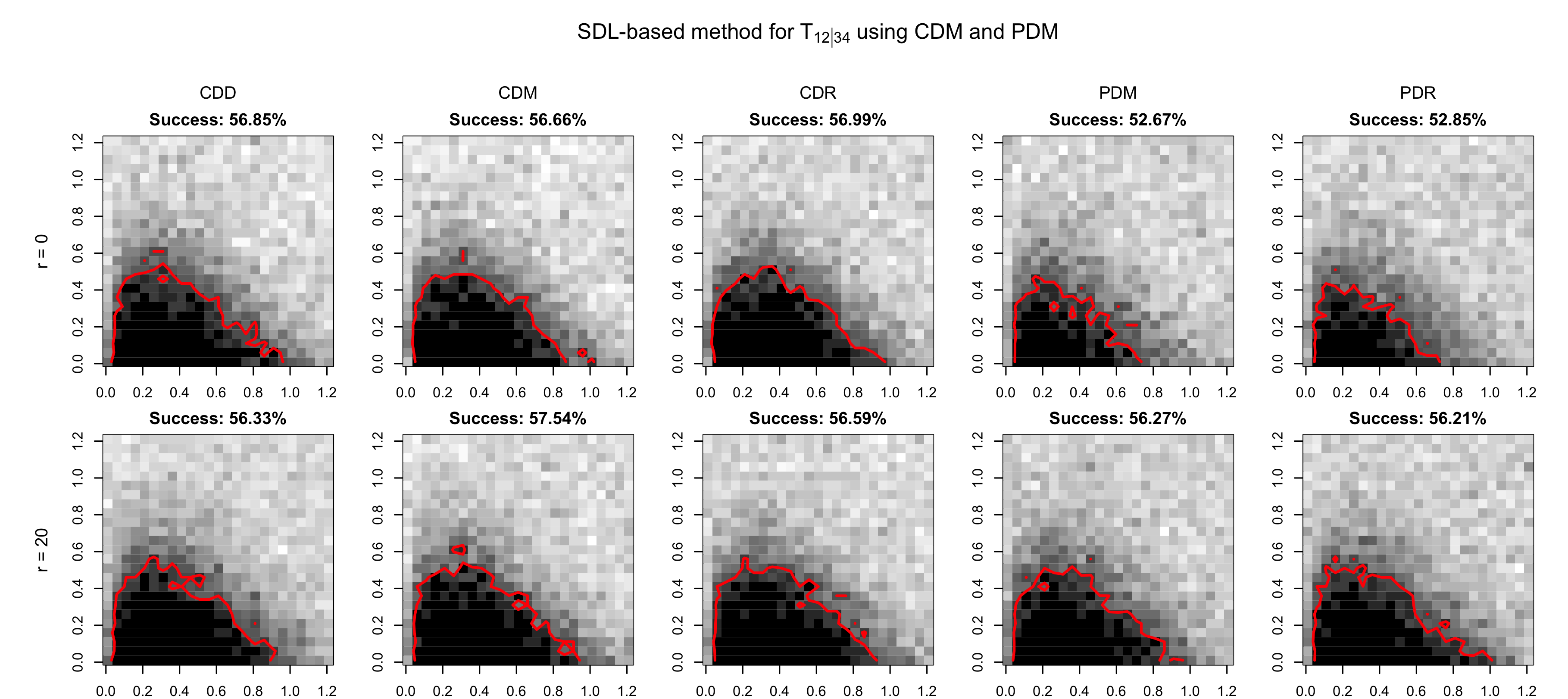}
  \caption{Performance of the SDL test for inferring the tree topology $T_{12|34}$. Columns correspond to
    different CFN model constraints (\cdd, \cdm, \cdr, \pdm, \pdr), and rows represent the number of convex
    combinations used, $r=0$ and $r=20$. 
    Grey levels represent the frequency of correctly inferring the topology for edge length pairs $(a,b)$ (black 100\%, white 0\%).
    }
  \label{fig_app:treespace_dataset1_1234}
\end{figure}

We analyse the performance of the SDL phylogenetic inference method for all five constraint sets in \cref{fig_app:treespace_dataset1_1234}, which is analogous to the left part of \cref{fig:treespace_inequality}, but includes the additional  sets \cdd, \cdr, and \pdr.

The conclusions from this figure are comparable to those of \cref{sec:comparison}. First, in the case $r=0$, the use of completely distinguishing constraint sets yields better performance than partially distinguishing sets (viz., columns 1,2,3, which have larger dark region and higher success percentages than columns 4,5). The completely distinguishing sets \cdd, \cdm, and \cdr all performed similarly: the differences in $p$-value distributions among them observed in \cref{fig_app:combined-pvalue-histogram-12-34} appeared to have no bearing on their performance for inference in this setting. 

The second important conclusion from \cref{fig_app:treespace_dataset1_1234} is that the performance of the partially distinguishing generating sets \pdr and \pdm increased when $r$ was increased from $0$ to $20$. Indeed, as a result of this improvement, all five sets performed comparably in the $r=20$ case. This improvement in performance for \pdm and \pdr is consistent with our observations in \cref{fig_app:combined-pvalue-histogram-13-24}, that --- at least for partially distinguishing constraints --- increasing $r$ appeared to increase the power of the test.

Note that variations in performance using different algebraic constraint sets for inference were previously observed \cite{Rusinko2012}, with symmetrizing ideal generators improving model selection.

\subsection{{Lack of long branch attraction bias}} 
\label{sec:disappearance-of-lba}
In this section we analyse the SDL test's behaviour for trees in the Felsenstein zone (see \cref{fig:treespace}), showing it differs from that of common methods used for phylogenetic inference. 
In particular, maximum parsimony \cite{Felsenstein1978} exhibits a \emph{long-branch attraction} bias in this region, in which the false topology $T_{13|24}$, pairing the two taxa on long pendent edges, is most frequently inferred. Similar bias is observed for maximum likelihood  \cite{susko-roger-2021,parks-goldman-2014} and previous algebraic methods \cite{fernandez2016}.

\begin{figure}[ht]
  \centering
  \includegraphics[scale=0.10]{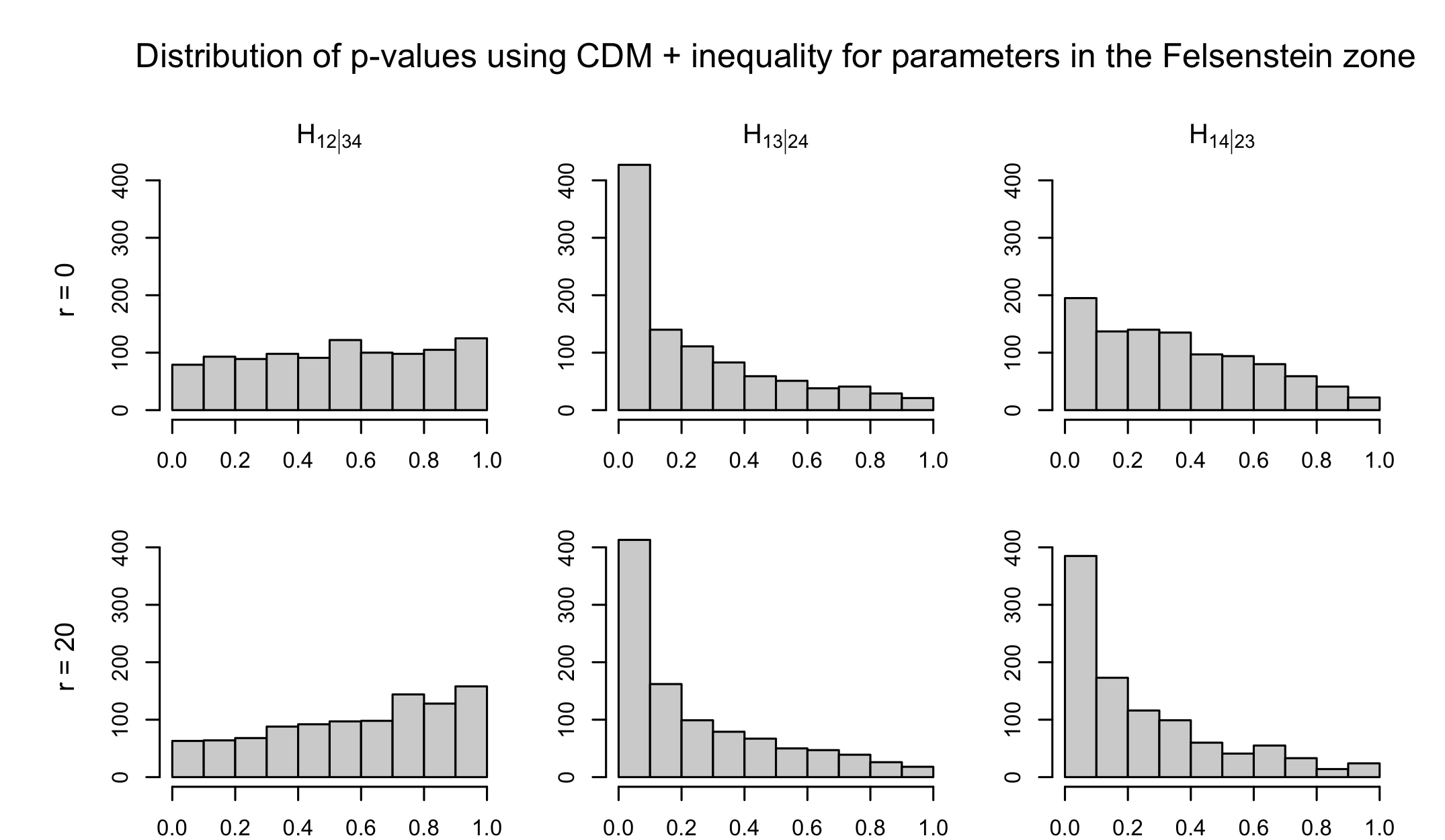}
  \includegraphics[scale=0.10]{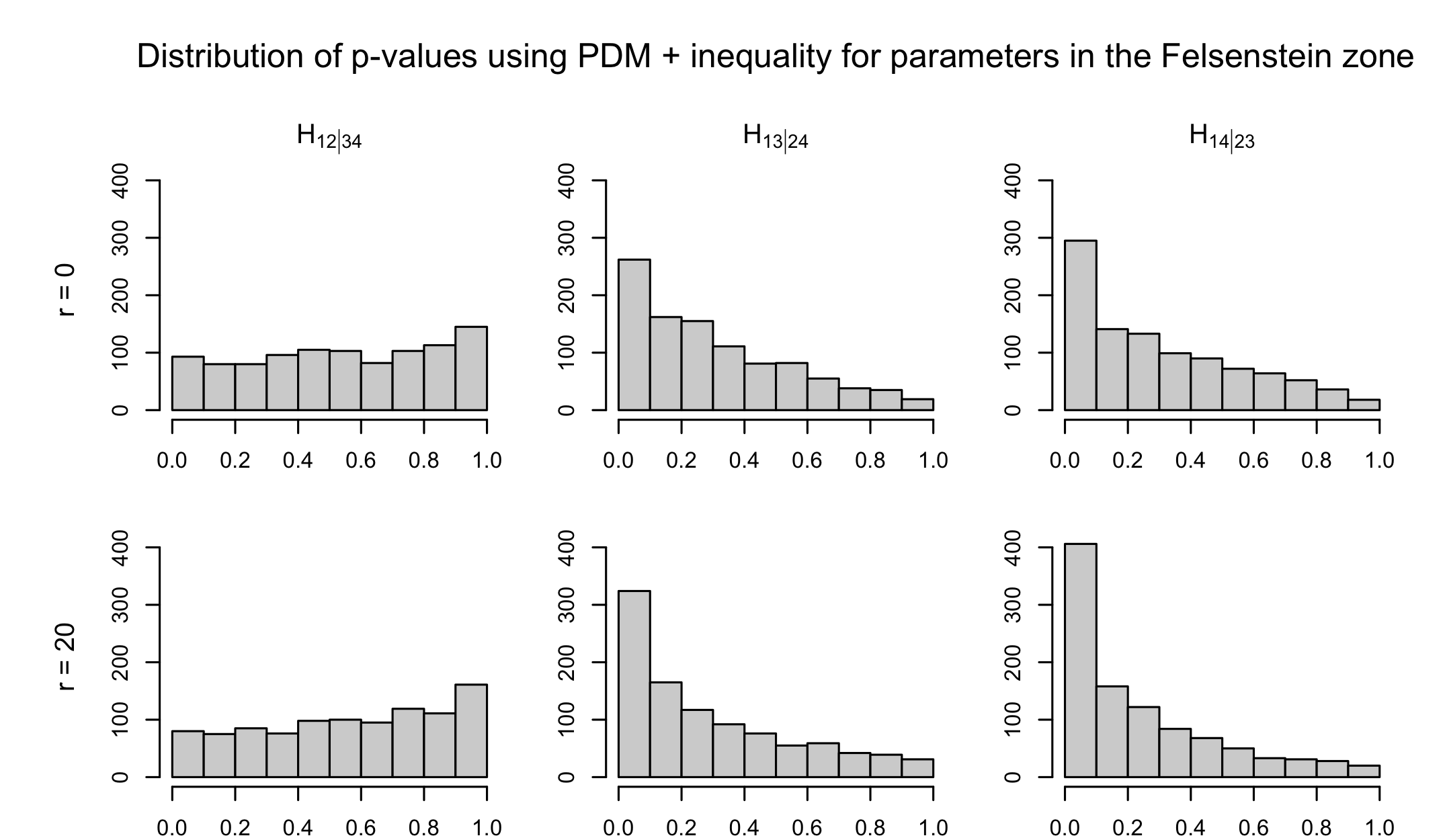}\\
  \caption{Histogram of $p$-values for \cdm (left) and \pdm (right) for a tree in the Felsenstein zone ($a=0.8$ and $b=0.05$) with $n=10000$ bp and $m=12$.}
  \label{fig:more-data-histogram-r-version}
\end{figure}

In \cref{fig:more-data-histogram-r-version}, we present $p$-values obtained from the SDL test using data generated from one tree with Felsenstein zone parameters $a=0.8$,  $b=0.05$, with $n=10,000$. We compared the SDL test using two different sets of constraints: \cdm (left plots) and \pdm (right plots); in both cases the internal edge inequality of \cref{eq:ineq_internal} was also used. We tested the three null hypotheses $H_{12|34}$, $H_{13|23}$, and $H_{14|23}$ (plot columns) for $r=0$ and $20$ (plot rows).

The choice of the \cdm versus \pdm constraints produces a marked discrepancy in test behaviour, especially for $r=0$. The first row of \cref{fig:more-data-histogram-r-version} ($r=0$) shows that the SDL test is much more likely to reject $H_{13|24}$ than $H_{14|23}$ for small test levels when using the \cdm constraints; on the other hand, the two false hypotheses are rejected at roughly equal frequency with the \pdm constraints. Both of these behaviours are in contrast with classical phylogenetic inference methods, which would tend to strongly support $H_{13|24}$ over $H_{14|23}$. 
Constraints \cdm and \pdm  produce almost-uniform distributions of $p$-values when testing $H_{12|34}$.

The second row of \cref{fig:more-data-histogram-r-version} shows that the addition of $r=20$ convex combinations for both the \cdm  and \pdm constraints reduced the asymmetry between test results of $H_{13|24}$ and $H_{14|23}$, and gave a more powerful test. Moreover, the test remained conservative for all values of $r$. However, a slight bias for $H_{13|24}$ appears, but only for the \pdm constraints.

In contrast to \cref{fig:combined-pvalue-histogram-12-34}, which showed increasing $r$ had little effect on the aggregated $p$-value distribution over a larger set of parameters,  \cref{fig:more-data-histogram-r-version} indicates that for certain parameter values, incorporating convex combinations can have a major effect --- in particular, by increasing the power of the SDL test. 

The general lack of bias toward $H_{13|24}$, together with the  overall conservativeness of the test, indicates that the  SDL test can perform quite well in the Felsenstein zone. Furthermore, the differing $p$-value distributions between \cdm and \pdm underscore how the choice of constraint sets can significantly impact SDL test performance.

The reduced long branch attraction bias for SDL is not unique to the parameters used for \cref{fig:more-data-histogram-r-version}, but persists across the Felsenstein zone.
In \Cref{table:lba-comparison}, we present the percentage of times that each of the three possible quartet topologies is inferred by the SDL, MLE and SVD methods from data in Collection 1, both across the whole treespace shown in \cref{fig:treespace} and only across the Felsenstein zone. 

These results show that for all three inference methods the topology $T_{13|24}$ was inferred more frequently than $T_{14|23}$ across the full parameter space, but especially in the Felsenstein zone. However, the SDL-based method showing the least susceptibility to this preference.
In the Felsenstein zone, there is an extreme bias for the SVD method, with $T_{13|24}$ inferred 46.07\% of the time, even more frequently than the 37.13\% for the true $T_{12|34}$.
For MLE, the effect was less pronounced, although $T_{13|24}$ was inferred noticeably more often than $T_{14|23}$ (15.74\% vs. 10.35\%). 
For the SDL-based method (using the \cdd constraints with $r=20$), the imbalance was proportionally smallest among the three methods (21.78\% vs. 18.02\%).

\begin{table}[ht]
\def\arraystretch{1.5}
\begin{tabular}{c|cccc|cccc}
    & \multicolumn{4}{c|}{Treespace}        & \multicolumn{4}{c}{Felsenstein Zone} \\ \cline{2-9} 
    & $12|34$ & $13|24$ & $14|23$ & \textsl{undecided} & $12|34$ & $13|24$ & $14|23$  & \textsl{undecided} \\ \hline
SDL & 69.51\% & 15.8\%  & 14.69\% & -         & 60.2\%   & 21.78\%  & 18.02\%  & -         \\
MLE  & 78.25\% & 8.86\%  & 6.95\%  & 5.94\%      & 71.88\%  & 15.74\%  & 10.36\%  & 2.01\%        \\
SVD & 65.82\% & 19.64\% & 14.54\% & -         & 37.13\%  & 46.07\%  & 16.8\%   & - 
\end{tabular}
\caption{Estimated tree topologies for the three methods SDL, MLE and SVD and the three topologies $12|34$, $12|34$ and $12|34$ in the entire treespace of \cref{fig:treespace} and in the Felsenstein Zone. The \textsl{undecided} column reports the percentage of times that MLE fails to distinguish between topologies. 
}
\label{table:lba-comparison}
\end{table}




\section{Technical Assumptions}\label{sec:mixed-degeneracy-appendix}

In order for the SDL test to be asymptotically valid for a particular hypothesis testing problem, there are a number of technical assumptions which need to be satisfied. In this section, we state the six conditions assumed in \cite{SDL2024}, and verify that they hold for the models considered in our paper. Despite their technical nature, these conditions are all straightforward to verify for the models we consider.

To state the conditions, let $X_1,\ldots,X_m \sim P_{\theta}$ be iid random variables, and let $\mu = (\mu_{1},\ldots,\mu_{p})^{\top}:= \E\left[h(X_{1},\ldots,X_{m})\right]$. In addition, define
$\sigma_{h,j}^{2} := \E\left[\left(h_{j}(X_1,\ldots,X_m)-\mu_{j}\right)^{2}
\right]$ and $\sigma_{g,j}^{2} := \E\left[\left(g_{j}(X_1,\ldots,X_m)-\mu_{j}\right)^{2} \right]$. For any positive $\beta$, define the function $\psi_{\beta}(x) = \exp(x^\beta)-1$, and for any random variable $Y$ define $\lVert Y \rVert_{\psi_{\beta}}:=\inf \left\{t>0 : \E\left[\psi_{\beta}\left( |Y|/t \right)  \right] \leq 1\right\}$.

The theoretical results in \cite{SDL2024} assume that there exists a constant $\beta\in (0,1]$ and a sequence $D_{1},D_{2},\ldots \geq 1$ such that:
\begin{enumerate}[label=(C\arabic*)]
  \item \label{item:C1} $\E\left[\left| h_{j}(X_1,\ldots,X_m)-\mu_{j} \right| \right] \leq
  \sigma_{h,j}^{2}D_{n}^{l}$ for all $j=1,\ldots,p$ and $l=1,2$.
  \item \label{item:C2} $\lVert h_j(X_1,\ldots,X_m)-\mu_{j} \rVert_{\psi_{\beta}}\leq D_{n}$ for
  all $j=1,\ldots,p$.
  \item \label{item:C3} There exists $\underline{\sigma}_{h}^{2}>0$ such that
  $\underline{\sigma}_{h}^{2}\leq \min_{1 \leq j \leq p}\sigma^{2}_{h,j}$.
  \item \label{item:C4} There exists $\underline{\sigma}_{g^{(1)}}^{2}>0$ such that
  $\underline{\sigma}^{2}_{g^{(1)}}\leq \min_{1 \leq j \leq p_{1}}\sigma_{j,g}^{2}$ for some positive integer $p_{1}<p$.
  \item \label{item:C5} There exists $k$ such that
  $\lVert g_{j}(X_{1})-\mu_{j}\rVert_{\psi_{\beta}}\leq n^{-k}D_{n}$ for all
  $j=p_{1}+1,\ldots,p$.
  \item \label{item:C6} $\E\left[\left| g_{j}(X_{1})-\mu_{j} \right|^{2+l}\right] \leq
  \sigma^{2}_{g,j}D^{l}_{n}$ for all $j=1,\ldots,p$ and $l=1,2$.
\end{enumerate}
In the above conditions, it is furthermore assumed that $2 \leq m \leq \sqrt{n}$,
$n \geq 4$, $p \geq 3$. (Note that in Model 1 of \Cref{sec:trinomial}, we have only $p=2$, but the assumption that $p\geq 3$ is not strictly necessary; for more detailed discussion of these assumptions, see \cite[Section 2.1]{SDL2024}).

Next, we check that conditions \ref{item:C1}-\ref{item:C6} hold:
\begin{itemize}
  \item First observe that condition \ref{item:C3} is satisfied whenever
  $h_{j}(X_1,\ldots,X_m)$ is not almost surely constant, which is straightforward
  to check for all the examples considered in the present paper, since in all our examples $X_1$ takes the form of a multinomial random variable with a single trial.
  \item Second, for the examples considered in the present paper, the state space $\mathcal{S}$ of $X_{1}$ is always a finite set, and hence
  $h_{j}(X_1,\ldots,X_m)$ is almost surely bounded. Together with \ref{item:C3}, this
  implies that we can choose finite $D_n$ satisfying
  \begin{equation*}
  D_n \geq \left( 
  \max_{\substack{1 \leq j \leq p \\ x_1, \ldots, x_m \in \mathcal{S}}} 
  \frac{\left| h_j(x_1, \ldots, x_m) - \mu_j \right|}{\sigma^2_{h,j}} 
  \right) \vee 1
  \end{equation*}
  for all $n \geq 1$. Moreover, for this choice of $D_n$, condition \ref{item:C1} holds
  \item Next we show that by possibly making each $D_n$ larger, it is possible to find $D_n$ large enough that \ref{item:C6} is also satisfied. On the one hand, if $\sigma^{2}_{g,j}=0$ then the inequality in \ref{item:C6} holds trivially with both sides equal to zero. On the other hand, for $j$ with $\sigma^{2}_{g,j}>0$, the inequality in \ref{item:C6} is satisfied if 
    \begin{equation*}
      D_{n} \geq \max_{l\in \left\{1,2\right\}}\left(\max_{\substack{j : \sigma^{2}_{g,j}>0 \\  }}\ \max_{ x_1\in \mathcal{S}} 
      \frac{\left| g_j(x_1) - \mu_j \right|^{2+l}}{\sigma^2_{g,j}}\right)^{\frac{1}{l}},
  \end{equation*}
  and without loss of generality we can assume this inequality holds since right-hand side is finite (due to the maximums being taken over finite sets).
  
  \item Furthermore, we will show that the terms of the sequence $D_1,D_2,\ldots$ can also be chosen large enough to satisfy \ref{item:C2}. To see this, write $Y = h_{j}(X_1,\ldots,X_m)$ and observe that since $Y$ has finite state space,  there exists finite $C_j$ such that $|Y| \leq C_{j}$ almost surely. It then follows by definition of $\lVert\cdot\rVert_{\psi_{\beta}}$ that $\lVert Y \rVert_{\psi_{\beta}}\leq C_{j}/\sqrt[\beta]{\log(2)}$. Again, without loss of generality, $D_n$ may be chosen so that $D_{n} \geq C_{j}/\sqrt[\beta]{\log(2)}$ which is sufficient to imply \ref{item:C2}.

  \item Finally, it remains to consider conditions \ref{item:C4} and \ref{item:C5}, which together are referred to as the \textit{mixed degeneracy} conditions in \cite{SDL2024}. In fact, there is nothing to show: for all the examples considered in the present paper, we have $p=O(1)$ as $n \to \infty$, and as a consequence of this, conditions  \ref{item:C4} and \ref{item:C5} hold trivially, as discussed in \cite[Section
  2.1]{SDL2024}.

\end{itemize}

\printbibliography

\end{document}